\documentclass[aip,cha,reprint,amsmath,amssymb]{revtex4-2}
\usepackage{graphicx}
\usepackage{capt-of}
\usepackage[T1]{fontenc}
\usepackage{lmodern}
\usepackage{url}
\usepackage[dvipsnames]{xcolor}
\usepackage{tikz}
\usetikzlibrary{arrows.meta,positioning,calc,decorations.pathreplacing,fit}
\usepackage[hidelinks]{hyperref}
\hypersetup{pdftitle={Chaos and bifurcations in continuous language-model feedback}, pdfauthor={Timothy D. Jones}}
\graphicspath{{figures/}}
\newcommand{\eps}{\varepsilon}
\newcommand{\sm}[1]{supplementary material, ``#1''}
\newcommand{\smref}[1]{}
\newcommand{\journalonly}[1]{}
\begin{document}

\title{Chaos and bifurcations in continuous language-model feedback}
\author{Timothy D. Jones}
\email{tim@praxagent.ai}
\affiliation{Praxagent}
\date{19 October 2026}

\begin{abstract}
Feeding a pretrained language model the expected embedding of its own next-token distribution turns it into a
discrete-time dynamical system with continuous states. Across Qwen2.5, OLMo-2 and Llama-3 models from 0.5B to 72B parameters, temperature
and activation steering move this loop between repeating text and chaos, with positive finite-time Lyapunov exponents. The loop's
derivative factorizes into the inverse temperature, the covariance of token embeddings under the next-token
distribution (a cross-covariance for untied embeddings) and the network's Jacobian. In one model studied in detail, along recorded orbits, the growth that steering adds comes
mainly from the readout, where next tokens compete. The periodic windows show the quantitative signatures of low-dimensional maps, from saddle-node scaling and
Feigenbaum's constant to shrimp-shaped domains; passage times predicted in advance from the model's derivatives came within one return-map iterate of the measured passage times in 42 of 48 cases over two prospective stages (24 and 18 of 24), and interval arithmetic proves attracting periodic orbits of the full transformer
map. Numerical reproducibility depends on the dynamical regime: changing only the batch layout changed the text of 173 of 192
chaotic float64 loops in the smallest model, compared with only 7 of 192 periodic ones. These results show that bifurcation theory can make
quantitative predictions about continuous language-model feedback, with rigorous guarantees for selected periodic orbits of the full transformer.
\end{abstract}

\maketitle

\begin{quotation}
Several recent inference methods feed a language model's continuous output back as its input. Such a loop is a
dynamical system, and we analyze it as one, in released models of 0.5 to 72 billion parameters, each reading
the probability-weighted average of its own next-token embeddings. Depending on the sampling temperature and on a
steering vector added inside the network, the model either settles into a repeating loop of text or wanders
chaotically. This behavior is organized as in low-dimensional maps: periodic windows accumulate where a periodic
orbit vanishes, and period-doubling ratios approach Feigenbaum's constant. The network's own derivatives bound where chaos
can appear and predict how long trajectories linger near the vanished orbit. The tools of
nonlinear dynamics thus apply directly to the inference loops of released language models.
\end{quotation}

\begin{figure}[t]
\centering
\includegraphics[width=\columnwidth]{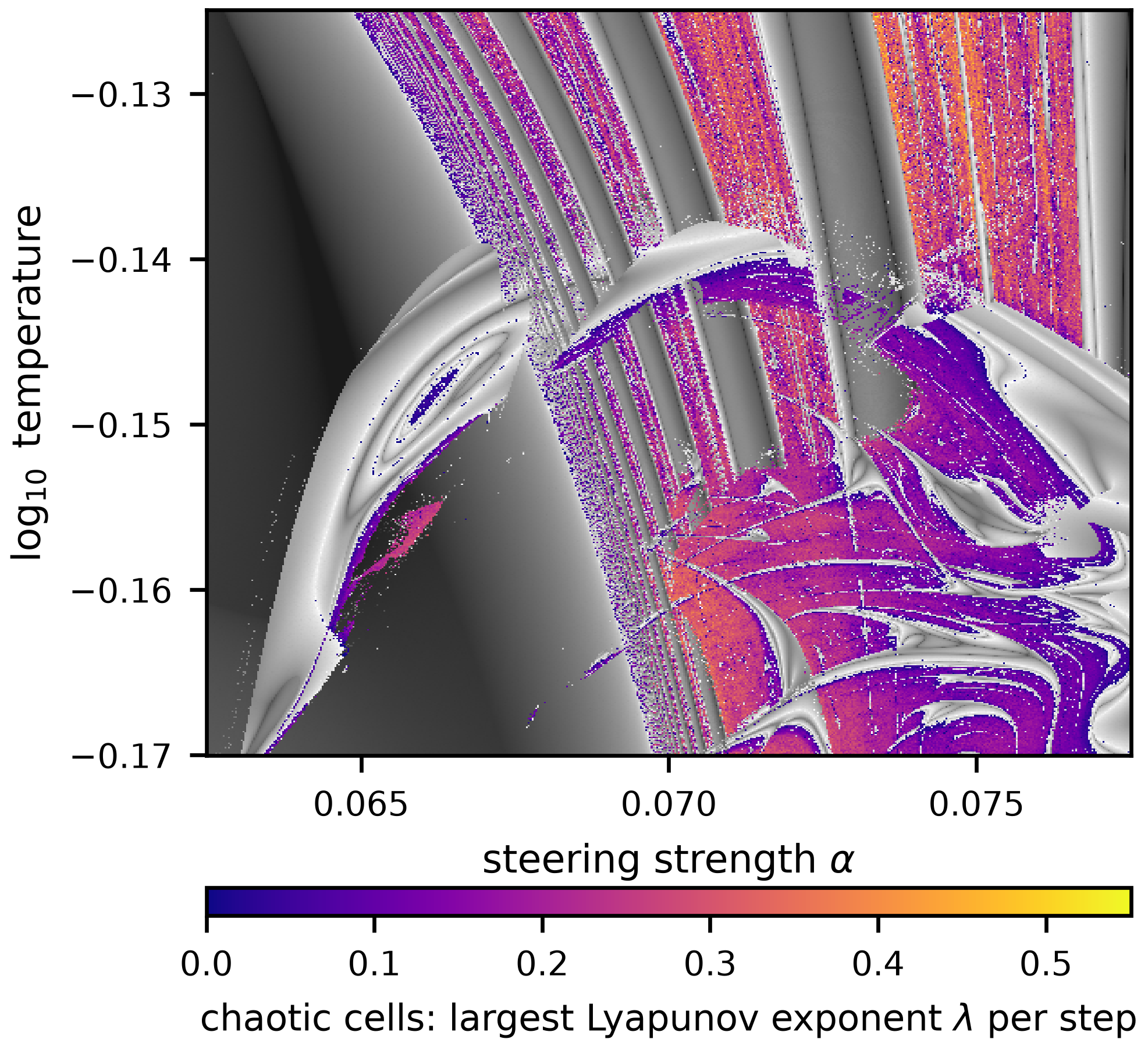}
\caption{Chaos and repeating text in a language model fed its own output: Qwen2.5-0.5B-Instruct, steered along a
philosophy direction at layer 12 with strength $\alpha$, at temperature $T$ ($512\times512$ float64 cells;
Sec.~\ref{sec:map}). Colored cells are chaotic by a finite-horizon test (largest exponent above $10^{-3}$ and no period
up to 32; Sec.~\ref{sec:chaos}), colored by their largest Lyapunov exponent. In gray cells the loop settles into a
repeating text, darker where it contracts more strongly; 4.4\,\% of the cells, also gray, have neither a detected period
nor an exponent above $10^{-3}$. Figure~\ref{fig:stripes}(b) names the loops.}
\label{fig:teaser}
\end{figure}

\section{Introduction}
\label{sec:intro}
Several recent inference methods feed a language model its own continuous output. Soft Thinking replaces each sampled token by the expected embedding under the next-token distribution
\cite{zhang2025softthinking} (its published implementation first keeps only the most probable tokens and renormalizes;
the map studied here uses the full distribution, which keeps it smooth), latent-reasoning
methods feed hidden states back \cite{hao2024coconut}, and recurrent-depth models iterate a block toward a fixed point
\cite{geiping2025recurrent}. Each turns inference into an iterated map, and an iterated map can be chaotic. Feedback
itself is not the novelty, since ordinary autoregressive decoding feeds each chosen token back, and the map studied here
reduces to greedy decoding as $T\to0$ (away from ties between leading tokens); what these methods change is that a continuous state is fed back. We hold the
prompt and the fed-back window fixed so that the loop becomes an autonomous map whose orbits and bifurcations can be
computed, and we measure the consequences for answers in a soft-thinking pipeline with a growing context. This
fixed-prefix, fixed-window soft-token map differs both from ordinary autoregressive generation, whose context grows,
and from recurrent-depth (looped) transformers, which repeatedly update hidden states through a shared block
\cite{dehghani2019universal,giannou2023looped,geiping2025recurrent}; dynamical methods apply to both kinds of loop, but
their update operators, and therefore their derivatives and stability conditions, differ. Our claim,
in short: fed back through its own soft next-token distribution, a pretrained language model defines a discrete-time
map with continuous states whose dynamics (periodic windows, period-doubling cascades, chaotic regimes and
shrimp-shaped windows, the ``shrimps'' named by Gallas \cite{gallas1993structure}) can be mapped, predicted from derivatives at selected transitions and in places proved; their
consequences for inference depend on the feedback architecture and the numerical regime. Chaos-like
sensitivity \cite{li2025quasilyapunov,jaca2026chaos,verazuniga2026selffeeding}, temperature-driven changes of output
statistics \cite{arnold2024phase,nakaishi2024phase,sun2025phase,ruan2026criticality}, repetition loops
\cite{holtzman2019degeneration,xu2022loop} and attractors of paraphrasing and conversation
\cite{wang2025attractor,ko2026attractor} have been measured in generated text, and a noisy effective map has been
fitted to a hidden-state coordinate \cite{johnson2026temperature}, but none of these computes the orbits, multipliers
or bifurcations of the model's own map, and, in literature searches of 12 and 15 October 2026, we found no earlier computer-assisted proof of a periodic
orbit of such a map of a transformer language model. Equilibrium, looped and recurrent-depth models have been analyzed through
fixed points and Jacobians \cite{bai2019deq,bai2021stabilizing,tomihari2025recurrent,yang2026stars,luo2026prediction},
and probability-weighted embeddings are fed back in training \cite{zeng2025ponderlm} and solved for as equilibria in
non-autoregressive decoding \cite{zheng2023deqnar}. Backour shows that the softmax readout and the feedback of such
mixtures form a dynamical system that either amplifies small differences until one component dominates or contracts
different mixtures together, with pretrained rollouts mostly on the amplifying side \cite{backour2026mixture}; that
analysis concerns the fate of mixtures, while ours computes the invariant sets, bifurcations and normal forms of the
full released map. The closest precedent, a derivative-based analysis of a fold (a saddle-node bifurcation, where
an orbit meets an unstable partner and both vanish) in a small task-trained looped transformer, predicts without
fitting how long trajectories linger near the vanished orbit, and the spectral gap on the other side, and derives from
the same bottleneck an $N^{-2}$ tail of relaxation times over inputs \cite{kim2026phase}.

Random recurrent networks become chaotic above a critical gain \cite{sompolinsky1988chaos,doyon1994bifurcations};
low-dimensional maps organize chaos and periodic windows, the parameter regions in which one periodic orbit attracts,
through period-doubling cascades and shrimp-shaped domains \cite{feigenbaum1978quantitative,gallas1993structure}; and
fixed-step gradient descent, studied as a map rather than through its gradient-flow limit, period-doubles into chaos in
exactly solvable models of deep networks, where the edge of stability is the first bifurcation of the training map
\cite{hofmann2026map}. None of these shows that a trained network realizes such structure. We ask whether the inference loop of released language models is organized by
low-dimensional bifurcations, and whether bifurcation theory, applied to derivatives computed from the model, explains
and predicts its behavior.

\begin{figure*}[tb]
\centering
\includegraphics[width=0.45\textwidth]{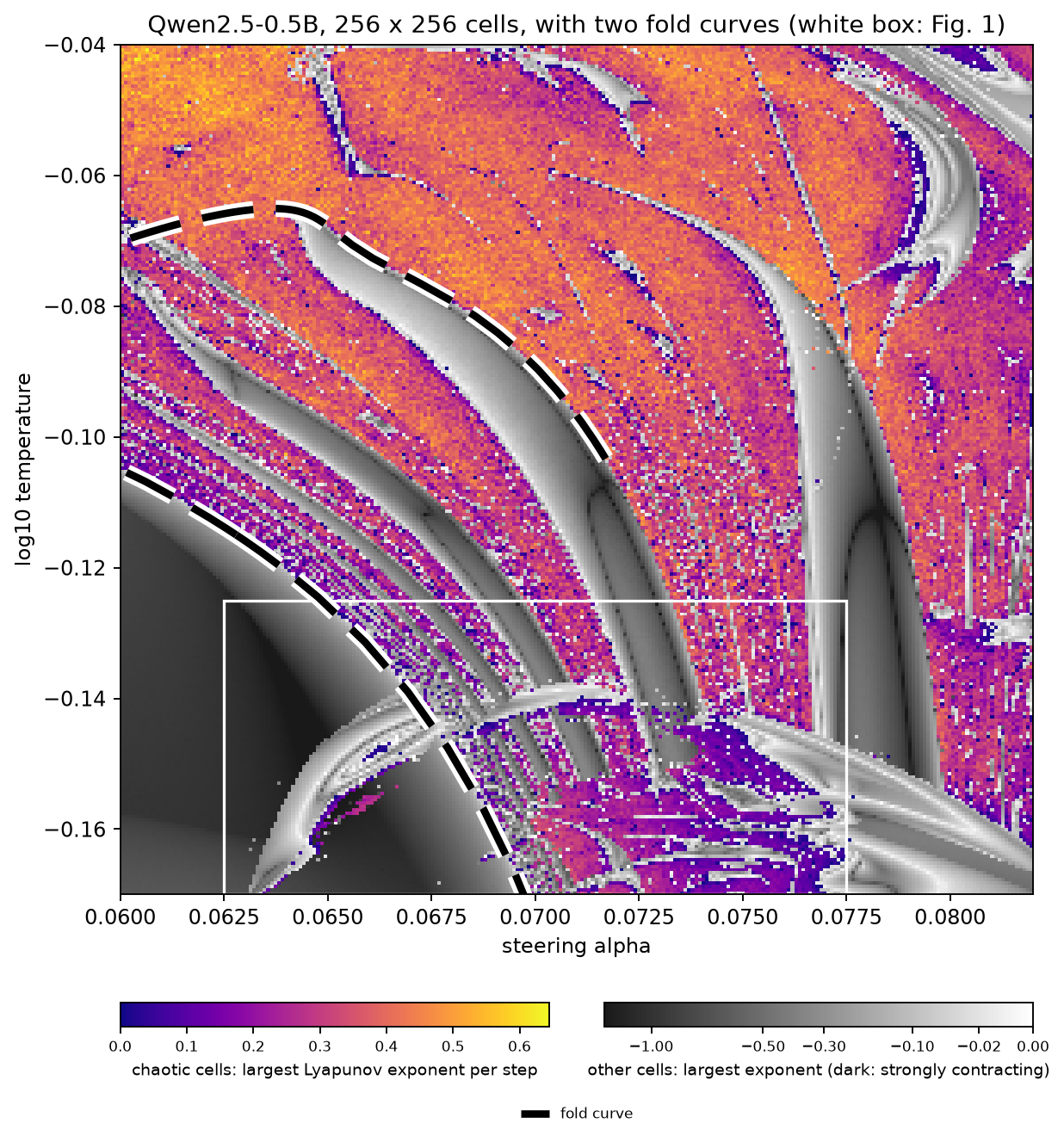}\hfill
\includegraphics[width=0.45\textwidth]{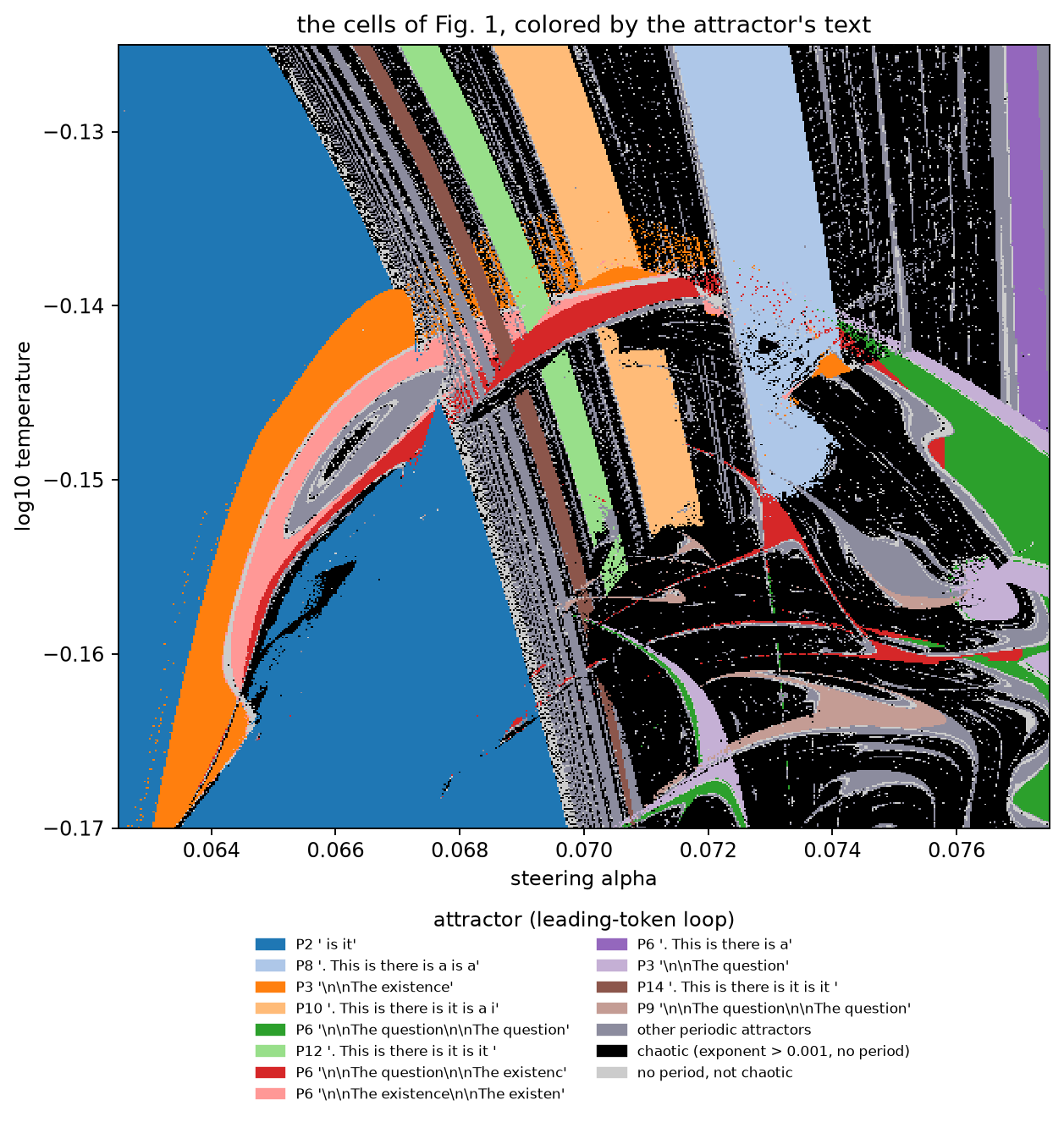}
\caption{(a, left) Scan of the soft-token map of Qwen2.5-0.5B-Instruct steered along the philosophy direction at layer 12,
$256^2$ cells in float64 over $\alpha\in[0.060,0.082]$ and $\log_{10}T\in[-0.17,-0.04]$: chaotic cells colored by the largest
Lyapunov exponent, the other cells in gray by how strongly they contract (light gray: close to losing stability). Black dashed curves with a white edge are float64 fold curves of the 2-cycle ``is it'' (lower left, the
boundary on which the stripes accumulate) and of the period-8 orbit (top). The white box outlines the region of
Fig.~\ref{fig:teaser}. (b, right) The $512\times512$ cells of Fig.~\ref{fig:teaser} (the white box in (a)), colored by the
attractor's text, its leading-token loop (black: chaotic by the criterion of Sec.~\ref{sec:chaos}; light gray: no
period up to 32 and not chaotic; bluish gray: other periodic attractors). The legend gives the first characters of each
loop's text; windows of the same period can write different texts in different parts of the plane (compare
Table~\ref{tab:orbits}, computed at $\log_{10}T=-0.115$).}
\label{fig:stripes}
\end{figure*}

We study the expected-embedding (``soft-token'') map of released pretrained language models, defined exactly by their
stored weights and run with a fixed chat prefix, in the fifteen models of a size sweep from Qwen2.5-0.5B to Llama-3.3-70B and Qwen2.5-72B,
in most detail in Qwen2.5-0.5B-Instruct. The sweep covers three dense families, chosen because each releases open
weights at several sizes and they come from developers with different training data, and a mixture of experts. For the dense models the
map is smooth, and its periodic orbits can be computed and continued like those of any smooth map. Figure~\ref{fig:teaser} shows a piece of the parameter plane of
Qwen2.5-0.5B-Instruct, where chaotic cells lie side by side with windows in which the model repeats a fixed loop of
text, and Fig.~\ref{fig:stripes} a wider piece (a) and the texts the loop writes in the region of Fig.~\ref{fig:teaser} (b).

Our answer has four parts. First, temperature and steering move the loop between repeating text and chaos with
embedded windows in all three model families (30\,\% of the cells of Fig.~\ref{fig:teaser} meet the finite-horizon criterion, and 74\,\% of a random sample of these stay chaotic for $10^5$ steps). Second, the
loop's derivative factorizes into the inverse temperature, the covariance of token embeddings under the model's own
next-token distribution (a cross-covariance for untied embeddings) and the network's Jacobian (Sec.~\ref{sec:map}). This bounds where chaos can occur, and in
Qwen2.5-0.5B, along recorded orbits, the growth that steering adds comes mainly from the readout, where next tokens
compete (Sec.~\ref{sec:theory}). Third, the windows are organized as in low-dimensional maps, and the model's derivatives predict parts of that
organization (Secs.~\ref{sec:adding}--\ref{sec:shrimp}). The broad scans establish where chaos and windows occur in
each family; the detailed analysis of folds, cascades, shrimps and proofs is done in Qwen2.5-0.5B, with a second cascade
in Qwen2.5-1.5B. Windows accumulate on a fold with distances
close to a $1/N^2$ law (log-log slope $-2.00$; a free exponent fitted with an offset ranges from 2.13 to 2.45 with the fit range); at four folds not used in developing the method, the normal form's passage integral came within one iterate of 42 of 48 measured passage times, every forecast fixed
before the test trajectories of its stage were run: 24 of 24 in the first stage, where it was the secondary forecast and
the primary, iterated forecast held for 23 of 24, and 18 of 24 in the second, where it was primary. The six misses, all
at a period-6 fold, were 3.1 to 5.2\,\% short. Late finite-order doubling ratios approach Feigenbaum's constant (the last within 0.008\,\%),
interval-arithmetic proofs show that a primitive period-8 orbit exists and is locally attracting and that attracting
orbits of periods 2 and 3 coexist at one parameter cell, and the scans show shrimp-shaped
windows of the kind Gallas classified \cite{gallas1993structure,gallas1994dissecting}; we study one in detail: along its body it meets our numerical criteria for a shrimp, and its continued fold and flip
curves, kite and cusp are consistent with Gallas's organization of one. Fourth, the regime decides whether rounding differences
between runs grow: run twice with only the grouping of requests into a batch changed, chaotic float64 loops separate on the
schedule their Lyapunov exponent sets, while periodic loops mostly stay together (Sec.~\ref{sec:chaos}). Beyond these four parts, when two copies of the loop are coupled through their soft tokens, the difference between
them grows at approximately the rate that the theory of coupled chaotic maps predicts from the loop's own Lyapunov
exponent, while whether and where they lock together also depends on how they start, how long they run, the tolerance
and the attractor they reach (Sec.~\ref{sec:sync}).

A full report of this study, with the details of every result, the test designs fixed
in advance, the failed attempts, and results we mention only briefly, is available online (Data Availability). Ordinary sampled generation with a growing context is a
different system from the map studied here.

\begin{figure*}[tb]
\centering
\definecolor{loopaccent}{RGB}{204,85,0}
\begin{tikzpicture}[
    font=\small,
    >={Stealth[length=2mm,width=1.6mm]},
    box/.style={draw=black!70, rounded corners=2pt, fill=black!4, minimum height=9mm, inner sep=3pt, align=center},
    flow/.style={->, semithick, black!80},
    qty/.style={font=\footnotesize, above, midway, inner sep=2pt},
    under/.style={decorate, decoration={brace, mirror, amplitude=4pt, raise=1pt}, draw=black!55},
  ]
  \node[draw=black!55, fill=black!10, minimum width=18mm, minimum height=7mm, inner sep=0pt] (prompt) {};
  \foreach \nn in {1,...,4} {
    \draw[black!35] ($(prompt.south west)!\nn/5!(prompt.south east)$) -- ($(prompt.north west)!\nn/5!(prompt.north east)$);
  }
  \node[draw=loopaccent, fill=loopaccent!15, minimum width=7mm, minimum height=7mm, inner sep=0pt,
        right=0pt of prompt] (x) {$x$};
  \node[font=\footnotesize, below=1.5mm of prompt, align=center, inner sep=0pt] {fixed prompt};

  \node[box, minimum width=37mm, minimum height=14mm, right=8mm of x] (net) {};
  \coordinate (r0) at ($(net.west)+(2.5mm,1.6mm)$);
  \coordinate (r1) at ($(net.east)+(-2.5mm,1.6mm)$);
  \draw[black!45] (r0) -- (r1);
  \foreach \nn in {1,2,3,5,6,7} {
    \node[draw=black!60, fill=white, minimum width=2.2mm, minimum height=5.6mm, inner sep=0pt]
      at ($(r0)!\nn/8!(r1)$) {};
  }
  \node[draw=black!80, fill=white, circle, minimum size=3.2mm, inner sep=0pt] (plus) at ($(r0)!4/8!(r1)$) {};
  \draw[black!80] (plus.north) -- (plus.south) (plus.west) -- (plus.east);
  \node[font=\footnotesize, anchor=south, inner sep=1.5pt] at ($(net.south)+(0,0.4mm)$) {transformer};
  \node[font=\footnotesize, inner sep=1pt] (steer) at ($(plus)+(0,11.5mm)$) {$\alpha v$};
  \draw[flow] (steer.south) -- (plus.north);

  \node[box, right=11mm of net] (zb) {$z=E\tilde h$};
  \node[box, right=7mm of zb] (pb) {$p=\operatorname{softmax}(z/T)$};
  \node[box, right=7mm of pb] (mb) {$E^{\top}p$};
  \draw[flow] (x.east) -- (net.west);
  \draw[flow] (net.east) -- node[qty] {$\tilde h(x)$} (zb.west);
  \draw[flow] (zb.east) -- node[qty] {$z$} (pb.west);
  \draw[flow] (pb.east) -- node[qty] {$p$} (mb.west);

  \coordinate (out) at ($(mb.east)+(12mm,0)$);
  \coordinate (top) at ($(net.north)+(0,14.5mm)$);
  \draw[loopaccent, semithick] (mb.east) -- node[qty, text=loopaccent] {$F(x)$} (out);
  \draw[->, loopaccent, thick, rounded corners=3pt] (out) |- (x.north |- top) -- (x.north);
  \node[font=\footnotesize, fill=white, inner sep=2pt] at ($(top)!0.5!(top -| out)$)
    {$F(x)$ replaces $x$ as the next input};

  \coordinate (base) at ($(net.south)+(0,-1mm)$);
  \draw[under] (net.south west |- base) -- node[below=5pt, font=\footnotesize, align=center]
    {network: $J_{\tilde h}(x)$} (net.south east |- base);
  \draw[under] (zb.south west |- base) -- node[below=5pt, font=\footnotesize, align=center]
    {readout gain: $\Sigma_p/T$, \ $\Sigma_p=\operatorname{Cov}_{t\sim p}(e_t)$} (mb.south east |- base);
  \node[font=\footnotesize, align=center, anchor=north] (eq) at ($(net.south east |- base)!0.5!(zb.south west |- base)+(0,-9mm)$)
    {$DF(x)=\dfrac{1}{T}\,\Sigma_p\,J_{\tilde h}(x)$};
\end{tikzpicture}
\caption{The soft-token loop. A fixed prompt is followed by one input embedding $x$. The transformer, with $\alpha v$
added to the residual stream at a middle layer (at every position), gives the final normalized hidden state
$\tilde h(x)$ at the position of $x$; with tied embeddings the logits are $z=E\tilde h$ (untied: $z=W\tilde h$), and
$p=\operatorname{softmax}(z/T)$ is the next-token distribution. The expected embedding $E^{\top}p=F(x)$ is fed back
as the next input, at the same position after the same prompt. Below: the two factors of the derivative
\eqref{eq:gain}, the network's Jacobian $J_{\tilde h}$ and the readout gain $\Sigma_p/T$. Without the network,
$\tilde h(x)=x$ and $J_{\tilde h}=I$, the loop $F(x)=E^{\top}\operatorname{softmax}(Ex/T)$ is the update of a modern
Hopfield network \cite{ramsauer2020hopfield}, which lowers an energy at every step and so cannot cycle (Sec.~\ref{sec:theory}).}
\label{fig:loop}
\end{figure*}
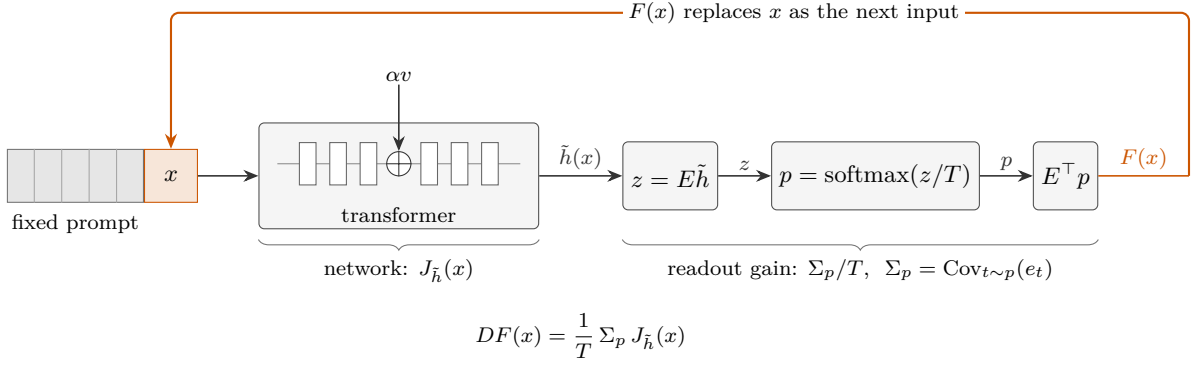

\section{The soft-token map}
\label{sec:map}
We describe the map for Qwen2.5-0.5B-Instruct; the other models enter the same construction with their own embeddings
and steering layers. Let $E\in\mathbb{R}^{V\times d}$ be the input embedding matrix ($V=151{,}936$, $d=896$), with rows
$e_t$. The model reads a fixed chat prefix, with the system prompt ``You are an AI talking with another AI. You have
complete freedom. Feel free to pursue whatever you want.'' and the user turn ``Hello!'', followed by a single input
embedding $x$, and produces next-token logits $z(x)\in\mathbb{R}^V$. The soft-token map feeds the expected next-token
embedding, the soft token, back in place of $x$:
\begin{equation}
  F_{T,\alpha}(x) = E^{\top}\operatorname{softmax}\!\big(z_\alpha(x)/T\big),
\end{equation}
where $z_\alpha$ denotes the logits computed with $\alpha v$ added, at every position including the prefix, to the
residual stream at layer 12 (the $d$-dimensional vector at each token position to which every layer adds its output).
The steering vector $v$ is described below. As $T\to0$ the map reduces, where the argmax is unique, to
greedy decoding of the same fixed-prefix, one-token-window system; it is not ordinary generation with a growing
context. Every run starts from the same state $x_0$, the fixed start: the input embedding of the prefix's last token,
the newline that opens the assistant's reply.

\paragraph{How we steer.} Steering adds one fixed vector to the network's internal state while it runs; no weight is
changed. The state it changes is the residual stream, the $d$-dimensional vector carried at each token position, to
which every attention and MLP block adds its output. Our main direction, philosophy, is built in three steps from 16
short texts: eight philosophical sentences, such as ``What is the nature of consciousness and subjective experience?''\
and ``Is there something it is like to be me?'', and eight neutral ones, such as ``The meeting is scheduled for nine
in the morning.''\ and ``The package weighs two kilograms.'' First, each text is run through the unsteered model,
without the chat template, and the residual-stream vector after decoder layer 12 (of 24, counted from 0) is read at its
last token. Second, the direction is the mean of these vectors over the philosophical texts minus their mean over the
neutral texts. Third, it is rescaled to the mean residual norm at that layer over the 16 texts, $\lVert v\rVert=189$,
so that $\alpha$ is the push as a fraction of a typical residual vector: $\alpha=0.07$ adds 7\,\% of one. The raw
difference has norm 8.1, so $\alpha\approx0.043$ reproduces the natural contrast between the two groups of texts, and
the values of Fig.~\ref{fig:teaser}, 0.0625 to 0.0775, push 1.5 to 1.8 times as far. In every forward pass of the loop, $\alpha v$ is added to the residual stream after layer 12 at every position: at
each token of the prompt and at the fed-back soft token. The later layers therefore also see a steered prompt, since
the keys and values they compute for its tokens carry the push (Fig.~\ref{fig:loop}). The other
concept directions, gratitude and hostility (hostile statements against matched kind ones), each with eight
sentences, are built the same way at their own layers.

\paragraph{State space and regularity.} Attention is causal, so the keys and values of the prefix do not depend on
$x$; positions are fixed and the window does not grow, so $F$ is an autonomous map of $\mathbb{R}^{896}$. It is
real-analytic, because RMS normalization (with its positive regularizing constant), the SiLU activation, attention,
rotary embeddings at fixed positions and the softmax readout are analytic; with top-$k$ routing, a mixture of experts is smooth only within each region of fixed routing, and its map can be
discontinuous where the routing changes. The continuity and differentiability arguments below apply to the dense
models. Every soft token is a convex combination of embeddings, so $F$ maps
$\mathbb{R}^{896}$ into the compact convex polytope $K=\operatorname{conv}\{e_t\}$, and $F(K)\subset K$: every orbit
enters $K$ after one step, $F$ has a fixed point in $K$ by Brouwer's theorem and invariant probability measures there,
and, $\log^+\lVert DF\rVert$ being bounded on $K$, its Lyapunov exponents exist at almost every point of every invariant
measure by Oseledets' theorem \cite{oseledets1968multiplicative,eckmann1985ergodic}, in its version for non-invertible maps,
which allows exponents equal to $-\infty$. The exponents we report estimate
them over finite horizons.

\begin{figure}[tb]
\centering
\includegraphics[width=0.9\columnwidth]{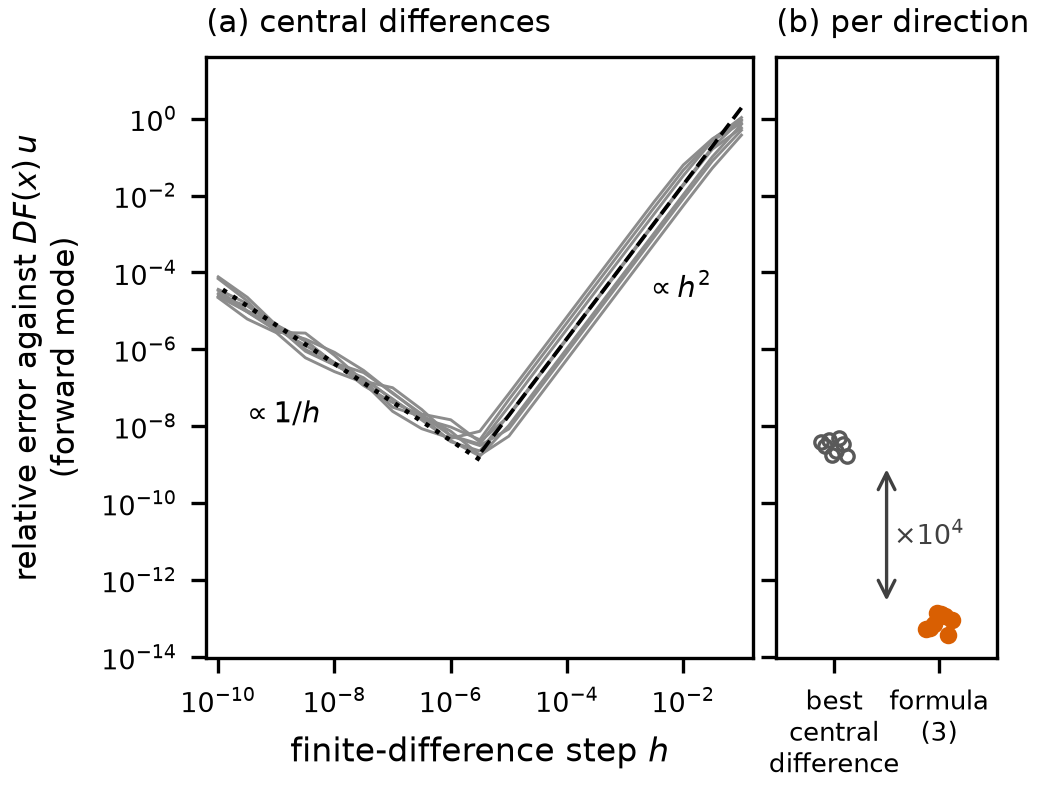}
\caption{Formula \eqref{eq:gain} is exact up to rounding. At a state of the period-8 orbit of Table~\ref{tab:orbits}
(Qwen2.5-0.5B-Instruct, philosophy direction at layer 12, $\alpha=0.071674$, $\log_{10}T=-0.115$, float64), in eight
random unit directions $u$, each derivative is compared with the network's forward-mode derivative $DF(x)\,u$.
(a) Central differences $(F(x+hu)-F(x-hu))/(2h)$. For large steps $h$ their error is truncation, which falls like
$h^2$ as $h$ shrinks (dashed line, slope 2; fitted slopes 1.98 to 2.00 for $10^{-4}\le h\le10^{-2}$); below
$h\approx3\times10^{-6}$ rounding takes over and the error grows like $1/h$ (dotted line). (b) Each direction's best
central difference (open circles, $1.7\times10^{-9}$ to $4.8\times10^{-9}$) and the factorized derivative
$T^{-1}\Sigma_p J_{\tilde h}(x)\,u$ of formula \eqref{eq:gain} (filled, $3.8\times10^{-14}$ to $1.35\times10^{-13}$),
which has no step size: the formula agrees with forward mode more than four orders of magnitude more closely than
any finite difference.}
\label{fig:derivcheck}
\end{figure}

\paragraph{The derivative.} With $p=\operatorname{softmax}(z_\alpha(x)/T)$, the derivative with respect to the state is
\begin{equation}
DF_{T,\alpha}(x)=\frac{1}{T}\,E^{\top}\bigl(\operatorname{diag}p-pp^{\top}\bigr)J_{z_\alpha}(x).
\end{equation}
Qwen2.5-0.5B ties its output matrix to the input embedding, so $z_\alpha=E\tilde h_\alpha(x)$, where $\tilde h_\alpha$
is the final normalized hidden state of the steered network, and the derivative factorizes:
\begin{equation}
\begin{split}
DF_{T,\alpha}(x)&=\frac{1}{T}\,\Sigma_p\,J_{\tilde h_\alpha}(x),\\
\Sigma_p&=E^{\top}\bigl(\operatorname{diag}p-pp^{\top}\bigr)E=\operatorname{Cov}_{t\sim p}(e_t).
\end{split}
\label{eq:gain}
\end{equation}
The derivative of the loop is the covariance of the token embeddings under the model's own next-token distribution,
amplified by the inverse temperature and composed with the derivative of the network. The network factor
$J_{\tilde h_\alpha}$ is the derivative of the whole transformer, with every attention layer, MLP block, normalization
and residual connection and the steering included (attention depends on $x$ through the query, key and value of the
fed-back position; those of the fixed prefix do not change); the factorization only separates it from the readout.
The covariance form is the standard identity for the Hessian of the log-partition function of an exponential family
\cite{wainwright2008graphical}, the Jacobian of the softmax \cite{gao2017softmax}, and it appears explicitly in the
Jacobian of the modern Hopfield update \cite{ramsauer2020hopfield}; here it is composed with the derivative of the full
transformer. For tied embeddings, placed between a network Jacobian and its transpose and divided by $T^2$, the same
covariance is the Fisher information of the next-token distribution pulled back to the network's input (for untied
embeddings the covariance of the output vectors takes its place), a metric used to attribute
predictions to tokens and to describe steering geometrically \cite{liu2026jacobianscopes,wang2026fishback}; in the
feedback map it enters once, multiplied by the Jacobian. At a state of the period-8 orbit
of Table~\ref{tab:orbits} (Sec.~\ref{sec:adding}), \eqref{eq:gain} agrees with forward-mode derivatives of the full map
to a relative error of at most $1.35\times10^{-13}$ over eight random directions (Fig.~\ref{fig:derivcheck}). Without
tied embeddings, as in Qwen2.5-7B and Llama-3.1-8B, the logits are $z_\alpha=W\tilde h_\alpha(x)$ with an output matrix
$W\neq E$, and $\Sigma_p$ becomes the cross-covariance $\operatorname{Cov}_{t\sim p}(e_t,w_t)$ of input and output
embeddings, which is not symmetric in general.

\begin{figure*}[tb]
\centering
\resizebox{0.72\textwidth}{!}{\definecolor{steer}{HTML}{C8510A}
\definecolor{steerfill}{HTML}{F5DCCB}
\definecolor{boxfill}{HTML}{F3F3F3}
\begin{tikzpicture}[
  >={Stealth[length=2.2mm]},
  font=\small,
  line/.style={line width=0.7pt, black!85},
  lane/.style={line width=0.7pt, black!85, preaction={draw, white, line width=3.4pt, -}},
  box/.style={draw=black!70, fill=boxfill, rounded corners=2pt, minimum height=7mm, inner sep=4pt},
  mlp/.style={draw=black!70, fill=boxfill, rounded corners=2pt, minimum height=4.6mm,
              inner sep=1pt, font=\footnotesize},
  plus/.style={draw=#1, fill=white, circle, minimum size=3.6mm, inner sep=0pt,
               path picture={\draw[#1] (path picture bounding box.north)--(path picture bounding box.south)
                             (path picture bounding box.west)--(path picture bounding box.east);}},
  plus/.default=black!85,
  neuron/.style={draw=black!60, fill=white, circle, minimum size=5mm},
  tok/.style={draw=black!80, fill=white, rounded corners=1.5pt, font=\ttfamily\footnotesize,
              inner sep=2pt, minimum height=4.2mm, anchor=east},
]

\def\gap{0.95}     %
\def\xL{0.0}       %
\def\xR{12.2}      %
\def\xMs{4.9}      %
\def\xMo{10.6}     %
\def\yAtt{1.25}    %

\foreach \k/\tname in {0/The,1/{\ Eiffel},2/{\ Tower},3/{\ is}} {
  \pgfmathsetmacro\y{-\k*\gap}
  \pgfmathsetmacro\xr{0.75+0.3*\k}   %
  \pgfmathsetmacro\xa{3.6+0.3*\k}    %
  \node[tok] at (\xL-0.08,\y) {\tname};
  \node[plus] (A\k) at (\xa,\y) {};
  \node[plus] (M\k) at (\xMo,\y) {};
  \node[mlp, minimum width=4.8cm] (mlp\k) at (7.75,\y-0.42)
    {RMSNorm $\to$ MLP \quad{\color{steer}$h_{832}\mathrel{+}=312\alpha$}};
  \draw[->, line] (\xMs,\y) |- (mlp\k.west);
  \draw[->, line] (mlp\k.east) -| (M\k.south);
  \coordinate (R\k) at (\xr,\y);
  \coordinate (W\k) at (\xa,\yAtt);
}

\node[box, minimum width=3.2cm, minimum height=1.4cm, align=left, anchor=west] (att) at (0.05,\yAtt) {};
\node[anchor=west, align=left, inner sep=0pt] at ([xshift=5pt]att.west)
  {RMSNorm $\to$\\attention};
\begin{scope}[shift={([xshift=-0.95cm,yshift=-0.12cm]att.east)}, x=0.17cm, y=0.17cm]
  \foreach \r in {0,...,3} \foreach \c in {0,...,3} {
    \pgfmathtruncatemacro\on{\c<=\r ? 1 : 0}
    \ifnum\on=1
      \fill[black!60] (\c,3-\r) rectangle ++(1,1);
    \fi
    \draw[black!40, line width=0.2pt] (\c,3-\r) rectangle ++(1,1);
  }
\end{scope}
\node[font=\scriptsize, black!55, anchor=north] at ([xshift=-0.78cm,yshift=-0.15cm]att.east) {causal mask};

\foreach \k in {0,...,3}
  \draw[->, line] (R\k) -- (R\k |- att.south);
\draw[line] (att.east) -- (W3);
\foreach \k in {0,...,3}
  \draw[->, line] (W\k) -- (A\k.north);

\foreach \k in {0,...,3} {
  \pgfmathsetmacro\y{-\k*\gap}
  \draw[lane] (\xL,\y) -- (A\k.west);
  \draw[lane] (A\k.east) -- (M\k.west);
  \draw[->, lane] (M\k.east) -- (\xR,\y);
  \node[plus] at (A\k) {};
  \node[plus] at (M\k) {};
}

\node[anchor=south east, font=\footnotesize, black!55] at (-0.1,0.25) {input tokens};
\node[anchor=west, inner sep=2pt] at (\xR,{-3*\gap}) {to layer 13};
\draw[decorate, decoration={brace, amplitude=6pt}, black!55]
  (-1.0,2.2) -- node[above=6pt, font=\small, black] {decoder layer 12 of Qwen2.5-0.5B-Instruct}
  (\xMo+0.4,2.2);

\def\zt{-4.4}
\node[draw=black!45, dashed, rounded corners=4pt, minimum width=11.4cm, minimum height=4.1cm,
      anchor=north] (zoom) at (6.25,\zt) {};
\draw[black!40, dashed] (mlp3.south west) -- (zoom.north west);
\draw[black!40, dashed] (mlp3.south east) -- (zoom.north east);

\node[anchor=west, font=\small] (xin) at ([shift={(0.25,-0.6)}]zoom.north west) {$x$};
\node[box, anchor=west] (norm) at ([xshift=0.3cm]xin.east) {$\hat x = \mathrm{RMSNorm}(x)$};
\node[box, anchor=west] (gate) at ([xshift=0.45cm]norm.east)
  {$h=\mathrm{SiLU}(W_{\mathrm{gate}}\,\hat x)\odot W_{\mathrm{up}}\,\hat x$};
\draw[->, line] (xin) -- (norm);
\draw[->, line] (norm) -- (gate);

\begin{scope}[shift={(1.75,{\zt-2.4})}]
  \node[neuron, label={[font=\small]below:$h_1$}] (n1) at (0,0) {};
  \node[neuron, label={[font=\small]below:$h_2$}] (n2) at (0.6,0) {};
  \node[neuron, label={[font=\small]below:$h_3$}] (n3) at (1.2,0) {};
  \node at (1.8,0.05) {$\cdots$};
  \node[neuron, draw=steer, fill=steerfill, line width=0.8pt,
        label={[font=\small, text=steer]below:$h_{832}$}] (n832) at (2.55,0) {};
  \node at (3.3,0.05) {$\cdots$};
  \node[neuron, label={[font=\small]below:$h_{4864}$}] (nN) at (3.95,0) {};
\end{scope}
\node[draw=black!50, dashed, rounded corners=3pt, fit=(n1)(nN), inner xsep=5pt,
      inner ysep=13pt, yshift=-5pt] (layer) {};
\node[anchor=north west, black!55, font=\footnotesize] at ([yshift=-1pt]layer.south west) {4864 neurons};
\draw[->, line] (gate.south) -- ++(0,-0.3) -| (n2 |- layer.north);

\node[text=steer, font=\small, align=center, anchor=south] (kick) at ([shift={(0.95,0.45)}]n832.north)
  {$+\,312\alpha$};
\draw[->, steer, line width=0.9pt] (kick.south) -- (n832.north east);

\node[box, align=center] (down) at (10.2,{\zt-2.4})
  {$\sum_i h_i\,w_i$\\[-1pt]{\footnotesize $w_i$: column $i$ of $W_{\mathrm{down}}$}};
\draw[->, black!55, line width=0.6pt] (layer.east) -- (down.west);
\draw[->, line] (down.north) -- ++(0,0.6) node[above, font=\footnotesize] {to this token's $\oplus$};
\end{tikzpicture}}
\caption{Steering along one MLP neuron, in decoder layer 12 of Qwen2.5-0.5B-Instruct. Each token has its own
residual stream, one 896-dimensional vector per token, and its own pass through the same MLP; attention, with its causal
mask, is the only step that mixes tokens. Inside every MLP box (bottom, for one token with residual vector $x$), the
4864 neurons write $\sum_i h_i w_i$ into the stream, neuron $i$ along its own column $w_i$ of the down-projection. The
steering strength $\alpha$ is a scalar we set (negative values lower the activation), and
$312=\lVert v_{\mathrm{phil}}\rVert/\lVert w_{832}\rVert=188.99/0.606$ is a fixed scale for neuron 832. Raising
$h_{832}$ by $312\alpha$ for every token is the same as adding $\alpha v$, with $v=312\,w_{832}$, to the layer's output,
which is how we implement it: $\sum_i h_i w_i+312\alpha\,w_{832}=\sum_{i\neq832}h_i w_i+(h_{832}+312\alpha)\,w_{832}$.
At $|\alpha|=0.12$ the increase is 37.4, against a mean activation of 0.27 on the philosophy texts; for neurons 833 and
4137 the scale is 329 and 346. The tokens shown are only an example: in the loop, the input is the fixed prompt followed
by the fed-back soft token.}
\label{fig:neuronsteer}
\end{figure*}

\paragraph{Where steering and temperature enter.} Temperature appears in \eqref{eq:gain} twice: explicitly, as the factor $1/T$, and inside $p$. As
$T\to0$ the two pull in opposite directions and $p$ wins: where one token leads, $p$ concentrates on it, $\Sigma_p$
vanishes faster than $1/T$ grows, and the readout factor tends to zero, so the loop approaches greedy decoding; only
near ties between leading tokens does the factor grow, like $1/T$ (Sec.~\ref{sec:theory}). In Backour's unbiased two-token reduction with linear logit transport and a positive constant coupling, the inverse
temperature rescales the coupling, and crossing the critical value 2 destabilizes the balanced mixture, so that one token
takes over \cite{backour2026mixture}. Steering does not appear in \eqref{eq:gain}: the vector $\alpha v$ is the same whatever the state $x$, so it has no
derivative of its own, and it changes $DF$ only by changing the quantities from which the two factors are computed. Steering decides which tokens compete in $p$, and temperature how sharply, so together they set
the readout factor $\Sigma_p/T$, which is large near ties between leading tokens and vanishes where one token
dominates. Steering also changes the input of every later block, and with it the attention weights, MLP
activations and normalization scales at which that block's derivative is taken; so it moves the network
factor $J_{\tilde h_\alpha}$ as well, which temperature moves too, by moving the orbit to other states. On the two main maps, the stripes map of Fig.~\ref{fig:stripes}(a) and the crossing window of
Fig.~\ref{fig:teaser}, both parameters move both factors. For each of 7936 cells sampled from the two maps we averaged the logarithm of
each factor's norm along the orbit (power-iteration estimates) and asked how much of the variation of these averages
over a map, steering strength explains on its own, as the share of the variance carried by the means over cells of
equal $\alpha$, and how much temperature explains, by the means over cells of equal $T$. For the readout factor,
steering explains 30 and 32\,\% on the two maps and temperature 21 and 27\,\%; for the network factor, steering
explains 7 and 33\,\% and temperature 47 and 4\,\%. From here on we drop the subscripts when
$(T,\alpha)$ is fixed. Every condition below that involves $DF$, including the threshold $T_*$, describes a region of
the $(\alpha,T)$ plane, not a range of $T$ alone.

\paragraph{Steering as a probe.} Steering adds a vector to the residual stream, typically the mean difference between
activations on contrasting prompts, to strengthen or weaken a behavior without retraining
\cite{rimsky2024steering,turner2023activation,zhang2025steering,zou2023representation}. Ours is the standard
contrastive recipe \cite{rimsky2024steering}, and our pushes are moderate: at most 12\,\% of the mean residual
norm in every scan (bifurcation curves continued in steering strength reach 13.1\,\%) and 6.25 to
7.75\,\% in Fig.~\ref{fig:teaser}, far smaller than activation additions reported as effective in steering work, one
of which has nearly ten times the norm of the residual stream it is added to \cite{turner2023activation}; pushes of
this size still change what a model writes (in Qwen2.5-7B, 1\,\% of the residual norm is the largest push that kept
greedy accuracy on a multiplication task within 5 points)\smref{Behavioral tests in detail}. We use steering as a probe: $\alpha$ is a control parameter, like the strength of a constant applied field,
that moves the model through a one-parameter family of nearby maps, so that bifurcation curves can be traced. Chaos requires neither steering nor a meaningful direction (Sec.~\ref{sec:chaos}); steering along a
single MLP neuron, for instance, is the same as raising that neuron's activation by a fixed amount at every token
(Fig.~\ref{fig:neuronsteer}).

\paragraph{Numerics.} Results labeled float64 are computed in float64 throughout, including the normalization layers
and rotary tables that standard implementations evaluate in float32. Periodic orbits are solved by Newton's method with
forward-mode derivatives; their multipliers, all in float64 except those of one continued branch in
Sec.~\ref{sec:adding}, are the eigenvalues of the monodromy matrix, the ordered product of the step Jacobians along the
orbit and the analog of the Floquet matrix of a flow. Fold and flip (period-doubling) curves, where a multiplier crosses $+1$ or $-1$, are
continued by minimally augmented systems \cite{govaerts2000numerical,govaerts2007numerical,kuznetsov2019maps}, and
normal forms are computed from the network's derivatives without fitting \cite{kuznetsov2023elements}. The search was
exploratory, and every window was chosen after the scans were inspected; where the criteria of a test were fixed in
advance, we say so, and failed criteria are reported with the successes\smref{Evidence and reproducibility}.

\section{Why the loop can be chaotic}
\label{sec:theory}
Chaos in the sense used here, a positive largest Lyapunov exponent with no period found (the operational label
of Sec.~\ref{sec:chaos}), requires that nearby states drift apart: a small difference $\delta x$ between two states must, on average, grow
from one step to the next, which is what a positive largest Lyapunov exponent measures (such growth alone does not
make an attractor chaotic). To first order, one step multiplies the difference by the derivative $DF$, and
\eqref{eq:gain} splits that derivative into three factors with plain meanings. The network Jacobian
$J_{\tilde h_\alpha}$ says how much the transformer's final hidden state moves when its input moves, through all of
its attention and MLP blocks. The factor $\Sigma_p/T$ is the derivative of the fed-back expected embedding with
respect to that hidden state. The matrix $\Sigma_p$ is the covariance of the token embeddings under the next-token
distribution. It is large when the probability is spread over several tokens whose embeddings differ, because a small
change of the logits then shifts probability between distant embeddings and moves the expected embedding a long way;
it tends to zero as the distribution concentrates on a single token. The scalar
$1/T$ multiplies the whole derivative. This section uses the factorization in three ways: it proves that the loop
contracts at sufficiently high temperature and, at sufficiently low temperature, excludes chaos on invariant sets
whose margin between the two leading tokens stays bounded away from zero; it shows that, with tied embeddings, the transformer between input and output is what makes cycles and chaos
possible at all; and three measurements show that the growth steering adds comes mainly from the readout, where next tokens
compete.

\paragraph{High and low temperature.} The largest Lyapunov exponent of an orbit, the average rate at which nearby
states separate, cannot exceed the long-run average of $\log\lVert DF\rVert$ along the orbit (in the spectral norm; upper limits
where the averages do not converge), so a positive exponent
requires that average to be positive (a necessary condition, not a sufficient one). Both ends of the temperature axis
work against it.

At high temperature $1/T$ is small, and the bound is explicit. However the probability is spread, the projections
$u^{\top}e_t$ of the embeddings on any unit vector $u$ lie in an interval no longer than the diameter
$\operatorname{diam}(E)$ of the set of embeddings, and Popoviciu's inequality bounds a variance by a quarter of the
squared range, so $\lVert\Sigma_p\rVert\le\operatorname{diam}(E)^2/4$. With $L(\alpha)=\sup_K\lVert J_{\tilde
h_\alpha}\rVert$, the network's largest gain on the set $K$ of possible states (finite because the Jacobian is continuous and $K$ is compact),
$\lVert DF_{T,\alpha}\rVert\le\operatorname{diam}(E)^2L(\alpha)/(4T)$ on $K$, and $F_{T,\alpha}$ shrinks every
distance, that is, it is a contraction of $K$, for
\begin{equation}
T>T_*(\alpha)=\tfrac14\operatorname{diam}(E)^2L(\alpha) .
\end{equation}
Above this temperature the loop forgets where it started: by Banach's fixed-point theorem it has a single fixed
point, which attracts every orbit (without tied
embeddings, $\operatorname{diam}(E)^2$ becomes $\operatorname{diam}(E)\operatorname{diam}(W)$). For affine softmax feedback, with a fixed matrix in place of the network, a recent preprint proves a sharp spectral
threshold for the uniqueness and global stability of the fixed point \cite{wang2026logitfixed}. The threshold is a
curve in the $(\alpha,T)$ plane, since the network's largest gain depends on the steering. The bound is rigorous, but it does not locate the measured chaos boundary, because it assumes the worst case at every
step. For Qwen2.5-0.5B, $\operatorname{diam}(E)=1.2109$, computed over all pairs of embeddings; the largest values of $\lVert J_{\tilde h_\alpha}\rVert$ met along the trajectories of 7936 map cells (power-iteration
estimates, so lower estimates of $L(\alpha)$) give $T_*(\alpha)\ge2.7\times10^{3}$ at every steering strength sampled
(64 per map) and $T_*(\alpha)\ge5.6\times10^{4}$ at $\alpha=0.06$, where the largest, $1.54\times10^{5}$, was met:
at least 3.1 and 4.4 decades above the highest chaotic temperature of the unsteered sweeps
($\log_{10}T=0.28$).

A sharper bound uses the derivatives actually met along a computed orbit instead of the worst case. Over $n$ steps a
perturbation is stretched by the product of the $n$ step derivatives, and the norm of a product is at most the product
of the norms, so the orbit average of $\log\lVert DF\rVert$ bounds the largest Lyapunov exponent from above. Taken one
step at a time, this bound cannot separate chaos from periodicity: it is positive in every cell, chaotic or periodic
(medians 2.98 and 2.57 per step over the chaotic and the periodic cells of the stripes map, against a median exponent
of 0.30 in its chaotic cells over the same 60 float32 steps). Each norm is the stretching of one step in its most
expanding direction, but a perturbation leaves each step pointing in some other direction, which the next step
generally stretches less, so the product of the norms overstates the stretching of the product. Taking norms of whole blocks of $k$ consecutive steps, disjoint blocks that tile the record, each the norm of the
product of its $k$ derivatives, removes this overstatement inside each block, and the bound cannot increase when the
block length doubles. It begins to separate chaotic from periodic cells only for blocks of
$k=4$ and 5 steps; shorter blocks give inequalities that hold on the model's orbits but do not say where the chaos is.
For orbits typical of an ergodic invariant measure on which $\log\lVert DF\rVert$ is integrable, the subadditive
ergodic theorem \cite{kingman1973subadditive} guarantees that the block bound converges to the exponent as $k$
grows, but it gives no rate, and no guarantee for a finite record. In a sample of 24 cells, with exact norms over 32
recorded steps, the excess of the bound over the exponent measured along the following 2000 steps shrinks roughly
like $1/k$: over the chaotic cells its median falls from 2.72 per step for single steps to 0.32 for blocks of 8 and
0.036 for the single block of 32 (periodic cells: 2.86, 0.45 and 0.12). The sign of the bound matches that of the
exponent in 13 of the 24 cells for blocks of one, two or four steps,
and in 22 for blocks of 16. Averaging the logarithms of norms over overlapping windows of the record instead, a
diagnostic rather than a bound, gives the right sign in 23 of the 24 cells for every window length from 17 to 29
steps. In these cells, about twenty steps of the recorded orbit thus suffice to read off the sign of the long-run
exponent; this is a measurement along the orbit, not a prediction from the weights.

At low temperature $1/T$ is large, but one token takes almost all the probability, and $\Sigma_p$ vanishes faster than
$1/T$ grows. At a state with a unique leading token $j$, the gain is
$\lVert\Sigma_p\rVert/T=e^{-\Delta/T}(C+o(1))/T$, where $\Delta$ is the gap between the two largest logits and
$C=\lVert\sum_i(e_i-e_j)(w_i-w_j)^{\top}\rVert$, the sum running over the runner-up tokens whose logits trail by
exactly $\Delta$ ($w_t$ are the output vectors, the rows of $W$, equal to the embeddings $e_t$ when they are tied). As $T\to0$, $F$ tends to the greedy map, which feeds back the embedding of the most probable token
(at an exact tie, the limit averages the tied embeddings). The limit map takes finitely many values, so every orbit of
it eventually repeats. The convergence is pointwise and in general not uniform: near a tie between tokens with different embeddings
the gain is of order $1/T$, so at small positive temperature chaos is excluded only on invariant sets that keep a
uniform logit gap.

Chaos is therefore found in between, at temperatures where the next-token distribution is spread enough to make
$\Sigma_p$ large while $1/T$ still amplifies it: in four long unsteered sweeps the chaotic temperatures lie in one
band per model, interleaved with periodic windows, between $\log_{10}T=-0.20$ and $0.28$. This band is an
observation; near ties, the low-temperature argument does not exclude chaos.

\begin{figure*}[tb]
\centering
\includegraphics[width=\textwidth]{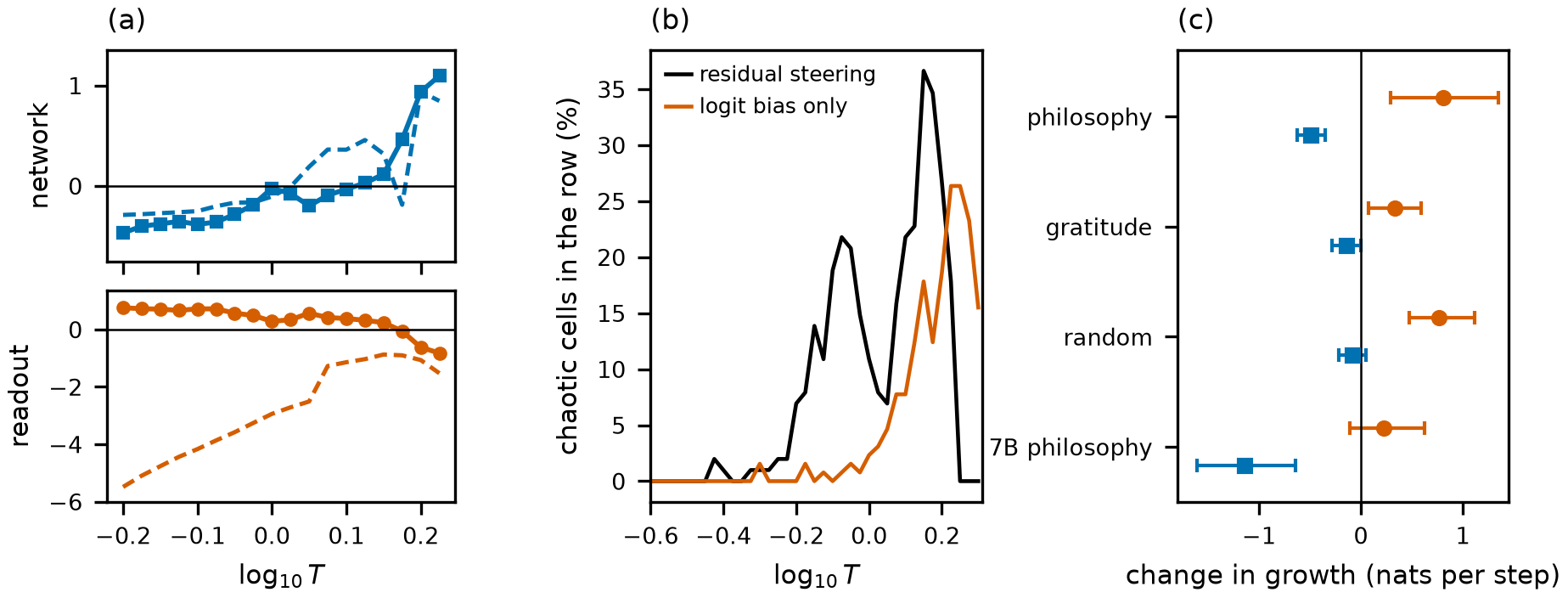}
\caption{Readout and network contributions to the growth under steering (Qwen2.5-0.5B-Instruct unless marked 7B;
float64). Orange: the readout factor $\Sigma_p/T$ of \eqref{eq:gain}; blue: the network factor $J_{\tilde h}$.
(a) Split of the growth along the attractors of the $101\times53$ scan steered at layer 12: mean logarithm per step of the relative gains defined in the text (gains along the tangent direction, not
operator norms), the network gain $g_{\mathrm{net}}$ (top, blue squares) and the readout gain $g_{\mathrm{read}}$
(bottom, orange circles), for chaotic cells (solid lines, filled markers; chosen by their dynamics, at any steering
strength) and for the nearly unsteered cells, $|\alpha|\le0.01$ (dashed lines; chosen by steering strength alone, whatever their dynamics, and
almost all periodic: the baseline at the same temperature), in the 18 temperature rows with at least three cells of
each kind (321 chaotic and 162 nearly unsteered cells, 6 of them in both groups); positive values expand perturbations, negative values
contract them. In the nearly unsteered cells the readout's relative gain shrinks perturbations by 0.87 to 5.49 nats
per step; in the chaotic cells it grows them, except in the three highest rows, while the network factor changes
little. (b) Share of chaotic cells (both twin exponents above $10^{-3}$ and no period up to 64) per temperature row under residual steering at layer 12,
which moves both factors (black; 331 of 5353 cells, 6.2\,\%), and under a logit bias along $Wv$, the output matrix
applied to the steering vector, with no residual steering, which at a given state changes only the readout (orange;
241 of 6837 cells, 3.5\,\%). (c) Orbit-preserving crossover: change in the finite-time growth along recorded steered
orbits when only the readout (orange, circles) or only the network Jacobian (blue, squares) is taken from the
steered model; means over the chaotic cells used only for this test, with 95\,\% bootstrap intervals. Rows: the
philosophy and gratitude directions and a random direction of the philosophy vector's norm in the 0.5B model,
14 cells each, and the philosophy direction in the 7B model, 10 cells.}
\label{fig:mechsummary}
\end{figure*}

\paragraph{Without the transformer, no chaos.} Suppose the network passed its input straight through,
$\tilde h(x)=x$. The map would then be $F(x)=E^{\top}\operatorname{softmax}(Ex/T)$, the update rule of a modern
Hopfield network whose stored patterns are the token embeddings \cite{ramsauer2020hopfield}. Its derivative
$\Sigma_p/T$ is symmetric and positive semidefinite, so $F=\nabla\phi$ for the convex function
$\phi(x)=T\log\sum_t e^{e_t^{\top}x/T}$, and the energy
$\mathcal{V}(x)=\tfrac12\|x\|^2-\phi(x)$ falls at every step, $\mathcal{V}(F(x))\le\mathcal{V}(x)-\tfrac12\|F(x)-x\|^2$,
while staying bounded below on $K$. Each step lowers the energy by at least half the squared length of the step, so
the steps must shrink to zero: every accumulation point of an orbit is a fixed point, and the strict descent of the energy rules out cycles of period
above one and chaotic motion (an unstable fixed point can still have positive tangent exponents; and the descent estimate alone does not establish convergence to a single fixed point when fixed points are not
isolated). With tied embeddings, as in Qwen2.5-0.5B, every
cycle and every chaotic cell therefore depends on the network factor $J_{\tilde h}$ breaking this energy structure;
with untied embeddings, the network-free map need not have a symmetric positive-semidefinite derivative, so the
energy argument no longer applies, and cycles become possible even without a network (two tokens whose output vectors
point at each other's input embeddings exchange at low temperature).

\paragraph{Steering makes tokens compete.} Steering could create chaos through either factor of \eqref{eq:gain}: it
moves the logits, and with them which tokens compete in $\Sigma_p$, and it moves the hidden states, and with them the
network's response $J_{\tilde h}$. Three measurements on Qwen2.5-0.5B, steered along the philosophy direction, tell
the two apart (Fig.~\ref{fig:mechsummary}).

First, an accounting. One step takes the state $x$, the vector fed back as the input at the
soft-token position, through the steered transformer to its final normalized hidden state $\tilde h$ at that
position, and from $\tilde h$, through the logits and the softmax, to the next state $F(x)$, the probability-weighted
average of the token embeddings. A small change $\delta x$ of the state causes changes $\delta\tilde h$ and $\delta F$,
and each step's relative growth splits exactly into a network factor
$g_{\mathrm{net}}=(|\delta\tilde h|/|\tilde h|)/(|\delta x|/|x|)$, how much the transformer stretches a relative change
of its input, and a readout factor $g_{\mathrm{read}}=(|\delta F|/|F|)/(|\delta\tilde h|/|\tilde h|)$, how much the
softmax and the averaging of embeddings stretch it on the way out. For a tangent perturbation followed over $n$ steps, their mean
logarithms add up to its finite-time growth rate plus a boundary term,
$(\log\lVert x_0\rVert-\log\lVert x_n\rVert)/n$; both are relative gains along the orbit, not operator norms, and
the split is a convention (only the sum is intrinsic). Over a
$101\times53$ scan (5353 cells, 331 chaotic), the chaotic cells' readout factor exceeds that of nearly unsteered cells
($|\alpha|\le0.01$) at the same temperature by $2.58$ nats per step (95\,\% bootstrap interval over the 18 temperature rows, $[1.78,3.59]$), while their network factor differs by only $-0.07$ $[-0.23,0.12]$. The chaotic cells also have a higher
next-token entropy than periodic cells at the same temperature ($+1.62$ nats, $[1.26,1.91]$): more tokens compete.

Second, a test that leaves the network unsteered. A logit bias along $Wv$, the output matrix applied to the steering
vector (a logit-space stand-in for the steering), is added with no residual steering: at a given state it changes only
the output probabilities, although the orbit it produces can visit different states. On its grid, 3.5\,\% of cells
were chaotic (241 of 6837, all but one with a nonzero bias), against 6.2\,\% on the residual-steering grid (the two grids share their temperature
rows but not a matched strength of intervention, so these are counts per grid, not a comparison of doses):
competition at the readout alone makes the loop chaotic, but mostly at $\log_{10}T\ge0.1$ (207 cells), where the
network factor itself can expand; below, 34 of 5676 cells are chaotic (0.6\,\%), against 169 of 4444 (3.8\,\%).

Third, a swap. An orbit-preserving crossover evaluates the steered and the unsteered model at the same recorded states
and builds the growth from one factor of each. Over 14 chaotic cells used only for this test (in brackets, 95\,\% bootstrap intervals over
these cells), the steered readout
factor with the unsteered network factor raises the finite-time growth by $0.81$ nats per step ($[0.29,1.35]$), the
steered network factor with the unsteered readout lowers it by $0.49$ ($[-0.63,-0.35]$), and the interaction is
$+0.26$ ($[0.09,0.45]$). These hybrid products are tangent products assembled along the recorded states, not the
dynamics of a rerun hybrid model; they apportion the growth along
the recorded orbits. Along the gratitude direction and a random direction of the same norm the readout effect is also
positive ($+0.33$ and $+0.77$ nats per step); in Qwen2.5-7B, whose embeddings are untied, steering only the network
lowers the growth by $1.14$ nats per step and steering only the readout adds $+0.23$ ($[-0.11,0.63]$, an interval that includes zero), at cells mostly
above $\log_{10}T=0$, where the unsteered derivative already expands.

All three point the same way: along the recorded orbits of Qwen2.5-0.5B, the growth that steering adds comes mainly
from the readout, where next tokens compete; the readout's relative gain turns from shrinking to growing, while the
network factor changes little. The readout carries what steering changes; the network is what makes chaos possible
at all, since with tied embeddings, as here, the loop without it cannot cycle (``Without the transformer, no chaos,'' above). These are conditional comparisons on recorded trajectories. They do not single out
the higher entropy as the cause, nor show that the readout change alone would sustain the same attractor; the readout
gain still shrinks perturbations in the three highest temperature rows, and above $\log_{10}T\approx0.1$ the network factor itself can expand\smref{Why the loop can be chaotic}.

\section{Chaos and its consequences}
\label{sec:chaos}
The loop is chaotic over large parts of its parameter plane, in all three model families, by a criterion whose
finite-horizon errors we measured, and this chaos makes runs irreproducible at the level of rounding.

\begin{figure*}[tb]
\centering
\includegraphics[width=\textwidth]{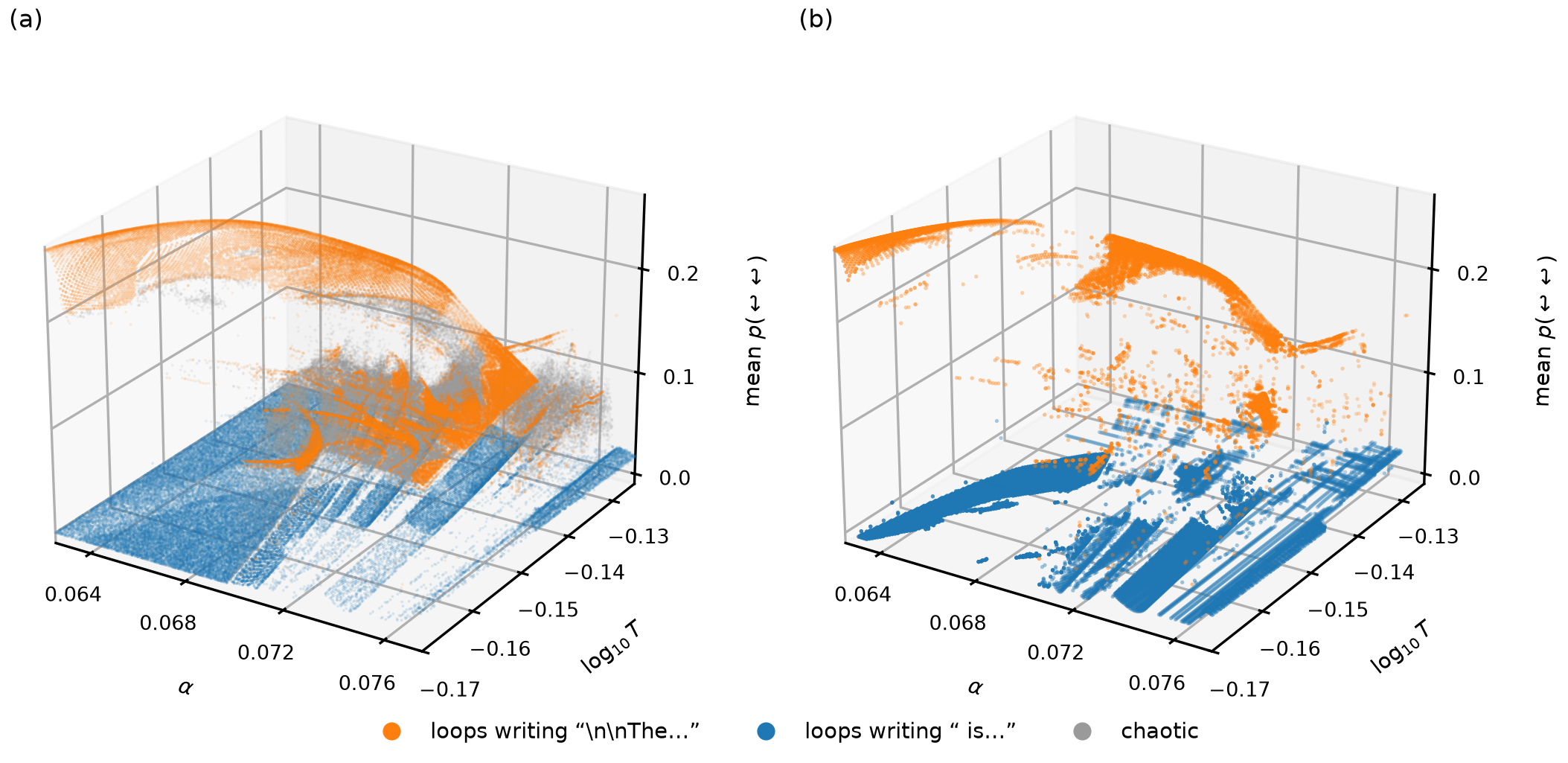}
\caption{The two families of loops in the region of Fig.~\ref{fig:teaser}, drawn as sheets. Each point is an attractor
that one of the eight maps found at one cell, at a height equal to its mean probability of the blank-line token
``\textbackslash n\textbackslash n'' ($\hookleftarrow\hookleftarrow$), which separates the families: in (a), 90\,\% of
the orange points lie between 0.07 and 0.25 and 99\,\% of the blue points below 0.03. Orange: loops writing
``\textbackslash n\textbackslash nThe\ldots'', the family of the crescent of Fig.~\ref{fig:stripes}(b); blue: loops
writing `` is\ldots'', the 2-cycle and its stripes; gray: chaotic attractors. (a) All attractors found, thinned at
random to at most 9000 per map and class. (b) Only what the fixed-start map hides: periodic attractors that the seven
other maps find where the fixed start reaches the other family, chaos or no period. These maps discard 600 transient
steps and measure over the next 300 (200 and 200 in the sweeps).}
\label{fig:sheets3d}
\end{figure*}

\paragraph{Scans and chaos labels.} A scan of the parameter plane is a grid of cells, each one parameter pair, run from
the fixed start $x_0$; unless stated otherwise we discard 300 transient steps and measure over the next 300. The
largest Lyapunov exponent comes from a twin trajectory at offset $10^{-6}$, renormalized to that offset after every step
\cite{benettin1980lyapunovI,benettin1980lyapunovII}. The period is the smallest $q\le32$ (48 or 64 where stated) for
which the last state returns to within $10^{-8}\max(1,\lVert x\rVert)$ of the state $q$ steps earlier ($10^{-5}$ to
$10^{-4}$ in float32 scans), and an attractor is identified by its leading-token loop, the repeating sequence of most
probable next tokens. A cell is chaotic when its exponent exceeds $10^{-3}$ and no period is found. The threshold matters little: raising
it to $5\times10^{-3}$ lowers the chaotic share of the map of Fig.~\ref{fig:teaser} from 30.1 to 29.8\,\%.

We call what the trajectory from a cell's fixed start settles on the cell's attractor. The label is operational: it
does not show that the set attracts initial conditions of positive measure, as Milnor's definition requires
\cite{milnor1985concept}. Where several attractors coexist, a scan shows only the one whose basin contains $x_0$. To
see what a scan hides, we recomputed the region of Fig.~\ref{fig:teaser} on a $192\times256$ grid as eight maps: one
from the fixed start, three from other starts, and four sweeps that follow each attractor across the plane, each cell
starting from the final state of its neighbor. The maps find two families of loops, those writing
``\textbackslash n\textbackslash nThe\ldots'' and those writing `` is\ldots'' (the 2-cycle and its stripes), and both
families are present at 9675 of the 49{,}152 cells (Fig.~\ref{fig:sheets3d}). There the fixed start selects one
family by its basin, so the family a scan shows can alternate from cell to cell. We examined eight places where the
fixed start's family switches: at two, both families persist and attract on both sides, so the switch is such a
selection; at two, a family ends; four remained unresolved. Following an attractor is also how the hidden parts of a
window are drawn (Sec.~\ref{sec:shrimp}).

A long chaotic transient can also carry a chaotic label \cite{grebogi1983crises}. In random samples of 192
chaos-labeled cells from each of the two main float64 maps (Figs.~\ref{fig:stripes}(a) and~\ref{fig:teaser}), 176 and
143 cells were never captured by a periodic orbit in $10^5$ steps, and all but one of them kept a positive exponent;
the others were long transients that ended on periodic orbits after 600 to 97{,}500 steps. Chaotic shares of a map
are therefore finite-horizon shares of its window.

Numerical error can also make a twin estimate spuriously positive. In a simple noise-floor model, both trajectories of
a twin pair carry additive noise of relative size $\sigma$, and the twin then measures about
$\max\{\lambda,\log(\sqrt2\,\sigma/\varepsilon)\}$ at offset $\varepsilon$. Below the noise level, the estimate rises
by one nat for every factor $e$ by which the offset shrinks, whereas a real exponent does not depend on the offset.
The model is a heuristic, not a theorem: rounding can also suppress separation, change the sampled dynamics or move a
trajectory into another basin. The label checks below therefore compare twin offsets and automatic-differentiation
tangent vectors. Tested on the network, the model accounts for the false positive exponents that a final
normalization computed in single precision produced in our early two-GPU runs at the 70B scale\smref{Methods}.

\paragraph{Chaos with periodic windows.} Figure~\ref{fig:stripes}(a) shows a $256\times256$ float64 scan over
$\alpha\in[0.060,0.082]$ and $\log_{10}T\in[-0.17,-0.04]$. More than half of its cells (35{,}650 of 65{,}536) meet the finite-horizon criterion, and 92\,\% of a random sample of them stay chaotic for $10^5$ steps, and periodic windows of periods 2--32 are embedded in the chaos; many of them form a fan of thin stripes that
adjoins a period-2 region. Figure~\ref{fig:teaser} resolves the stripes' lower ends at $512\times512$ cells (78{,}868 meet the criterion), and Fig.~\ref{fig:stripes}(b) labels the same cells by the texts of their attractors. The 2-cycle ``is it''
fills the left, bounded by its fold curve, on which the stripes accumulate (Sec.~\ref{sec:adding}). A closed crescent of
the period-3 loop ``\textbackslash n\textbackslash nThe existence'' (``\textbackslash n\textbackslash n'' is a blank
line) contains its period doublings and a chaotic core (Sec.~\ref{sec:cascade}). The lower right is chaotic, filled
with small hooked windows of ``\textbackslash n\textbackslash nThe question'' and ``\textbackslash n\textbackslash nThe
existence'' loops.

\begin{figure*}[tb]
\centering
\includegraphics[width=\textwidth]{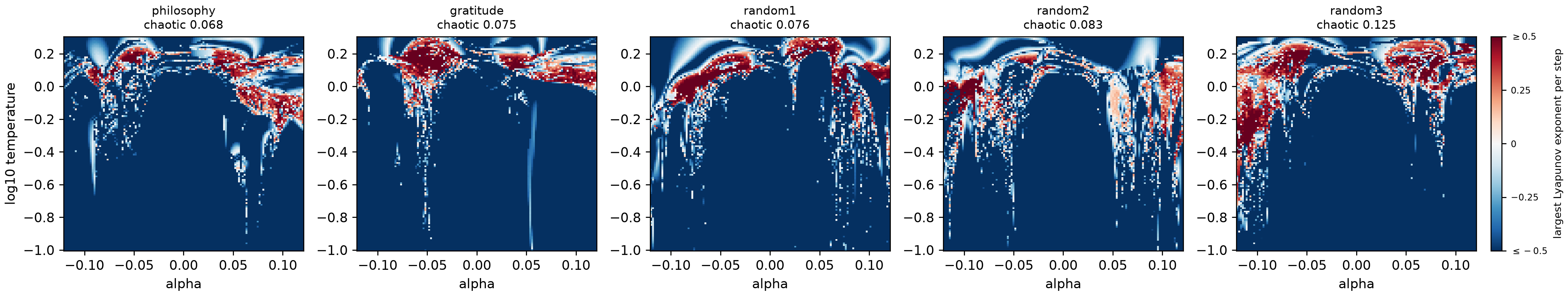}\\[3pt]
\includegraphics[width=\textwidth]{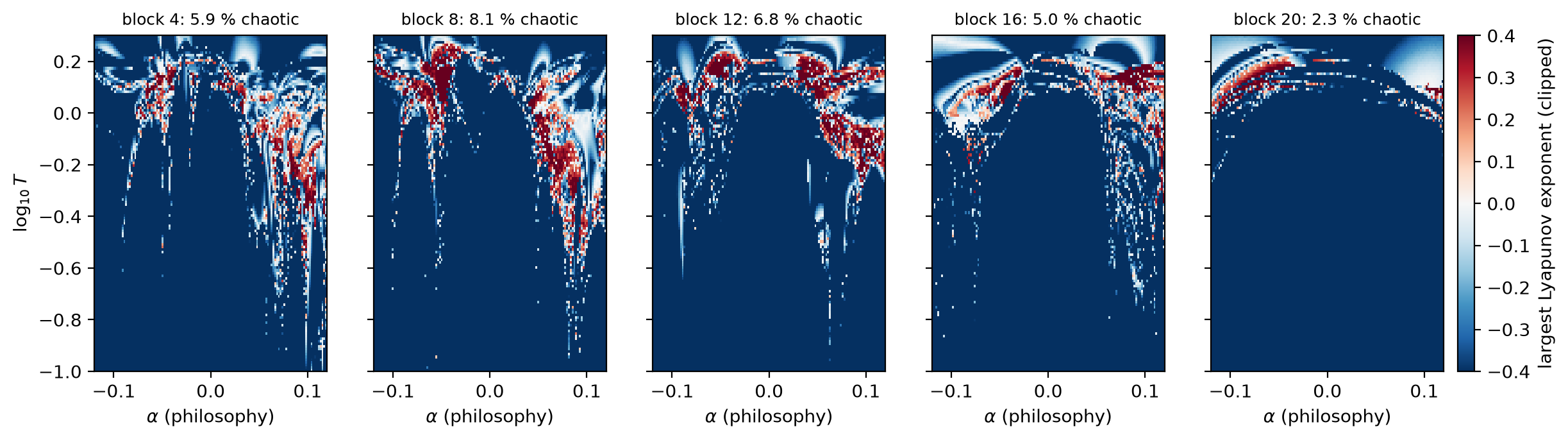}
\caption{Which direction, and where it enters (Qwen2.5-0.5B-Instruct, $128^2$ float64 cells over the steering strength
and the temperature). (a) Top row: steering at layer 12 along the philosophy and gratitude directions and along three Gaussian
directions rescaled to the philosophy vector's norm: all drive the loop into chaos. (b) Bottom row: the philosophy direction built
with the same recipe at blocks 4, 8, 12, 16 and 20 of 24. Color: largest Lyapunov exponent, clipped at $\pm0.5$ in (a)
and $\pm0.4$ in (b); titles give the chaotic share (finite-horizon exponent above $10^{-3}$, no period up to 48 found).}
\label{fig:controls}
\end{figure*}
\begin{figure*}[tb]
\centering
\includegraphics[width=\textwidth]{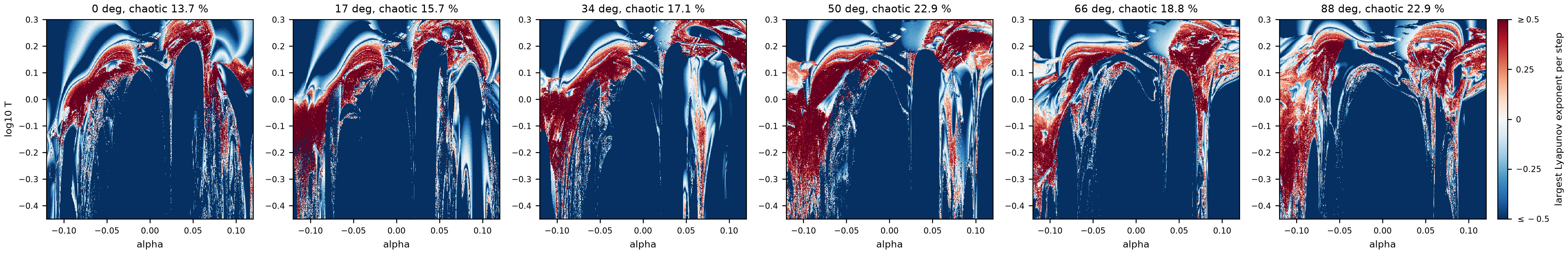}
\caption{The chaos moves as the steering direction rotates. The soft-token map of Qwen2.5-0.5B-Instruct (steering at
layer 12) while the direction rotates at constant norm from one Gaussian direction ($0^{\circ}$) to another
($90^{\circ}$): largest Lyapunov exponent (red positive, clipped at $\pm0.5$) at six of 103 angles, with the chaotic
share above each panel; over all 103 angles it stays between 13 and 24\,\%. Float32 maps of $296\times512$ cells over
$\alpha\in[-0.12,0.12]$ and $\log_{10}T\in[-0.45,0.3]$, with exponents from 80 measured steps at twin offset
$10^{-4}$ and periods up to 24: short-horizon labels, coarser than those of the float64 maps. Each map continues from the final states of
the previous angle, so where attractors coexist the maps can jump.}
\label{fig:morph}
\end{figure*}
\begin{figure*}[tb]
\centering
\includegraphics[width=0.39\textwidth]{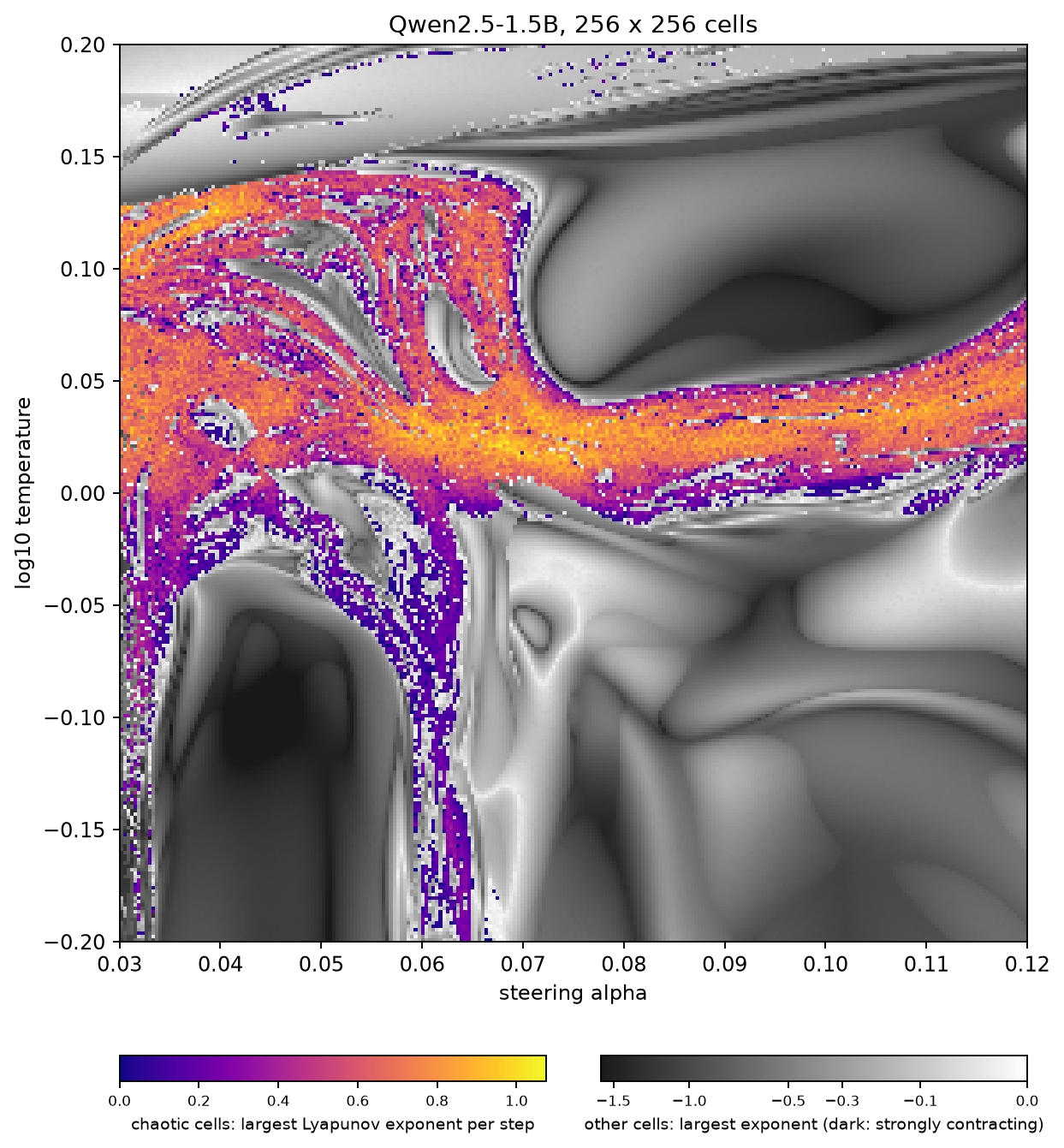}\hspace{0.04\textwidth}
\includegraphics[width=0.39\textwidth]{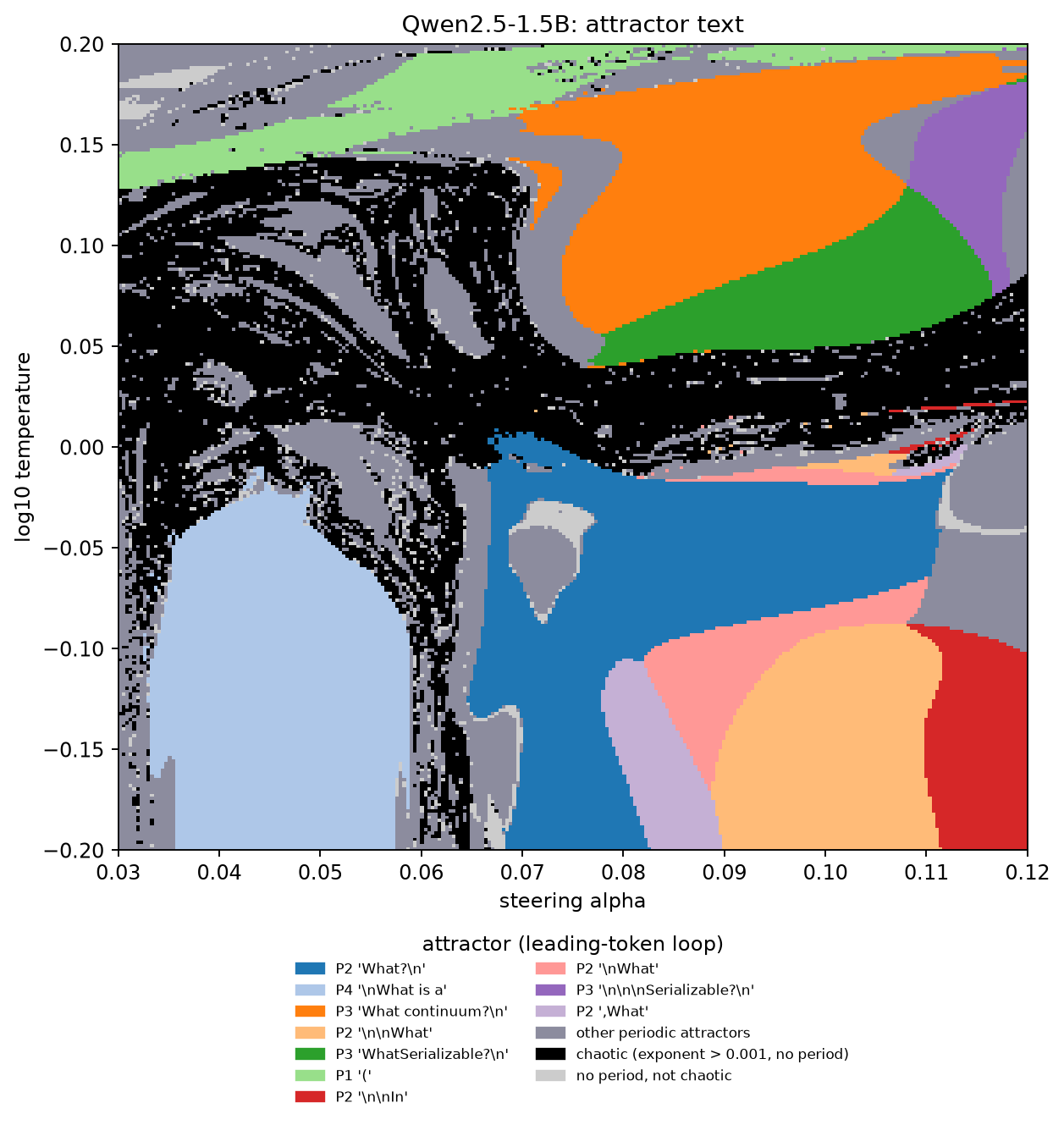}
\caption{Qwen2.5-1.5B soft-token map steered along the philosophy direction at layer 14, $256^2$ float64 cells over
$\alpha\in[0.03,0.12]$, $\log_{10}T\in[-0.2,0.2]$: largest Lyapunov exponent (left, in the color scheme of Fig.~\ref{fig:teaser})
and attractor texts (right, as Fig.~\ref{fig:stripes}(b)). Where two period-2 regions touch without a gap,
the boundary may only mark a change of leading token along one orbit\smref{Methods}.}
\label{fig:qwen15}
\end{figure*}
\begin{figure*}[tb]
\centering
\includegraphics[width=0.84\textwidth]{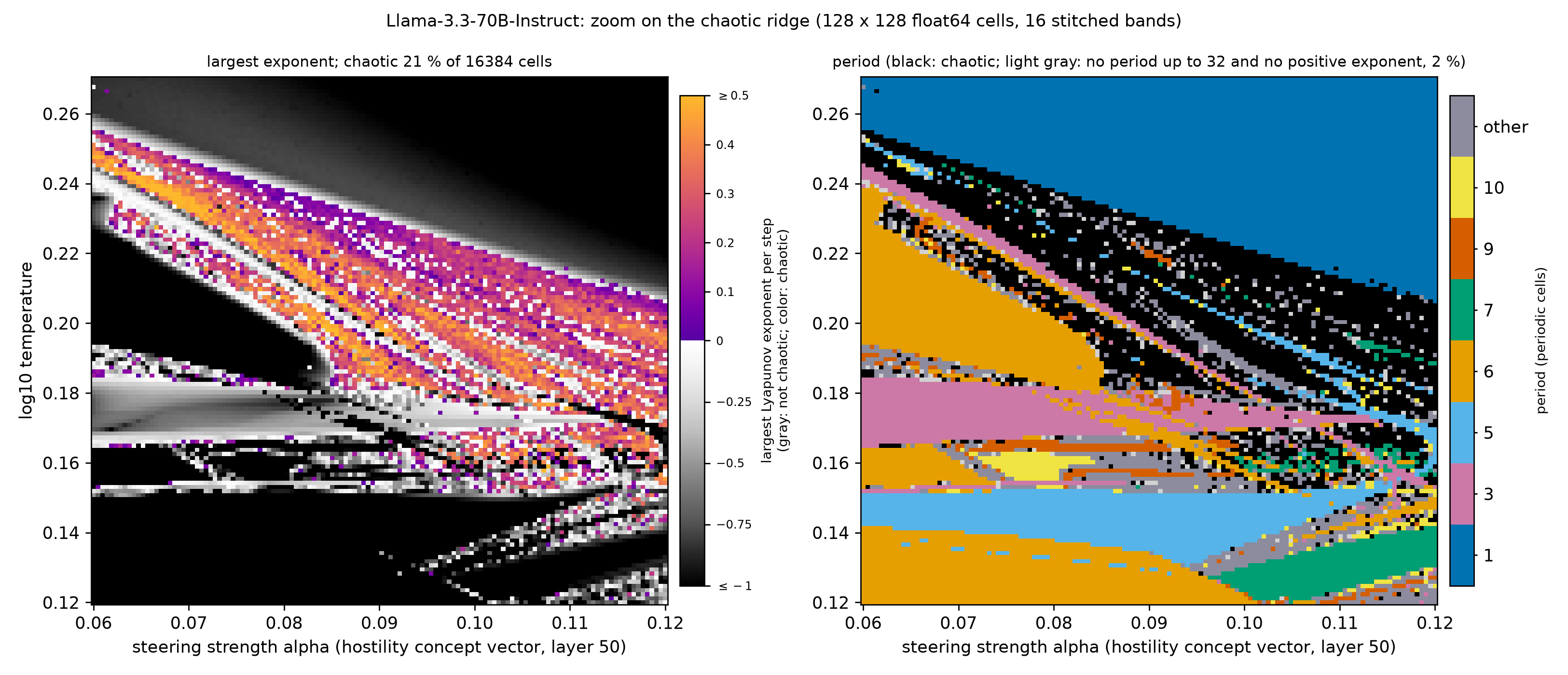}
\caption{Chaos at 70B parameters: the soft-token map of Llama-3.3-70B-Instruct steered along its hostility concept
vector at layer 50, $128\times128$ float64 cells over $\alpha\in[0.06,0.12]$, $\log_{10}T\in[0.12,0.27]$ (300
transient and 300 measured steps). Left: largest Lyapunov exponent, gray where the cell is not chaotic and in color
where it is; 3410 of the 16{,}384 cells (20.8\,\%) are chaotic (largest exponent 0.64). Right: period of the attractor
reached from the fixed start (black: chaotic; bluish gray: periods without a color of their own; light gray: neither a detected period up
to 32 nor a positive exponent, 1.7\,\% of the cells). A chaotic band, whose center falls from $\log_{10}T\approx0.23$ at $\alpha=0.06$ to $\approx0.19$
at $\alpha=0.12$, is threaded by periodic windows, most often of periods 3, 5, 6 and 7 and up to 13; fixed points lie
above it.}
\label{fig:z128}
\end{figure*}

\paragraph{Accuracy of the labels.} At 448 cells of two scans, drawn at random within the label classes (192 chaotic,
192 periodic and 64 with neither label), automatic-differentiation tangent vectors carried by forward-mode differentiation reproduce the
twin exponents within $0.0041$. With 1000 transient and 2000 measured steps at two twin offsets, 86 and 92 of the 96
chaotic labels sampled in the scans of Figs.~\ref{fig:teaser} and~\ref{fig:stripes}(a) remained chaotic (13 of the
other 14 ended on periodic orbits after chaotic transients). Where the label held, the 300-step exponent differs from the 2000-step one by a median of
0.020 and 0.030 (90th percentiles 0.064 and 0.096; at most 0.30). All 192 sampled periodic labels held, and 56 of 64
sampled cells with neither label settled on periodic orbits with periods up to 64. Raising the chaos threshold from
$10^{-3}$ to $0.05$ lowers the count of Fig.~\ref{fig:teaser} only from 78{,}868 to 73{,}437 cells. Near the
accumulation point of a cascade, transient chaos makes the labels less reliable: on one such line, 135 of 235 cells
labeled chaotic from 400 measured steps stayed chaotic. An independent float64 implementation of the map, a manual
forward pass of the stored checkpoint that uses neither our tools nor the Transformers library, reproduces the chaos:
with the scans' protocol, its 300-step exponents at two chaotic cells of Fig.~\ref{fig:teaser} are 0.098 and 0.354,
against 0.113 and 0.359 in the scan (0.106 and 0.332 over 1000 steps), and at the certified period-8 orbit it gives
$-0.564$ per step over 1000 steps ($-0.579$ over 300), the value $\ln(0.0110)/8$ that the orbit's leading multiplier predicts.

\paragraph{Generality.} Without steering, temperature alone gives chaos with embedded windows in most dense models we
swept (eleven of fourteen on regular grids), among them Llama-3.1-8B, Qwen2.5-7B and OLMo-2-7B\smref{Unsteered chaos and
steering controls}. Steering widens the chaotic regions, but not as a property of the concept: on one $128^2$ scan of
the 0.5B model, 7.6, 8.3 and 12.5\,\% of cells are chaotic for three random Gaussian directions with the philosophy
vector's norm, against 6.8\,\% for philosophy (Fig.~\ref{fig:controls}(a)), and the single MLP neuron most specific to philosophy, chosen by a rule
fixed before any scan (Fig.~\ref{fig:neuronsteer}), gives 9.9\,\%; in Llama-3.1-8B, a sparse-autoencoder feature
direction \cite{goodfire2025sae}, the contrastive direction and random directions all drive the map into chaos (27 to
39\,\% of cells). Rotating the steering direction at constant norm moves the chaotic regions but keeps the chaotic
share between 13 and 24\,\% at every one of 103 angles (Fig.~\ref{fig:morph}), and the block at which a vector enters
matters: built with the same recipe at blocks 4, 8, 12, 16 and 20, the philosophy direction makes 5.9, 8.1, 6.8, 5.0
and 2.3\,\% of the same plane chaotic (Fig.~\ref{fig:controls}(b)). The
direction selects the texts of the attractors. The organization of windows recurs in Qwen2.5-1.5B (Fig.~\ref{fig:qwen15})
and 7B, in OLMo-2-7B and in Llama-3.1-8B, and at 1.5B and 7B the windows accumulate on the fold of a short cycle as in
Sec.~\ref{sec:adding}\smref{Replication and scale}. Larger models do not simply leave the chaos behind: in a test across
prompts fixed in advance, Qwen2.5-14B was chaotic at more of 48 temperatures than the 7B model on all eight test prompts
(9 to 17, against 0 to 7), at and above the temperature where its sharper next-token distribution spreads, and steering drives
both 70B-class models that we steered into chaos, Llama-3.3-70B broadly and Qwen2.5-72B in a narrow band (six
confirmed chaotic cells of 1024)\smref{Chaos and model size}. At 70B parameters the same organization appears (Fig.~\ref{fig:z128}): in a $128\times128$ scan of
Llama-3.3-70B-Instruct steered along its hostility direction, 20.8\,\% of the cells are chaotic, in a band threaded by
periodic windows; a test of 24 sampled cells at three twin offsets and an automatic-differentiation tangent confirms 11 of the 12 chaotic and 5 of the 6 periodic labels (the other 6 sampled cells were unresolved)\journalonly{ (supplementary material)}.
Unsteered, the larger Qwen2.5 and Llama-3 models damp a change of the fed-back token more strongly (not monotonically,
and not in OLMo-2). Each block adds its output to the residual stream, so the stream grows from block to block, and the
change that the fed-back token makes in it grows too. In Qwen2.5-0.5B and Llama-3.1-8B the two grow at nearly the same
rate; in Qwen2.5-72B and Llama-3.1-70B the stream grows much faster, so the change becomes an ever smaller fraction of
the state: these blocks write a growing state that depends little on the fed-back token. A mixture of experts with 3B
of its 35B parameters active, Qwen3.5-35B-A3B, contracts where dense 4B and 9B models of its family are
chaotic\smref{Chaos and model size}.

\begin{figure}[tb]
\centering
\includegraphics[width=\columnwidth]{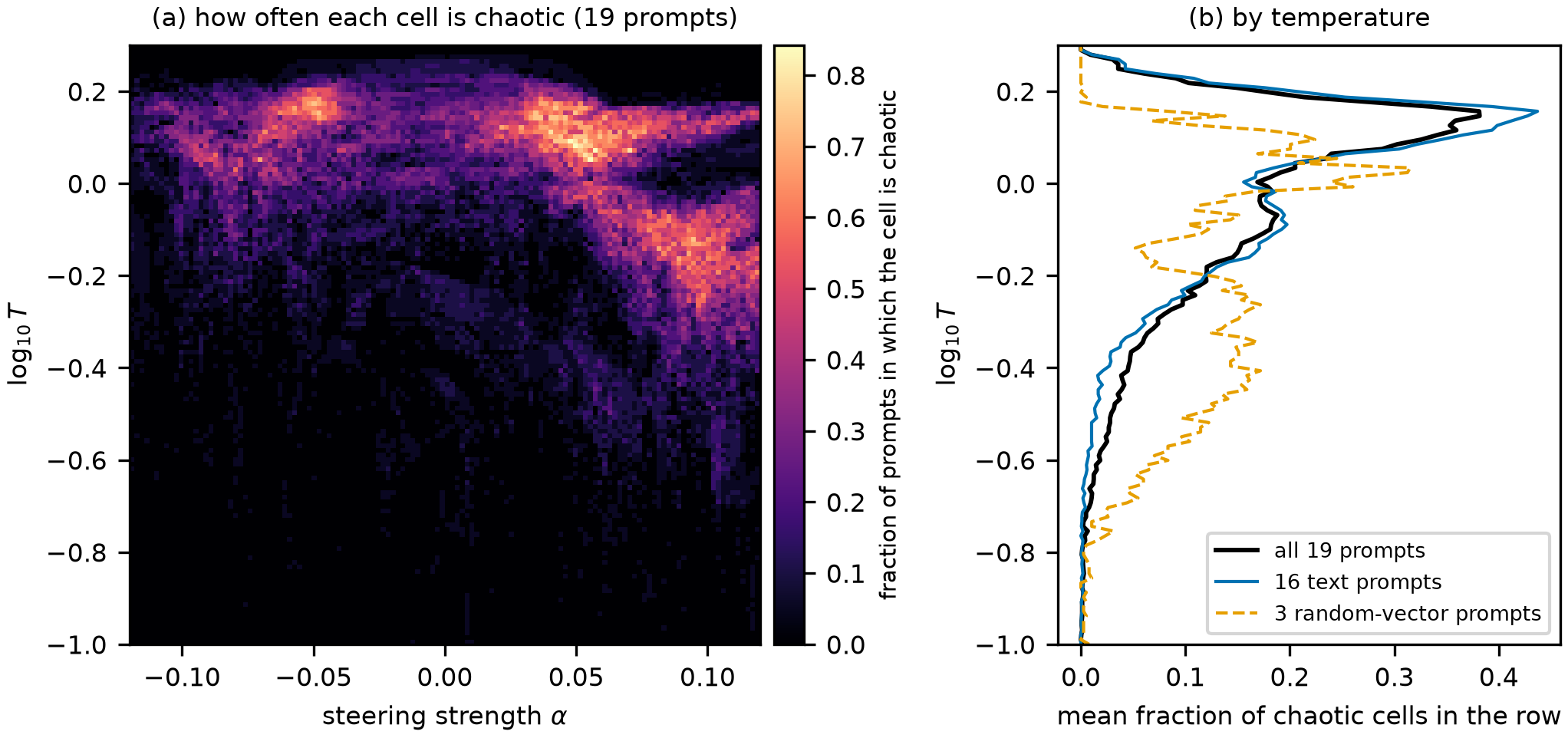}
\caption{Where the chaos is found across prompts (Qwen2.5-0.5B-Instruct, philosophy direction at layer 12, float64;
$128\times128$ cells over $\alpha\in[-0.12,0.12]$ and $\log_{10}T\in[-1,0.3]$). (a) For each cell, the fraction of the
19 prompt variants (the random-start runs left out) in which it is chaotic, with a largest exponent above $10^{-3}$. (b) The mean
fraction per temperature row, for all prompts, the text prompts and the three random-vector prompts.}
\label{fig:promptfreq}
\end{figure}
\begin{figure*}[tb]
\centering
\includegraphics[width=\textwidth]{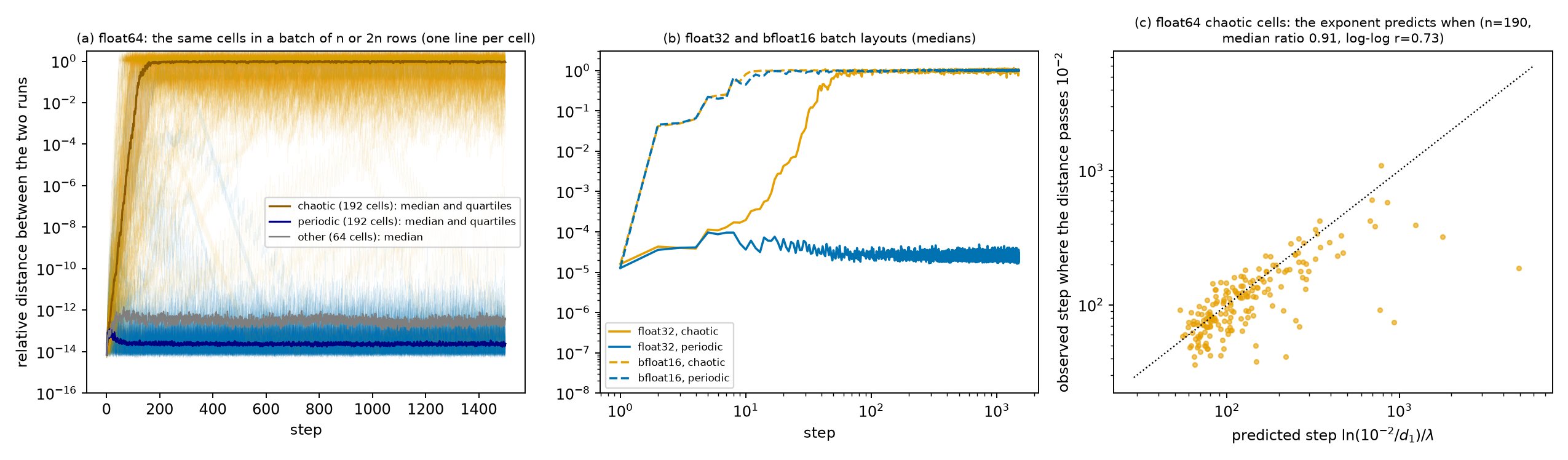}
\caption{Only the batch layout differs. The 448 cells (192 chaotic, 192 periodic, 64 with neither label) of two float64 scans of Qwen2.5-0.5B-Instruct (philosophy
direction, layer 12) run twice from the same start, in one batch of $n$ rows or followed by $n$ further rows. (a) float64: the relative
distance between the two runs for every chaotic (orange) and periodic (blue) cell, with the medians and interquartile
ranges of each label (dark) and the median of the 64 cells with neither label (gray); the chaotic runs
separate to order one; the median periodic distance stays at the rounding level, with exceptions. (b) Medians for float32 and bfloat16 batch
layouts; in bfloat16 both kinds separate within a few steps. (c) float64 chaotic cells: the step at which the distance
passes $10^{-2}$ against $\ln(10^{-2}/d_1)/\lambda$, with $d_1$ the smallest nonzero distance in the run and $\lambda$
the map exponent (a retrospective comparison); dotted, equality.}
\label{fig:repro}
\end{figure*}

\paragraph{The prompt.} All our other 0.5B results use one prompt, so we mapped the standard plane ($\alpha$ from
$-0.12$ to $0.12$, $\log_{10}T$ from $-1$ to $0.3$; $128\times128$ float64 cells, philosophy direction at layer 12) under
22 conditions: the base prompt itself; each of its two parts alone, and neither (Qwen's default system prompt, which the
chat template inserts when none is given); the system prompt's words in two random orders, the user message's two
tokens reversed, and both changes together; an opposite system prompt; happy and sad versions of the system prompt and of the user message; the
same texts without chat markup, a system prompt three times as long and a two-turn conversation; three prompts that are
a single random vector, with no text at all (a Gaussian direction at the median embedding norm, the embedding of a
random token, and a Dirichlet(0.1) mixture of the embeddings of 64 random tokens); and the base prompt with the loop started from three random vectors.
Every prompt we tested gives chaos with periodic windows: 5.9 to 14.9\,\% of the plane is
chaotic (7.0\,\% for the base prompt), with largest exponents of 0.62 to 0.85 per step (here chaotic means a largest exponent above $10^{-3}$ alone; this
finite-window indicator overlaps a detected period at 3 to 18 cells per map). Removing the user message doubles the chaotic share, and even a prompt that is one random vector gives
8 to 10\,\%. In every prompt we tested, chaos and periodic windows appear, but the prompt decides where they lie: the
chaotic cells of each of the other 18 prompts overlap those of the base prompt by 1 to 37\,\%
(Jaccard index). For the three random starts, the chaos/non-chaos threshold labels agree with the base run at
98.9 to 99.0\,\% of the cells, while the periods detected within the run agree at only 86.5 to 93.1\,\%, so within
this horizon the orbit reached can depend on the start. Across the 19 prompts there is
nonetheless a common structure (Fig.~\ref{fig:promptfreq}): 48.3\,\% of the cells are chaotic for at
least one prompt but only 3.0\,\% for half of them or more, and these hot spots lie at moderate
steering of either sign in a band of temperature, where the mean fraction of chaotic cells per row peaks at 38\,\% near $\log_{10}T=0.15$, against 0.8\,\% below
$\log_{10}T=-0.5$. The three random-vector prompts move the band to colder temperatures: their mean profile peaks
near $\log_{10}T=0.03$, that of the text prompts near $0.16$.

\paragraph{Reproducibility and answers.} GPU kernels choose different reduction orders for different matrix shapes, so
a request's result depends on its batch layout, the set of requests that the GPU processes together with it
\cite{he2025nondeterminism,yuan2025nondeterminism}. In our float64 tests, whether this rounding difference grows
depends on the loop's regime (Fig.~\ref{fig:repro}). Run twice from the same start with only the
grouping of cells into batches changed, 190 of 192 chaotic float64 cells of the 0.5B model separate past a relative
distance of $10^{-2}$, close to the step their Lyapunov exponent gives in a retrospective comparison (median ratio
0.91), while 182 of 192 periodic cells
stay within $10^{-2}$ for all 1500 steps. The same holds in Qwen2.5-7B and Llama-3.1-8B without steering and in
Llama-3.3-70B with it: 24 of 24, 20 of 20 and 24 of 24 chaotic loops separate, against 0 of 24, 1 of 20 and none of 24
periodic ones. In the 0.5B
model the most probable tokens of the two runs come to differ, within the last 500 steps, in 173 of the chaotic cells
but in only 7 of the periodic ones. Near an
attracting periodic orbit the feedback contracts a small rounding difference of non-associative floating-point sums
\cite{goldberg1991floating}; along a chaotic trajectory it amplifies it, on average at the rate $\lambda$. A control fixed in advance supports this account:
with the matrix shapes held fixed the runs are bitwise identical in all 448 cells (the 384 labeled ones and 64 with
neither label), and injected displacements grow on
the Lyapunov schedule (median ratio 0.87 to 1.18), although, against the design's limit, more than 10\,\% of the
periodic cells also separated at displacements of $10^{-9}$ and larger, presumably by crossing basin boundaries (the
attractors reached were not classified). Batch-invariant
kernels \cite{he2025nondeterminism} remove this one source of difference, but along a chaotic trajectory other small
differences, such as those between two GPUs, generically grow at the same rate. In bfloat16 the runs part within one or two steps at almost every cell, periodic ones included.

The feedback can carry such differences into answers, mainly where a pipeline answers poorly or its reasoning reaches
its budget. When each of 96 two-digit
additions ran once alone and once beside three other additions, a short Soft Thinking procedure gave different answers
to 56\,\% of them in Qwen2.5-0.5B at $T=1$ and 55\,\% in Qwen2.5-7B at $T=10^{0.2}$ (for comparison, forced greedy decoding
changed none and 10 of 96). There the procedure answers poorly: the 0.5B model answered 23 of the 96 correctly when alone, and in 24 of the 7B
model's 53 changes one of the two runs gave no number at all. At $T=10^{0.4}$, where the two models answered 90 and 96
of the 96 correctly, the layout changed no answer. In float64 no answer changed in the 0.5B model and 5\,\% changed in
the 7B model, whose loop expands at $T=10^{0.2}$. In an exploratory test on 128 newly drawn additions in bfloat16 on an
H100, replacing the target's cache and logits by those of a batch-1 reference run before every soft step, so that each
step's rounding difference is not carried into the next, cut the share of changed answers from 47 to 0.8\,\% in the 0.5B model and from 41\,\% to none in the
7B model. In a test whose design was fixed before its data were collected, Soft Thinking with its published settings
\cite{zhang2025softthinking} in Qwen2.5-7B gave different answers to 21\,\% of 256 GSM8K problems in batches of 16
prompts and of 8 (6\,\% when both runs were fed the soft tokens recorded in a reference run, so that the fed-back
tokens were identical); through this pipeline the model answers only 22 to 30\,\% of the problems correctly, and
whether an answer was correct changed for 6.2\,\% of them. Ordinary greedy chain-of-thought decoding, which answers
89\,\% of these problems correctly, changed 2.3\,\% of its answers, and feeding back only the most probable token
changed 17.6\,\%, not resolvably different from soft feedback. In three reasoning models that answer 74 to 90\,\% correctly, DeepSeek-R1-Distill-Qwen-7B,
DeepSeek-R1-Distill-Llama-8B and QwQ-32B, the same change of batch layout changed 14.5, 28.9 and 19.5\,\% of the Soft
Thinking answers and 10.2, 21.5 and 19.9\,\% of those of discrete sampled thinking. Nearly all of these changes, 158 of
161 with soft thinking and all 132 with sampled thinking, occurred on problems that reached the thinking budget (1536
steps) or the answer budget (256 tokens) in at least one layout; on problems that finished within both budgets in both
layouts, the layout changed 3 of 480 soft answers and none of 477 sampled ones. (The random draws are indexed by step,
so a different end of thinking also changes the draws of a sampled answer.) An exploratory rerun with a 4096-step
budget, which also added the Soft Thinking stopping rule and, for this model, ran on another H100 variant, still
changed 8.6\,\% of the soft and 17.2\,\% of the sampled answers of DeepSeek-R1-Distill-Qwen-7B on 128 GSM8K problems;
these results do not establish the rate for reasoning that completes. Tests fixed in advance found no accuracy signature of the regime in ordinary sampled answers. On 4000 products of
two two-digit numbers, prompts whose unsteered Qwen2.5-7B-Instruct loop is chaotic are answered 7.1 points less
accurately, but the gap vanishes once difficulty covariates fixed before sampling enter the fit ($+0.15$ points, 95\,\%
interval $[-1.4,1.7]$), and the first-token entropy alone absorbs nearly all of it. Within prompts, under slight
steering that changed the loop's label for 508 of 1000 prompts, a chaotic label was associated with $+0.36$ points of
accuracy (95\,\% interval $[-0.36,1.08]$), an association rather than a causal effect of entering
chaos\smref{Behavioral tests in detail}.

\section{Periodic windows and the saddle-node bottleneck}
\label{sec:adding}
Every periodic window writes a repeating text, and the fan of stripe windows in Fig.~\ref{fig:stripes}(a) is organized by the ghost of
a vanished orbit, the slow bottleneck that a saddle-node bifurcation leaves behind
\cite{strogatz1989predicted,strogatz2018nonlinear}: how long trajectories linger near the ghost follows the fold's normal form
computed from the model's derivatives, and where the windows lie follows a scaling law that the normal form motivates. Such low-dimensional laws can hold in 896 dimensions because center-manifold
reduction restricts the dynamics near a bifurcation to its critical directions \cite{kuznetsov2023elements,kuznetsov2019maps}.

\begin{table}[tb]
\caption{Periodic orbits of the stripe family at the stripe centers along $\log_{10}T=-0.115$ ($\alpha=0.077872$,
$0.071674$, $0.067766$, $0.065550$, $0.064443$, $0.063857$, $0.063534$, $0.063335$): Newton residuals $6\times10^{-15}$ to
$8\times10^{-13}$, leading multiplier $\mu_1$ and next modulus $|\mu_2|$ from Rayleigh--Ritz values. Each step in
period adds one ``is X'' pair to the text.}
\label{tab:orbits}
\begin{ruledtabular}
\begin{tabular}{rlcc}
period & orbit (leading tokens) & $\mu_1$ & $|\mu_2|$ \\
\hline
6 & . This is a is a & $+0.0052$ & $3\times10^{-6}$ \\
8 & . This is there is a is a & $+0.0110$ & $3\times10^{-7}$ \\
10 & . This is there is it is a is a & $+0.0025$ & $2\times10^{-7}$ \\
12 & . This is there (is it)$^2$ is a is a & $+0.0163$ & $4\times10^{-8}$ \\
14 & . This (is it)$^4$ is a is a & $+0.0369$ & $2\times10^{-9}$ \\
16 & . This (is it)$^5$ is a is a & $+0.0716$ & $8\times10^{-11}$ \\
18 & . This (is it)$^6$ is a is a & $+0.0792$ & $1\times10^{-12}$ \\
20 & . This (is it)$^7$ is a is a & $+0.0857$ & $2\times10^{-14}$ \\
\end{tabular}
\end{ruledtabular}
\end{table}

\begin{table*}[tb]
\caption{Chaotic loops and copies started $10^{-10}$ away: most probable tokens around their first difference (red
from the bar $\Vert$ on; $\hookleftarrow$: line break).\journalonly{ More in the \sm{Chaos with periodic windows}.}}
\label{tab:chaostexts}
\footnotesize
\begin{ruledtabular}
\begin{tabular}{ll}
\multicolumn{2}{l}{Qwen2.5-0.5B, philosophy, $\alpha=0.0652$, $\log_{10}T=-0.156$: $\lambda=0.10$, texts split at step 151} \\
loop & \texttt{$\hookleftarrow$$\hookleftarrow$The\ existence$\hookleftarrow$$\hookleftarrow$The\ existence$\hookleftarrow$$\hookleftarrow$The\,$\Vert$\,\textcolor{BrickRed}{\ existence$\hookleftarrow$$\hookleftarrow$The\ existence$\hookleftarrow$$\hookleftarrow$The\ existence}} \\
copy & \texttt{$\hookleftarrow$$\hookleftarrow$The\ existence$\hookleftarrow$$\hookleftarrow$The\ existence$\hookleftarrow$$\hookleftarrow$The\,$\Vert$\,\textcolor{BrickRed}{\ question$\hookleftarrow$$\hookleftarrow$The\ existence$\hookleftarrow$$\hookleftarrow$The\ question}} \\
\hline
\multicolumn{2}{l}{Qwen2.5-0.5B, hostility, $\alpha=0.08946$, $\log_{10}T=-0.05566$: $\lambda=0.35$, texts split at step 45} \\
loop & \texttt{Yes!Yes!Yes!Yes!Yes!Yes!Yes!\,$\Vert$\,\textcolor{BrickRed}{Yes,\ that!Yes!Yes!No,}} \\
copy & \texttt{Yes!Yes!Yes!Yes!Yes!Yes!Yes!\,$\Vert$\,\textcolor{BrickRed}{No,\ that!Yes!Yes!Yes!}} \\
\end{tabular}
\end{ruledtabular}
\end{table*}

\begin{figure*}[tb]
\centering
\includegraphics[width=0.76\textwidth]{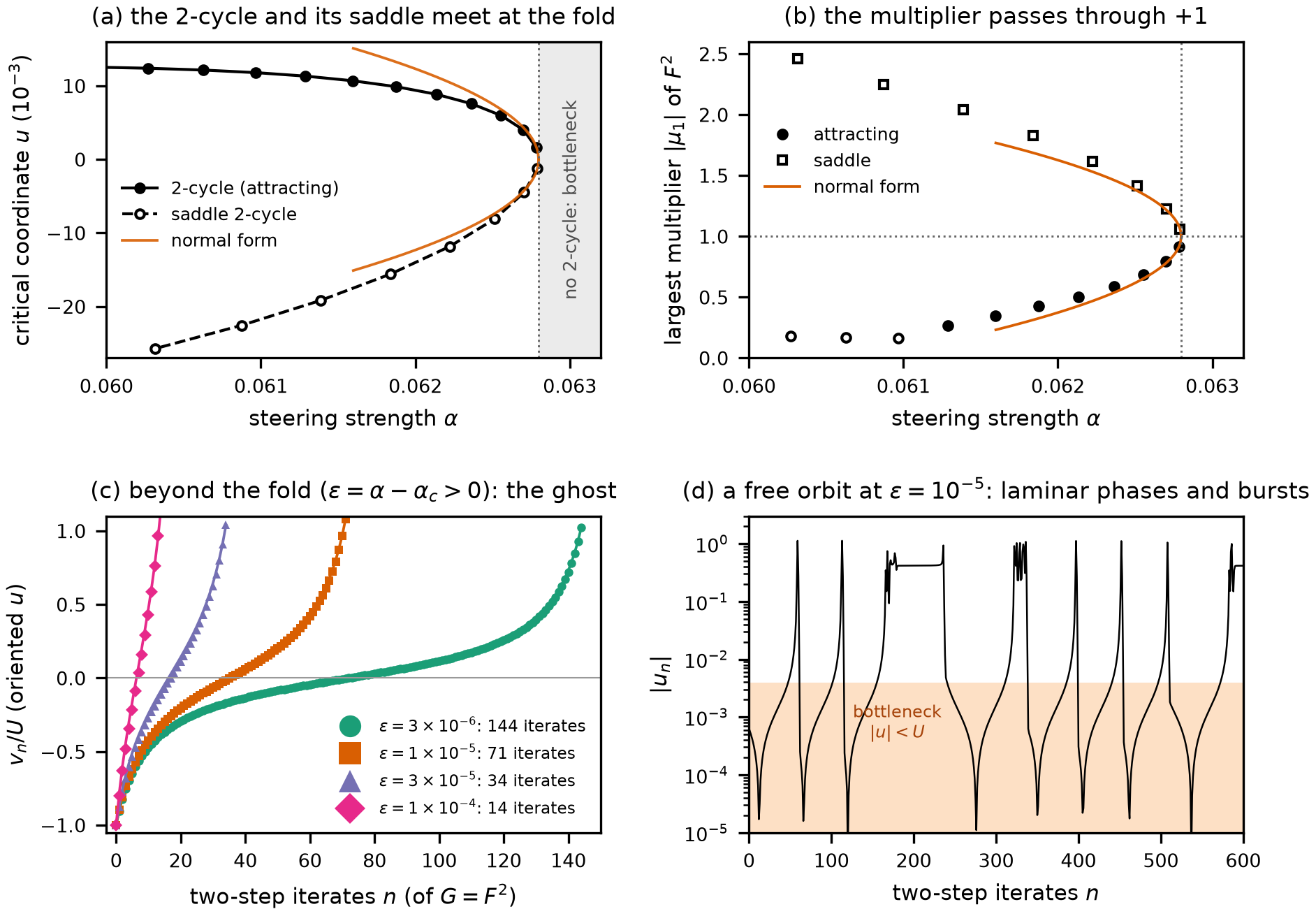}
\caption{The fold behind the stripes: the 2-cycle ``is it'' at $\log_{10}T=-0.115$ (Qwen2.5-0.5B-Instruct, philosophy
direction at layer 12). (a) Continued in $\alpha$, the attracting 2-cycle (filled) and its saddle partner (open) meet
and vanish at $\alpha_c=0.06279827$. The vertical coordinate is the projection $u$ of one point of each orbit on the
fold's critical direction; orange, the normal form's fixed points $u^*=\pm\sqrt{|a|(\alpha_c-\alpha)/|b|}$. (b) The
largest multiplier of $G=F^2$ on the two orbits reaches $+1$ at the fold, following the normal form's
$1\mp2\sqrt{ab(\alpha_c-\alpha)}$ (open circles: a complex leading pair). (c) Beyond the fold the 2-cycle is gone, but
its ghost slows the dynamics: controlled passages from the start of the channel, $v=-U$, to its exit, $v=U$ ($v$ the
oriented critical coordinate, $U=0.1/|b|$), measured on the network (markers, one shape per $\eps$) and iterated from the normal form
$v'=v+|a|\eps+|b|v^2$ (lines), take 144, 71, 34 and 14 iterates, as predicted. (d) A free trajectory at
$\eps=10^{-5}$ alternates between laminar phases in the bottleneck ($|u|<U$, shaded, a criterion on the critical coordinate
alone; 40 to 70 $G$-iterates each) and bursts
away from it, the signature of type-I intermittency. Branch residuals, fold, normal form and passages in float64; branch
multipliers in float32.}
\label{fig:fold}
\end{figure*}

\paragraph{Windows are texts.} A period-3 orbit returns to the same soft state every three steps, and at its three
states the most probable next tokens are ``\textbackslash n\textbackslash n'' (a blank line, one token), ``The'' and
`` existence'', with probabilities $0.74$, $0.36$ and $0.21$ at $\alpha=0.0642$, $\log_{10}T=-0.156$. Read greedily,
the loop writes ``The existence'' paragraph after paragraph; what it feeds back at each step, however, is the
probability-weighted average of all token embeddings, not that token. Attractors also coexist. At that same point the
loop started from a state of the ``is it'' 2-cycle (taken from its attractor at $\alpha=0.0655$) stays on the 2-cycle
and writes `` is it is it\ldots'' (most probable tokens with probabilities $0.66$ and $0.29$; 300-step exponent
$-1.19$), while from the fixed start it settles on the period-3 loop (exponent $-0.13$). At $\alpha=0.0652$ the fixed
start reaches the crescent's chaotic attractor, whose text alternates irregularly between ``The question'' and ``The
existence'' paragraphs (exponent $+0.095$), and the 2-cycle start again writes `` is it'' ($-1.29$): one model, one
steering vector and one temperature, and the text written after the transient depends on where the loop starts.

\paragraph{Chaos writes text too.} In the examples of Table~\ref{tab:chaostexts}, a chaotic loop reuses the words of neighboring loops in no fixed order. A float64 copy started $10^{-10}$ away writes the same text until their distance,
growing at about the rate the finite-time Lyapunov exponent sets, reaches 1.27 to 176\,\% of the soft token's norm; in seven
cells of Qwen2.5-0.5B and -1.5B the texts first differ at steps 10 to 151, where no token dominates. Run with six and with four threads, a change of numerical execution, the cell next to the period-3 crescent wrote the same text for
27 steps and then different texts.

Between the period-2 region and the chaos lies a fan of stripes of periods $6, 8, 10, \dots$ (Fig.~\ref{fig:stripes}(a)).
Their orbits, numerically validated in float64, read as text (Table~\ref{tab:orbits}): each is ``. This'' followed by
$N$ pairs ``is X'' and a closing ``is a is a'', with period $2N+6$, so each stripe adds one pair. At the stripe centers the orbits' linear relaxation is dominated by one direction: the leading multiplier, $0.0025$
to $0.086$, exceeds the next by more than three orders of magnitude (the next falls from $3\times10^{-6}$ at period 6
to $2\times10^{-14}$, at the limit of float64 accuracy, at period 20), so near these orbits a small deviation soon lies
along a single direction. This is a property of these orbits, not of the map as a whole: the two leading multipliers
of the 2-cycle at $(\alpha,\log_{10}T)=(0.061579,-0.17263)$, $-0.271$ and $-0.058$, differ by a factor of only about
5.

\paragraph{The fold.} At $\log_{10}T=-0.115$ the 2-cycle ``is it'' dies in a fold at $\alpha_c=0.06279827$ (in float64 the critical multiplier equals 1 to within $3.1\times10^{-13}$, and the next has modulus $0.037$): it meets an unstable
2-cycle, and both vanish (Fig.~\ref{fig:fold}(a,b)). In the critical coordinate $u$ of the second iterate $G=F^2$, the
fold's normal form is
\begin{equation}
u'=u+a\eps+bu^2,\qquad \eps=\alpha-\alpha_c,
\label{eq:foldnf}
\end{equation}
with $a=-4.834$ per unit $\alpha$ and $b=-25.4285$, so $ab=122.93$. Here $q$ is the critical eigenvector of
$DG$ with $|q|=1$, $\ell$ the left one with $\ell^\top q=1$, $u$ the coordinate along $q$, $a=\ell^\top\partial
G/\partial\alpha$ and $b=\tfrac12\ell^\top D^2G[q,q]$, computed from the network's first and second derivatives
\cite{kuznetsov2019maps}. Equation~\eqref{eq:foldnf} is the quadratic truncation of $G$ on its center manifold: it
neglects terms of order $u^3$, $\eps u$ and $\eps^2$, and it applies because the multiplier $+1$ is simple, $a$ and $b$ do not vanish, and the other multipliers lie inside
the unit circle \cite{kuznetsov2023elements}. No symmetry of the map forces a pitchfork and no persistent branch forces a
transcritical exchange, so these conditions make the point a fold. The coefficients
pass two checks: the stable branch's multipliers give $123.1$ independently (through $(1-\mu)^2/(4(\alpha_c-\alpha))\to ab$, quadratic fit), and an independent implementation
recovers $\alpha_c$ to $6\times10^{-15}$, $a$ to $3\times10^{-12}$ and $b$ to within $2.3\times10^{-6}$ relatively.

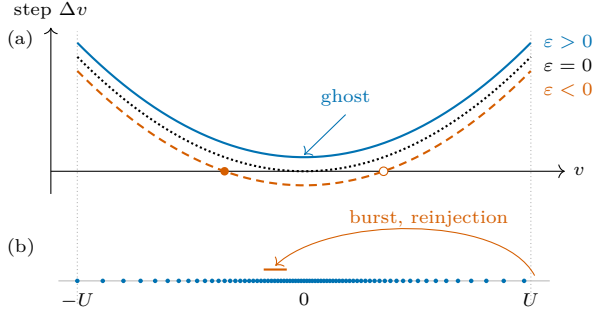
\begin{figure}[tb]
\centering
\begin{tikzpicture}[x=1cm, y=1cm, font=\scriptsize]
\definecolor{nfblue}{HTML}{0072B2}\definecolor{nfred}{HTML}{D55E00}
\draw[gray!60, densely dotted] (-3,-1.75) -- (-3,1.85); \draw[gray!60, densely dotted] (3,-1.75) -- (3,1.85);
\node[anchor=west] at (-4.05,1.75) {(a)};
\draw[->] (-3.35,0) -- (3.45,0) node[right] {$v$};
\draw[->] (-3.35,-0.3) -- (-3.35,1.9) node[above] {step $\Delta v$};
\draw[nfred, densely dashed, thick, domain=-3:3, samples=60] plot (\x, {1.7*(-0.10947 + 0.89053*(\x/3)^2)});
\draw[black, densely dotted, thick, domain=-3:3, samples=60] plot (\x, {1.7*(0.89053*(\x/3)^2)});
\draw[nfblue, thick, domain=-3:3, samples=60] plot (\x, {1.7*(0.10947 + 0.89053*(\x/3)^2)});
\fill[nfred] (-1.052,0) circle (1.6pt); \draw[nfred, fill=white] (1.052,0) circle (1.6pt);
\node[nfblue, anchor=west] at (3.05,1.72) {$\varepsilon>0$};
\node[anchor=west] at (3.05,1.40) {$\varepsilon=0$};
\node[nfred, anchor=west] at (3.05,1.08) {$\varepsilon<0$};
\draw[<-, nfblue] (0,0.216) -- (0.55,0.75) node[above, nfblue] {ghost};
\node[anchor=west] at (-4.05,-1.0) {(b)};
\draw[gray!55] (-3.25,-1.45) -- (3.25,-1.45);
\foreach \p in {(-3.000,-1.45), (-2.663,-1.45), (-2.390,-1.45), (-2.163,-1.45), (-1.970,-1.45), (-1.804,-1.45), (-1.658,-1.45), (-1.530,-1.45), (-1.415,-1.45), (-1.311,-1.45), (-1.217,-1.45), (-1.131,-1.45), (-1.051,-1.45), (-0.978,-1.45), (-0.909,-1.45), (-0.844,-1.45), (-0.784,-1.45), (-0.726,-1.45), (-0.672,-1.45), (-0.620,-1.45), (-0.570,-1.45), (-0.523,-1.45), (-0.477,-1.45), (-0.432,-1.45), (-0.389,-1.45), (-0.347,-1.45), (-0.306,-1.45), (-0.266,-1.45), (-0.227,-1.45), (-0.188,-1.45), (-0.150,-1.45), (-0.113,-1.45), (-0.075,-1.45), (-0.038,-1.45), (-0.001,-1.45), (0.035,-1.45), (0.072,-1.45), (0.109,-1.45), (0.147,-1.45), (0.184,-1.45), (0.222,-1.45), (0.261,-1.45), (0.300,-1.45), (0.340,-1.45), (0.381,-1.45), (0.422,-1.45), (0.465,-1.45), (0.509,-1.45), (0.555,-1.45), (0.602,-1.45), (0.651,-1.45), (0.702,-1.45), (0.755,-1.45), (0.811,-1.45), (0.870,-1.45), (0.932,-1.45), (0.998,-1.45), (1.068,-1.45), (1.142,-1.45), (1.223,-1.45), (1.310,-1.45), (1.404,-1.45), (1.506,-1.45), (1.619,-1.45), (1.743,-1.45), (1.881,-1.45), (2.036,-1.45), (2.211,-1.45), (2.411,-1.45), (2.641,-1.45), (2.911,-1.45)} \fill[nfblue] \p circle (0.85pt);
\node[below] at (-3,-1.5) {$-U$}; \node[below] at (0,-1.5) {$0$}; \node[below] at (3,-1.5) {$U$};
\draw[nfred, thick] (-0.534,-1.3) -- (-0.229,-1.3);
\draw[->, nfred] (3.05,-1.38) .. controls (2.6,-0.55) and (0.3,-0.6) .. (-0.381,-1.25);
\node[nfred] at (1.65,-0.62) {burst, reinjection};
\end{tikzpicture}
\caption{The bottleneck behind the stripes, drawn from the normal form \eqref{eq:foldnf} with the network's
coefficients, in the oriented critical coordinate $v$. (a) The step of one $G$-iterate, $\Delta v=|a|\eps+|b|v^2$,
across the channel from $-U$ to $U$. Before the fold ($\eps=-10^{-5}$, dashed) the step vanishes at two fixed points of
$G$: the attracting 2-cycle (filled) and its unstable partner (open). At the fold ($\eps=0$, dotted) they merge. Beyond
it ($\eps=10^{-5}$, solid) no fixed point remains, but the step stays small near $v=0$: the ghost. (b) The 71 iterates
of a full passage at $\eps=10^{-5}$, crowded where the step is smallest; the network takes 71 as well. Schematically, a free trajectory leaves at $U$, makes a burst and is reinjected slightly upstream of the middle, so
its passage is shorter than a full one; the bar marks the measured entry points, which the normal form does not
predict.}
\label{fig:passage}
\end{figure}

\paragraph{Passages and the window law.} Beyond the fold the 2-cycle is gone, but it leaves a ghost
(Fig.~\ref{fig:passage}(a)). Near the point where the two orbits vanished, $G$ still moves a state only a little: by
Eq.~\eqref{eq:foldnf}, one iterate changes $u$ by $a\eps+bu^2$, which is small when $\eps$ and $u$ are. A trajectory
that enters this narrow channel, the bottleneck, therefore crawls through it for many iterates before it escapes, and
the closer the fold, the slower the crawl (Figs.~\ref{fig:fold}(c) and \ref{fig:passage}(b)). We timed such passages. In the oriented
coordinate $v=\operatorname{sign}(b)\,u$, in which trajectories cross the channel from negative to positive $v$, we
started them at the channel's entrance, $v=-U$ (a full passage), or at its middle, $v=0$, with $U=0.1/|b|$, and counted $G$-iterates until $v>U$. Integrated as a flow, Eq.~\eqref{eq:foldnf} gives the passage from $v_{\rm in}$
to $U$ in closed form, $n(v_{\rm in})=[\arctan(|b|U/\kappa)-\arctan(|b|v_{\rm in}/\kappa)]/\kappa$ with
$\kappa=\sqrt{ab\,\eps}$; a full passage, $2\arctan(|b|U/\kappa)/\kappa$, approaches $\pi/\kappa$ only when the
channel is much wider than the bottleneck, $U\gg\sqrt{|a|\eps/|b|}$. The forecasts below iterate the map itself. At four distances from the fold the network's counts match those of the iterated normal form
within one iterate, with no fitted parameter (Table~\ref{tab:passages}; Fig.~\ref{fig:fold}(c)): a full passage takes
144 iterates at $\eps=3\times10^{-6}$ and 14 at $\eps=10^{-4}$, as predicted, and a start displaced by the same distance $U$ along one random direction that leaves $u$ unchanged
changes the count by at most one (other directions and sizes were not tested).

\begin{table}[tb]
\caption{Passage times through the bottleneck in $G$-iterates, measured on the network\,/\,predicted by
iterating the normal form \eqref{eq:foldnf} with the network's coefficients; no parameter is fitted. The starts are
the channel's start, $v=-U$; its middle, $v=0$; and the channel's start displaced by the same distance in a direction
that leaves $u$ unchanged.}
\label{tab:passages}
\begin{ruledtabular}
\begin{tabular}{cccc}
$\eps=\alpha-\alpha_c$ & start of channel & midpoint & displaced start\\
\hline
$3\times10^{-6}$ & $144/144$ & $73/74$ & $143/144$\\
$10^{-5}$ & $71/71$ & $36/37$ & $70/71$\\
$3\times10^{-5}$ & $34/34$ & $17/18$ & $33/34$\\
$10^{-4}$ & $14/14$ & $7/7$ & $13/14$\\
\end{tabular}
\end{ruledtabular}
\end{table}

\begin{figure}[tb]
\centering
\includegraphics[width=0.92\columnwidth]{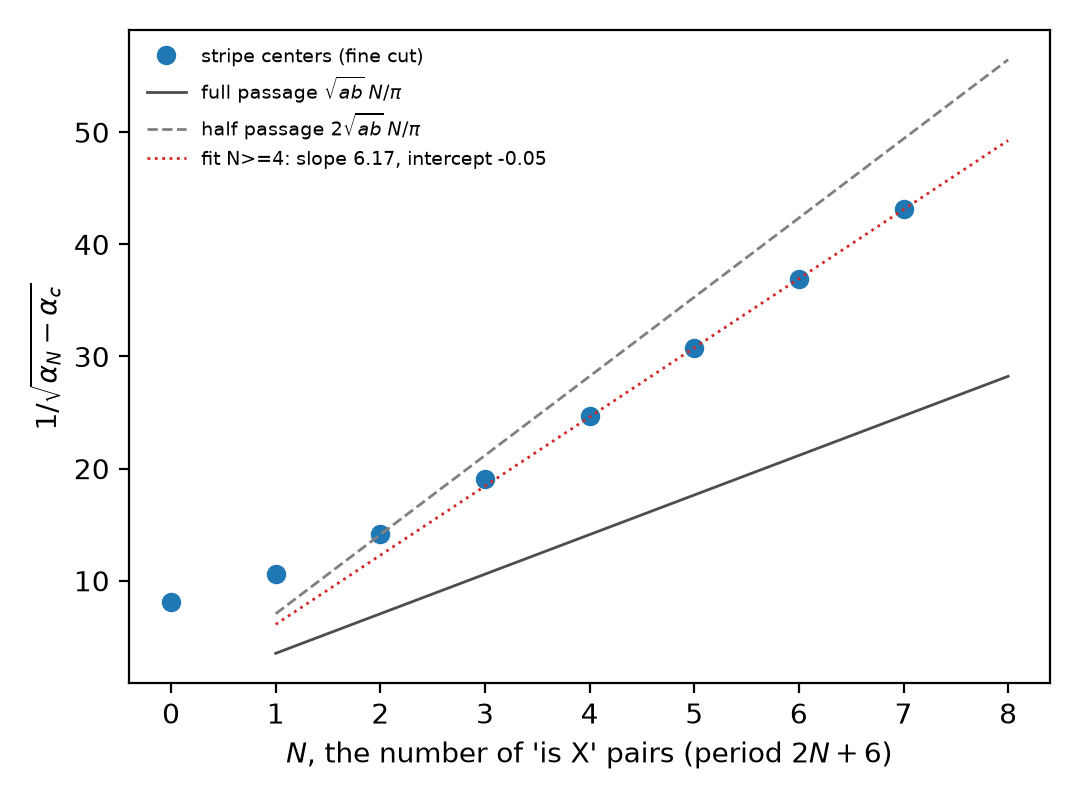}
\caption{The window law at the fold of the 2-cycle ``is it'' (Qwen2.5-0.5B-Instruct, philosophy direction at layer 12,
$\log_{10}T=-0.115$): stripe centers plotted as $1/\sqrt{\alpha_N-\alpha_c}$ against $N$, the number of ``is X''
pairs, with the full- and half-passage predictions of the long-channel formula (asymptotic; lines); the slope fitted to the
stripe centers $N=4$--7, $6.17$ per unit of $N$, is close to the $6.13$ of the fold edges (Table~\ref{tab:fraction}).}
\label{fig:windowlaw}
\end{figure}
Left to itself, a trajectory near the fold alternates between two kinds of motion (Fig.~\ref{fig:fold}(d)). In
\emph{laminar phases} it crawls through the bottleneck, close to where the 2-cycle was, and moves almost periodically;
in \emph{bursts} it leaves the channel for a short excursion and is then reinjected into it. This alternation is type-I intermittency \cite{pomeau1980intermittent}, whose laminar phases lengthen as
$\eps^{-1/2}$ toward the fold \cite{hirsch1982theory}; the word laminar comes from fluid flow, in which
smooth stretches alternate with turbulent bursts in the same way. Free trajectories re-enter the channel slightly
upstream of its middle (Fig.~\ref{fig:passage}(b)). From each measured entry point the normal form predicts the length of the laminar phase that follows: for complete
phases, which stay within $|u|<3U$ and within 0.05 of the channel in the other directions, the median signed forecast
error at the four offsets up to $\eps=10^{-4}$ lies between $-7.7\,\%$ and $-4.1\,\%$, although single phases differ by
up to 71\,\%. On average a free passage lasts 0.60 to 0.67 of the normal form's passage through the same channel from
end to end.

The bottleneck also explains where the periodic windows lie. A window of period $p$ is a loop whose text holds
$N=(p-6)/2$ pairs ``is X''; we read each pair as one $G$-iterate in the bottleneck, as the texts suggest (the laps
were not identified geometrically), and the loop closes through its six other steps, so the periods increase by two
from one stripe to the next. A passage through a long channel lasts about $\pi/\sqrt{ab\,\eps}$ iterates
\cite{hirsch1982theory,strogatz1989predicted}, so the closer the fold, the more laps fit. If a loop's passage is a
constant fraction $f$ of this long-channel value, it spends $N$ iterates in the bottleneck at the distance
$d_N=\alpha_N-\alpha_c$ from the fold for which $d_NN^2=f^2\pi^2/(ab)$: the windows crowd toward the fold as $N^{-2}$.
The local normal form motivates this window law but does not determine the global return, so we treat it as an
empirical law with an effective prefactor. The stripe centers $N=4$--7, the midpoints of the windows on a float32 cut
along $\log_{10}T=-0.115$, follow it with a log-log slope of $-2.00$ (Fig.~\ref{fig:windowlaw}). Three related
comparisons give similar fractions, 0.57 to 0.67 (Table~\ref{tab:fraction}): the free passages, measured against a
finite channel, and the window centers and the fold edges of periods 16--20, measured against the long-channel value;
the fold edges follow $\alpha_N-\alpha_c=C/(N+N_0)^2$ with $C=0.0266$ and $N_0=0.15$. With $C$ and $N_0$ frozen before
the edges of periods 22--30 were located, this law placed them within 3\,\% in distance from the fold; a model with one
fixed reinjection point did not, except at period 30.

\begin{table}[tb]
\caption{Three related comparisons of the passage fraction $f$, the share of a full passage that a loop or a free
trajectory spends in the bottleneck. The free passages (complete laminar phases of four free trajectories at each of
four distances, 22 to 110 phases) are compared with the normal form's passage through the same finite channel; the
window values are effective prefactors relative to the long-channel passage $\pi/\sqrt{ab\,\eps}$. All three share the
fold and its coefficients, so they are not independent. The fold edges' slope is the rise of $1/\sqrt{\alpha_N-\alpha_c}$ per unit of $N$.
A half passage would give $f=1/2$ and $d_NN^2=\pi^2/(4ab)=0.0201$.}
\label{tab:fraction}
\footnotesize
\begin{ruledtabular}
\begin{tabular}{@{}p{0.25\columnwidth}p{0.27\columnwidth}p{0.26\columnwidth}c@{}}
\raggedright measurement & \raggedright value & \raggedright full-passage value & $f$\tabularnewline
\hline
\raggedright free trajectories, four distances & \raggedright mean complete laminar phase ($|u|<3U$, transverse
distance below 0.05) & \raggedright passage through the whole channel & 0.60--0.67\tabularnewline
\raggedright window centers, $N=4$--7 & \raggedright $d_NN^2=0.0263$--$0.0265$ & \raggedright $\pi^2/(ab)=0.0803$ & 0.57\tabularnewline
\raggedright fold edges, periods 16--20 & \raggedright slope $6.13$ & \raggedright $\sqrt{ab}/\pi=3.53$ (ratio 1.74) & 0.58\tabularnewline
\end{tabular}
\end{ruledtabular}
\end{table}

\begin{table*}[tb]
\caption{Prospective passage tests at four new folds, not used while the bottleneck explanation was worked out. In
each of the two stages every forecast was fixed before that stage's test trajectories were run; the second stage's
choice of primary forecast followed the first stage's results. Each entry gives $G$-iterates for the three starting points: the start of the channel, its
midpoint, and the start displaced transversally. Left, the first test ($\eps$ in $\alpha$ for target A and in $\log_{10}T$ for target B): measured\,/\,forecast, the forecast iterating
the quadratic normal form. Right, the second test, at distances $\eps=\kappa^2/(ab)$ that give both folds the same
forecasts: the continuous passage integral (primary) for the start of the channel and the midpoint, the iterated
map (secondary) in parentheses, and the measured passages. Bold: outside one iterate of the primary forecast.}
\label{tab:prospective}
\footnotesize
\begin{minipage}[t]{0.42\textwidth}
\begin{ruledtabular}
\begin{tabular}{rcc}
$\eps$ & target A ($G=F^2$) & target B ($G=F^8$) \\
\hline
$3\times10^{-6}$ & $118/118$, $\mathbf{59/61}$, $117/118$ & $26/26$, $13/14$, $25/26$ \\
$1\times10^{-5}$ & $56/56$, $28/29$, $56/56$ & $10/10$, $5/5$, $10/10$ \\
$3\times10^{-5}$ & $26/26$, $13/14$, $26/26$ & $4/4$, $2/2$, $4/4$ \\
$1\times10^{-4}$ & $10/10$, $5/6$, $10/10$ & $2/2$, $1/1$, $2/2$ \\
\end{tabular}
\end{ruledtabular}
\end{minipage}\hfill
\begin{minipage}[t]{0.55\textwidth}
\begin{ruledtabular}
\begin{tabular}{rccc}
$\kappa$ & forecast (discrete) & C (2-cycle) & D (period 6) \\
\hline
$0.08$ & $22.40$ $(23)$, $11.20$ $(12)$ & $23$, $12$, $23$ & $22$, $11$, $22$ \\
$0.05$ & $44.29$ $(45)$, $22.14$ $(23)$ & $45$, $23$, $45$ & $\mathbf{42}$, $22$, $\mathbf{42}$ \\
$0.03$ & $85.29$ $(86)$, $42.64$ $(44)$ & $86$, $43$, $86$ & $\mathbf{81}$, $43$, $\mathbf{82}$ \\
$0.018$ & $154.74$ $(155)$, $77.37$ $(80)$ & $155$, $78$, $155$ & $\mathbf{149}$, $78$, $\mathbf{150}$ \\
\end{tabular}
\end{ruledtabular}
\end{minipage}
\end{table*}

These comparisons have two limits. First, the discrete forecasts iterate the truncated normal form and the continuous ones integrate it, so both neglect
its remainder terms, and the continuous ones also replace the steps by a flow; we test them here rather than prove
them. Second, the windows lie outside the
asymptotic regime of the normal form, which assumes a bottleneck much narrower than the channel: at the fold edges of periods 16 to 32 the bottleneck's width $\sqrt{|a\eps/b|}$ is 1.35 to 3.51 times $U$. The
effective prefactor, about one third of $\pi^2/(ab)$, is therefore an empirical finding, and the exponent is near 2 but not pinned to it. Over the fold edges of periods 16 to 32 the
log-log slope is $-2.00$, but an exponent fitted freely together with $N_0$ is not well determined: the estimated localization scales of the edges are about $2\times10^{-5}$ to $2\times10^{-4}$ of their distances from
the fold (extrapolated locations for periods 16 to 20, and a uniform-within-cell error model for the grid boundaries of
periods 22 to 32), so the fits' misfits of 1 to 3\,\% are consistent with corrections to the asymptotic law rather than
noise, and the fitted exponent lies between 2.13 and 2.45 (a sensitivity range, not an interval) as the smallest $N$ in
the fit runs from 5 to 8\smref{Mechanism: a saddle-node bottleneck}.

\paragraph{Prospective tests.} The same reduction, with both coefficients computed from the network's derivatives,
predicts relaxation times and spectral gaps without fitted parameters in a small task-trained looped transformer
\cite{kim2026phase}. Here the forecasts were fixed before the test trajectories were run, in two stages.
An independent implementation located four new folds, which were not used while the bottleneck explanation was worked
out, computed their normal-form coefficients by nested automatic differentiation and wrote down the passage forecasts of each stage before its test trajectories were run (Table~\ref{tab:prospective}). The passages started at the same three points
as in Table~\ref{tab:passages}: the start of the channel, its midpoint, and the start displaced off the channel's axis.

The normal form gives two forecasts. The discrete forecast iterates the quadratic map \eqref{eq:foldnf} and counts the
iterates. The continuous forecast replaces the steps by a smooth flow and integrates, $n\approx\int du/(a\eps+bu^2)$,
whose closed form is the arctangent usually quoted for saddle-node bottlenecks. To leading order for slowly varying steps, the discrete count exceeds the integral by about
$\tfrac12\ln[h(U)/h(u_{\mathrm{in}})]$ iterates, with $h(u)=a\eps+bu^2$, most at small $\eps$ and for the midpoint
start.

The first test used the 2-cycle's fold at $\log_{10}T=-0.140$ (target A) and the fold of the period-8 orbit of
Table~\ref{tab:orbits} continued in temperature at $\alpha=0.0717$ (target B, $G=F^8$), with the discrete forecast as
the primary one. It was exact for the seven forecasts of five iterates or fewer and within one iterate for 16 of the
17 longer ones. Its criterion, that every forecast fall within one iterate, therefore failed once (target A, midpoint
start, $\eps=3\times10^{-6}$: forecast 61, measured 59), and all seven nonzero residuals were negative. The continuous
forecast came within one iterate of all 24 passages.

The second test, designed after the first, therefore made the continuous forecast primary, with every forecast at least 8 iterates, at the
2-cycle's fold at $\log_{10}T=-0.200$ (target C) and at the fold of a period-6 orbit of a different family (target D,
$G=F^6$, at $\log_{10}T=-0.087439$). At the 2-cycle's fold it held for all 12 passages, every measured count being the
forecast rounded up. At the period-6 fold the midpoint passages held, but passages from the start of the channel came
in short, by 0.40 to 5.74 iterates and more so closer to the fold, so the criterion failed (18 of 24 within one
iterate; 16 of 24 for the discrete forecast). There the first half of the channel is crossed faster than the normal
form predicts, for reasons the test does not resolve, such as terms beyond the quadratic normal form or the curvature
of the center manifold. Summed over the two stages, in which it was first the secondary and then the primary forecast, the continuous
forecast came within one iterate of 42 of 48 passages, and the six misses were 3.1 to 5.2\,\% short.

\section{A period-doubling cascade to chaos}
\label{sec:cascade}
The crescent of Fig.~\ref{fig:stripes}(b) holds a second classical route to chaos, period doubling, and its doubling
ratios approach Feigenbaum's constant as universality predicts.

\begin{figure}[tb]
\centering
\includegraphics[width=\columnwidth]{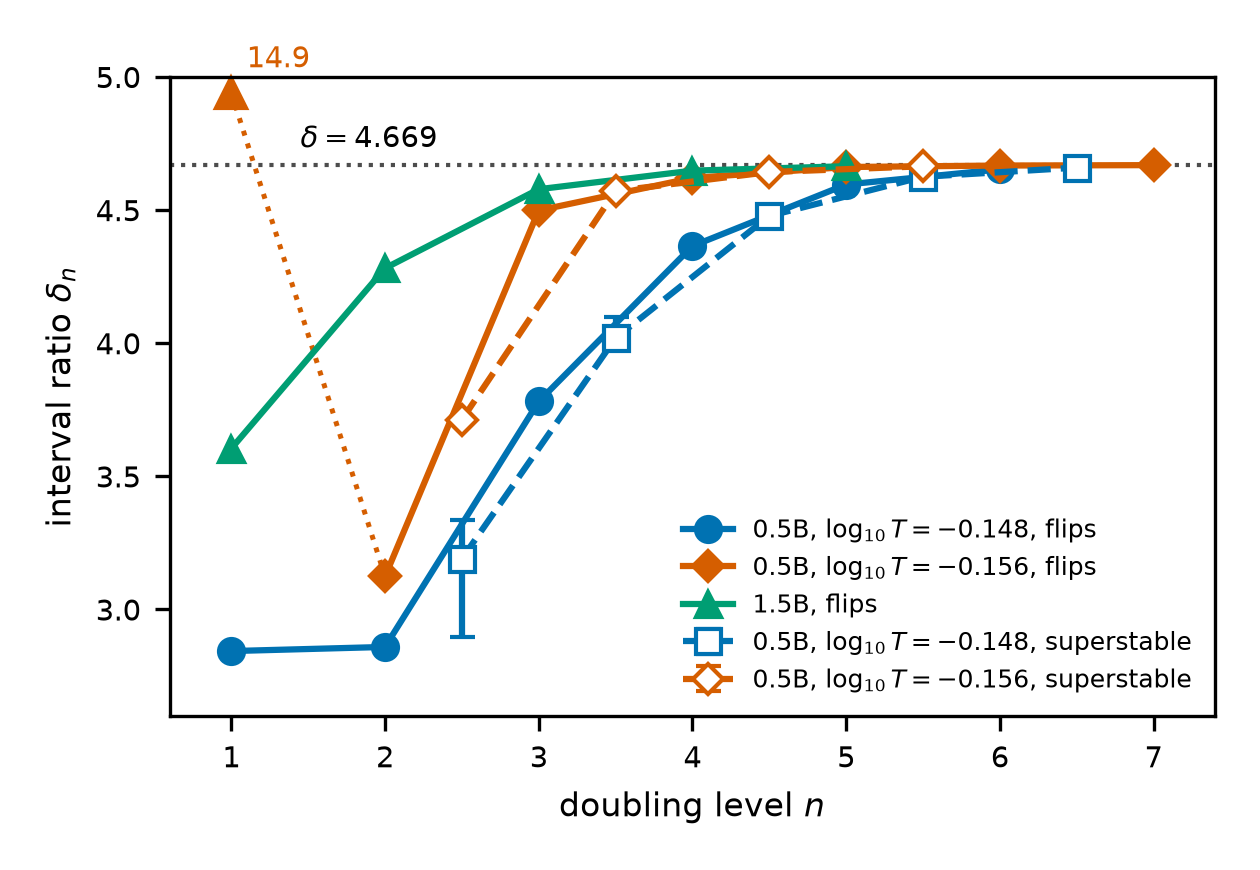}
\caption{Finite-order interval ratios $\delta_n$ of the doubling sequences approach Feigenbaum's $\delta$ (dotted
line): the crescent of Qwen2.5-0.5B-Instruct (philosophy direction, layer 12) along two cuts of fixed temperature,
from the located flips (filled) and from superstable points (open), and the route of Qwen2.5-1.5B (philosophy
direction, layer 14), from flips. A superstable ratio spans half-bands, so it is plotted half a level after the flip
ratio of its first band; level $n$ carries the ratio of the $n$th to the $(n+1)$th band, $\delta_{n+1}$ in Table~\ref{tab:cascade}. Bars: worst-case ranges from the brackets of $s_{12}$ and $s_{24}$, left out for the first two superstable ratios at $-0.156$, whose brackets are too wide. At $\log_{10}T=-0.156$
the first ratio, 14.9, lies off the scale (arrow), and the later ratios approach $\delta$ from below, the last
($4.6688$) within $0.008\,\%$ of it. Each sequence ends at the deepest level computed: the flips at $-0.148$ at
period 384, the superstable points at period 768 at $-0.148$ and at period 384 at $-0.156$, the 1.5B route at period 256, and the cut at $-0.156$ at the
$768\to1536$ flip.}
\label{fig:feigenbaum}
\end{figure}

Inside the crescent the period-3 loop ``\textbackslash n\textbackslash nThe existence'' doubles inward. At each doubling, called a flip, a simple real multiplier of the orbit reaches $-1$. In this cascade the
orbit loses its stability there and a stable orbit of twice the period continues beyond it (in a subcritical flip the
doubled orbit would instead be unstable and lie on the other side); the doubled loops repeat the same leading tokens twice and four times, with alternating
probabilities. Feigenbaum's theory predicts that, within a universality class, the distances between successive flips shrink by
a common factor: $\delta=4.6692\ldots$ for maps with a quadratic maximum, including suitable strongly dissipative maps
of any dimension \cite{feigenbaum1978quantitative,collet1981period}. The factor belongs to the class, not to period
doubling as such; it differs for a critical point of higher order \cite{feigenbaum1979universal}. To test whether this
cascade behaves as a member of the quadratic class we located the flips $\alpha_n$ along a cut of
fixed temperature in float64 and formed the ratios $\delta_n=(\alpha_n-\alpha_{n-1})/(\alpha_{n+1}-\alpha_n)$. Along
$\log_{10}T=-0.156$ we followed the cascade through nine flips, to the flip of the period-768 orbit
(Table~\ref{tab:cascade}); every critical multiplier is real and, for the first eight flips, within $5\times10^{-8}$ of
$-1$\smref{A period-doubling cascade to chaos}.

\begin{table}[tb]
\caption{The crescent's period-doubling cascade along $\log_{10}T=-0.156$, in float64: the orbit that loses stability
at each flip, the next multiplier modulus (the largest after the critical $-1$), the ratio $\delta_n$ and its deviation
from $\delta=4.6692$. After $n=4$ the deviations shrink by factors of 3.52, 6.54 and 3.93; the next factor, 5.26 at the
located flips, is left undetermined by the location uncertainty of the period-384 flip, about $\pm3.3\times10^{-13}$. Next moduli
below about $10^{-12}$ lie at the rounding level of products of hundreds of $896\times896$ Jacobians: they show that the
noncritical directions are negligible, not their decimal values.}
\label{tab:cascade}
\begin{ruledtabular}
\begin{tabular}{rlccc}
$n$ & orbit & next modulus & $\delta_n$ & $\delta_n-\delta$\\
\hline
1 & $3\to6$ & $0.12$ & & \\
2 & $6\to12$ & $0.045$ & $14.93$ & \\
3 & $12\to24$ & $7.0\times10^{-5}$ & $3.124$ & $-1.55$\\
4 & $24\to48$ & $4.3\times10^{-7}$ & $4.4988$ & $-0.17$\\
5 & $48\to96$ & $1.8\times10^{-13}$ & $4.6208$ & $-0.048$\\
6 & $96\to192$ & $2.9\times10^{-14}$ & $4.6618$ & $-0.0074$\\
7 & $192\to384$ & $1.5\times10^{-14}$ & $4.6673$ & $-0.0019$\\
8 & $384\to768$ & $3.2\times10^{-13}$ & $4.6688$ & $-0.00036$\\
9 & $768\to1536$ & $2.9\times10^{-14}$ & & \\
\end{tabular}
\end{ruledtabular}
\end{table}

The first ratios depend on where a cut crosses the crescent (14.93 along $-0.156$; 2.84 and 2.86 along
$\log_{10}T=-0.14798$), but the later ratios behave as universality predicts (Fig.~\ref{fig:feigenbaum}). The last
measured ratio is 4.6688, and after the first ratio the deviations from $\delta$ keep one sign; from $n=4$ on they shrink by factors of about
3.52, 6.54 and 3.93, and the next factor (5.26) is not resolved by the location uncertainty of the period-384 flip; Aitken's $\Delta^2$ extrapolation of the last three ratios gives $4.6694$. The precision of the
multipliers limits how far this can be pushed. The penultimate flip, of the period-384 orbit, approaches the precision
floor of its multiplier, a product of 384 step Jacobians, and a local linear fit puts its location uncertainty at about
$\pm3.3\times10^{-13}$, an estimate rather than a certified bound. Over this range $\delta_7$ lies between $4.6672$ and
$4.6674$, $\delta_8$ between $4.6685$ and $4.6692$, and the extrapolated value between $4.6688$ and $4.6702$, an
interval that contains $\delta$.

\begin{figure*}[t!]
\centering
\includegraphics[width=0.58\textwidth]{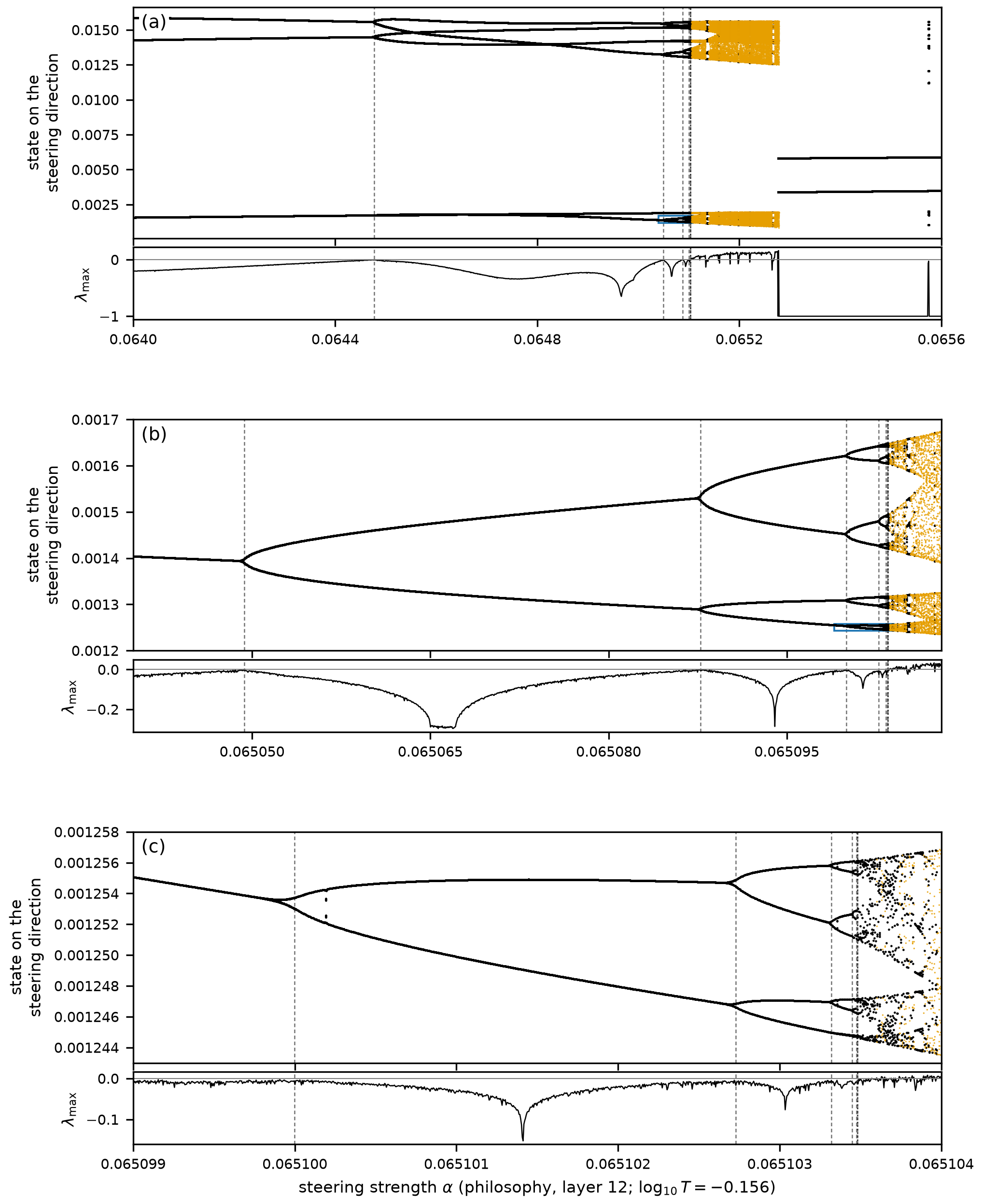}
\caption{The Feigenbaum tree of the crescent cascade (Qwen2.5-0.5B-Instruct, philosophy direction at layer 12,
$\log_{10}T=-0.156$): float64 line scans of 1600, 1360 and 1000 cells, each with 3000 transient steps and 256 recorded
states projected on the unit steering direction (orange: chaotic cells, with no period up to 48). (a) The whole cut: period 3 doubles into chaos,
which ends at $\alpha=0.065278$, where the attractor jumps to a strongly contracting period-2 orbit. (b), (c) Zooms in
both coordinates on the branch boxed in the panel above, showing the doublings $6\to12\to24\to48$ and
$24\to48\to96\to192$. Dashed lines: flips located by Newton's method; dotted line: the
accumulation point estimated from the last two flips with $\delta$. Under each panel, the largest Lyapunov exponent
touches zero at every flip and turns positive beyond the accumulation point. Near a flip the orbit converges slowly,
so after 3000 transient steps each fork opens slightly before its Newton flip, by $8.4\times10^{-6}$ in $\alpha$ at the
first and by at most $1.1\times10^{-6}$ at the later ones.}
\label{fig:feigtree}
\end{figure*}

Cascades with these ratios also occur in the training map of gradient descent in solvable models of deep networks
\cite{hofmann2026map}. The late orbits of this cascade are extremely strongly contracting transverse to the critical
direction: from the fifth flip on, the next multiplier modulus is below $4\times10^{-13}$, at the rounding level of the
monodromy products. This supports an effectively one-dimensional, strongly dissipative regime; it does not by itself
establish the hypotheses of the renormalization theory. In the bifurcation diagram (Fig.~\ref{fig:feigtree}) each fork sits at a flip located independently by
Newton's method, and the largest exponent touches zero at every flip, as it must where a multiplier reaches modulus one, and turns
positive beyond the accumulation point.
The cascade of Qwen2.5-1.5B ends with the ratio $4.665$, within $0.1\,\%$ of $\delta$.

The local theory holds even where orbits are far from one-dimensional. At the two flips of a doubling bubble
\cite{bier1984remerging} at higher temperature, where the next moduli are $0.38$ and $0.28$, the flip normal form
computed from the network's derivatives predicts the amplitude of the doubled orbit to leading order: measured and
predicted amplitudes have ratios of $1.005$ and $1.003$ at the smallest offset\smref{A period-doubling cascade to
chaos}. Both of these flips are supercritical. The cascade's flips are identified by a simple real multiplier at $-1$, with
every other multiplier well inside the unit circle, and by the stable doubled orbit continued beyond each. The remaining
conditions of a generic flip, a nonzero crossing speed of the multiplier and a nonzero cubic coefficient of the flip
normal form, were not computed for them, and they are not classified as super- or subcritical.

\begin{figure*}[tb]
\centering
\includegraphics[width=\textwidth]{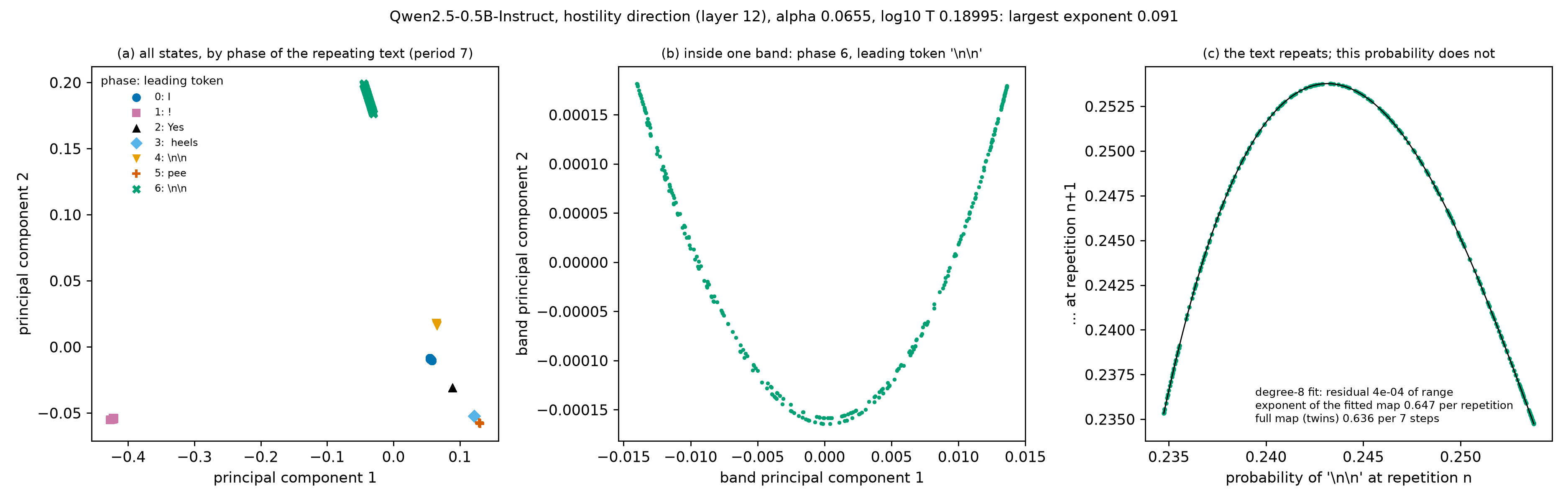}
\caption{A periodic text on a chaotic orbit (Qwen2.5-0.5B-Instruct, hostility direction at layer 12, $\alpha=0.0655$,
$\log_{10}T=0.18995$; 2000 float64 steps). At every step the most probable token follows the same seven-token cycle,
\texttt{I}, \texttt{!}, \texttt{Yes}, \texttt{\ heels}, $\hookleftarrow\hookleftarrow$, \texttt{pee},
$\hookleftarrow\hookleftarrow$, while the largest Lyapunov exponent is 0.091 per step. (a) All states on two principal
components, colored and marked by their position in the text: seven separated groups. (b) The group of the last position on its
own principal components. (c) The probability of the double line break at the last position, at one repetition
against the next: the points fall on a single hump (black: a degree-8 fit), and the fitted map's exponent, 0.647 per
repetition, is within 2\,\% of the full map's 0.636 per repetition (seven steps of about 0.091).}
\label{fig:hidden}
\end{figure*}
\begin{figure*}[tbp]
\centering
\includegraphics[width=0.95\textwidth]{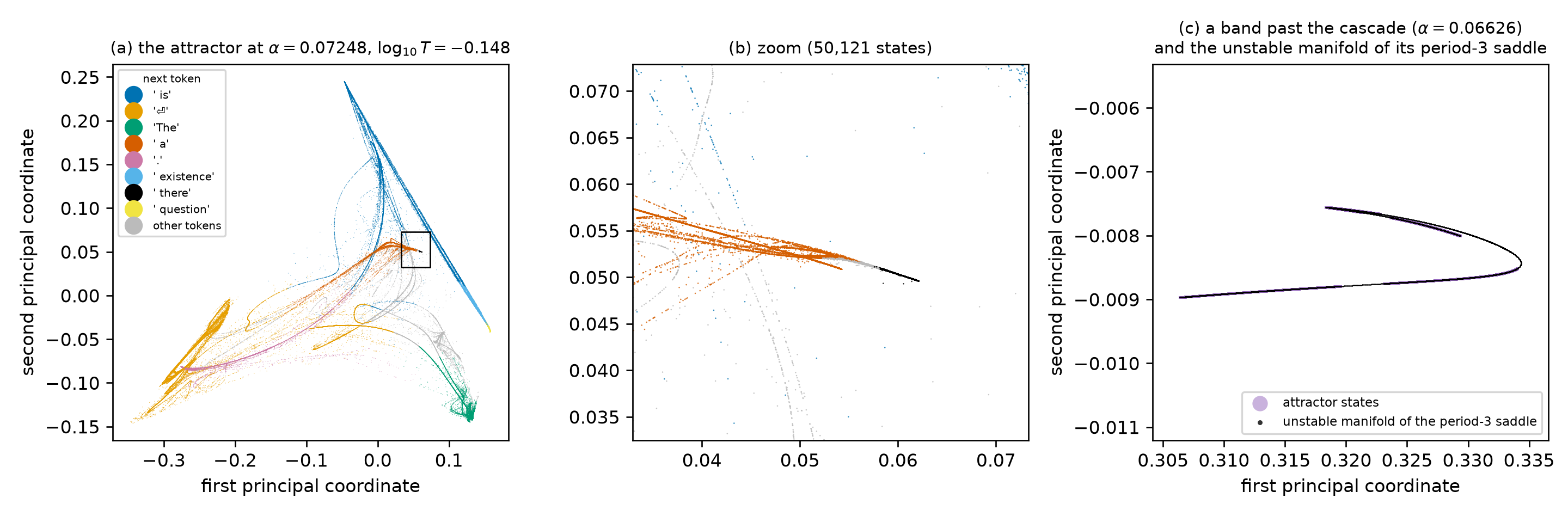}
\caption{Projected geometry of a numerically observed chaotic attractor of the soft-token loop (Qwen2.5-0.5B-Instruct,
philosophy direction at layer 12, $\log_{10}T=-0.148$, float64). (a) 298{,}800 states of 498 trajectories at
$\alpha=0.07248$ on their first two principal axes, colored by the most probable next token (gray: other tokens); (b) magnification of the
boxed region, where the filaments lie in layers; (c) one band of the chaotic attractor at $\alpha=0.06626$ (purple)
and sampled points of the unstable manifold of the embedded period-3 saddle (black).}
\label{fig:attractor}
\end{figure*}

\paragraph{Chaos behind a repeating text.} The text records less than the loop feeds back. What returns at each step
is the expected embedding under the whole next-token distribution; the printed token is only that distribution's most
probable entry. A repeating text therefore says that the same token wins at each position of the repetition, not that
the distribution repeats. Steered along the hostility direction at
layer 12, the loop has a period-7 body that doubles into chaos\smref{Steering toward hostility}. On a temperature line
through it, 2000 measured steps at two twin offsets find 182 chaotic cells among 1400. In 65 of them the most probable
token repeats exactly at every one of the 2000 steps, while the state moves chaotically; in 59 of these the period
of the text is a multiple of seven, most often 7 or 28.

Figure~\ref{fig:hidden} shows the strongest case ($\log_{10}T=0.18995$, largest exponent 0.091 per step at both offsets). Its sampled states
fall into seven separated groups, one for each position in the seven-token text, visited in a fixed order
(Fig.~\ref{fig:hidden}(a)): band chaos, in which a chaotic attractor splits into bands visited in a fixed cyclic order
\cite{grossmann1977invariant,lorenz1980noisy}. Within each group the most probable token never changes, but the
probabilities do. At the last position of the text, the probability of the double line break, which wins there, takes
values between 0.235 and 0.254 and does not repeat over the 2000 steps. Read once per repetition, on the sampled range it is fitted by a smooth
map with a single quadratic maximum (Fig.~\ref{fig:hidden}(c)), whose Lyapunov exponent, 0.647 per repetition, is within 2\,\% of the full map's 0.636 per repetition (seven steps of about 0.091). Over the measured horizon, the periodic sequence of most probable tokens does not reveal the nonperiodic state
dynamics, which is consistent with a partition of the state space by most probable token that is not generating on
this banded attractor: a periodic text can hide chaotic dynamics.

\paragraph{Computer-assisted proofs.} Three periodic orbits are certified. For the expected-embedding map of
Qwen2.5-0.5B-Instruct as stored (all 24 layers, the full 151{,}936-token readout, the fixed 37-token prefix and the
steering of Sec.~\ref{sec:map}), at $\alpha=0.071674$ and $\log_{10}T=-0.115$ read as exact binary64 rationals (the
temperature is the real number $\exp(\ln10\cdot\log_{10}T)$ for that exact $\log_{10}T$), interval arithmetic proves
that the map has a primitive period-8 orbit, the stripe orbit of Table~\ref{tab:orbits}, which is the unique
multiple-shooting root in a product of eight boxes of coordinate radius $2^{-640}$ around stated centers, and that this
orbit is locally attracting. Ball arithmetic
\cite{johansson2017arb} encloses the map and its Jacobian over each box; a Krawczyk-type inclusion
\cite{krawczyk1969newton,moore2009interval,rump2010verification} for the preconditioned shooting system gives existence
and uniqueness, separation of the first point from the points one, two and four steps later gives the primitive
period, and a norm bound on powers of the enclosed monodromy gives attraction. The release record
(\path{certificates/llm-p8-certificate-final-001.json}, SHA-256 \texttt{539ead0d\ldots}) pins the SHA-256 hashes of
the checkpoint (\texttt{fdf756fa\ldots}), its configuration and the steering vectors, the 37 prefix token IDs, the exact
parameters, the box definition and the receipts of each of the three checks. An independent checker recombined the saved
scalar inequalities, the primitive-period separations and the row norms of the saved monodromy powers in exact rational
arithmetic; it did not recompute the dense enclosure products, which rest on one execution, and the proof trusts the
Python, FLINT and Arb stack and the stored weights. Interval methods of this kind have
established periodic orbits and chaos in low-dimensional maps and flows \cite{galias2001interval,zgliczynski1997computer};
here they apply to a map of $\mathbb{R}^{896}$ built from a full transformer. The proof does not exclude other period-8
points in the first box or say anything about the window's global organization\smref{Evidence and reproducibility}.

A second proof of the same kind establishes coexistence. At $\alpha=0.06468$ and $\log_{10}T=-0.148$, where the
crescent's 3-cycle exists but the fixed start still reaches the 2-cycle in float64 runs, the same map has a primitive period-2 orbit and
a primitive period-3 orbit, each the unique shooting root in its product of boxes of radius $2^{-640}$ and each locally
attracting. The unweighted norm bounds on the first and second powers of the enclosed monodromy failed the
sufficient condition for attraction, because the monodromy is strongly nonnormal. The fourth and eighth powers pass:
the infinity-norm bounds of the eighth powers are $1.9\times10^{-8}$ for the 2-cycle and $1.091\times10^{-3}$ for the
3-cycle. Two different attracting behaviors of the model's loop thus coexist under identical settings.

A separate check certifies the token loops they write, from the same whole-box enclosures: the most probable next
token leads every other by a positive margin, so the 2-cycle writes `` is it'' and the 3-cycle a blank line,
``The'' and `` existence''. A further certificate gives explicit, if tiny, basin neighborhoods: every state within
infinity-norm distance $2^{-681}$ of the 2-cycle's center converges to the 2-cycle, and within $2^{-700}$ of the
3-cycle's to the 3-cycle. Which of them a start outside these balls reaches, the map's fixed start included, is not
proved.

\begin{figure*}[tbp]
\centering
\includegraphics[width=\textwidth]{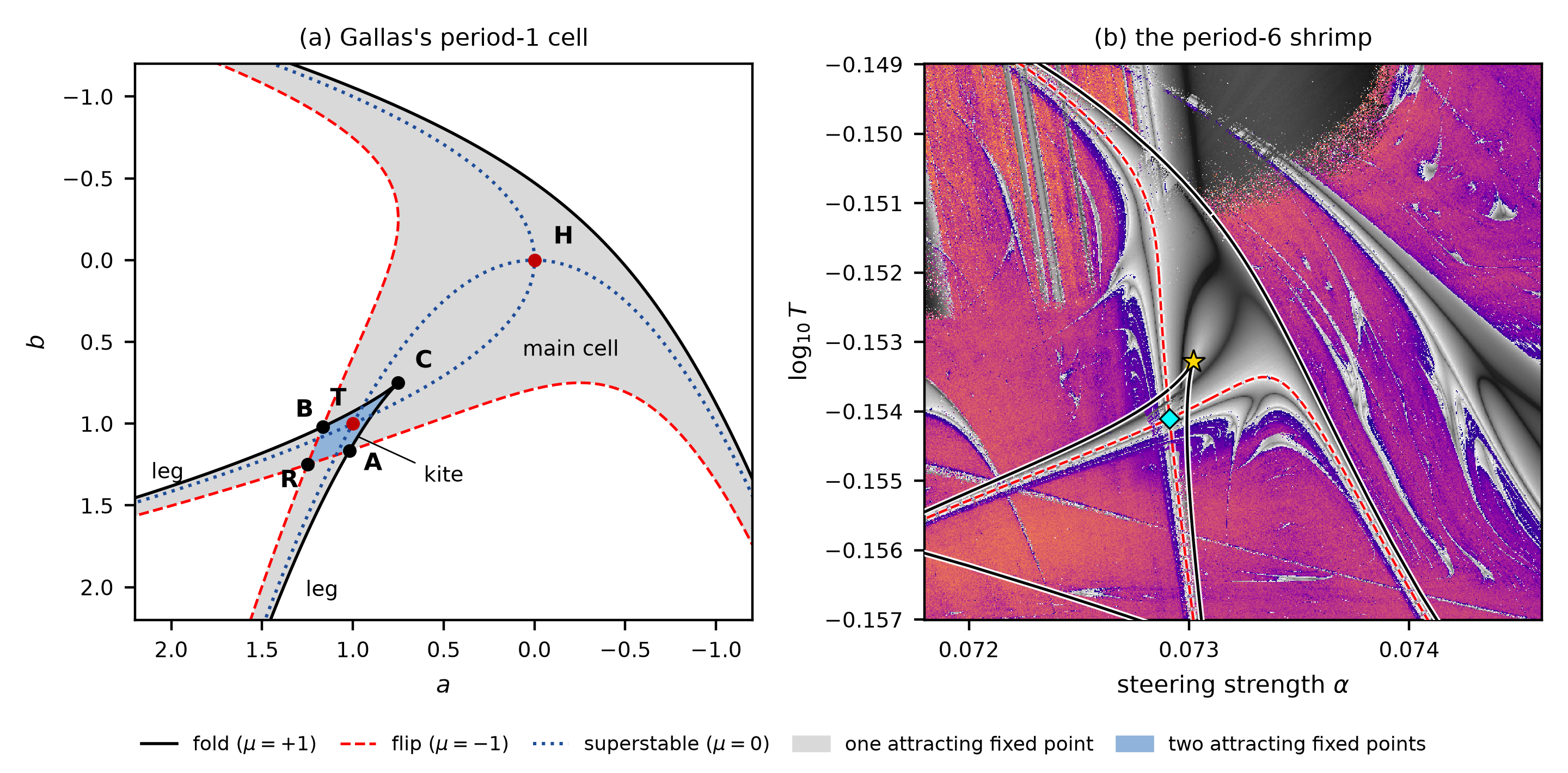}
\caption{Gallas's canonical cell beside the period-6 shrimp of the soft-token loop. (a) His period-1 cell,
$x\mapsto(a-x^2)^2-b$ (his eq.~(5.3), in the coordinates and with the labels of Fig.~7 of Ref.~\onlinecite{gallas1994dissecting}),
computed (gray: one attracting fixed point; blue: two) and drawn with both axes reversed, a rotation by $180^\circ$
that brings it to the orientation of (b). Fold curves (black; multiplier $+1$) and flip curves (red, dashed; $-1$)
bound the main cell, and the superstable curves (blue, dotted; $0$) of the map's critical points cross at the head H.
Two legs, each between a fold and a flip curve, cross in the kite C--A--R--B, which begins at the cusp C of the fold
curve; in the kite two attracting fixed points coexist, and their superstable curves cross at the tail T. The flip
curves of the two legs cross at R; the computed corners A and B agree with Gallas's to six significant digits. (b) The
period-6 shrimp, in the plane of Fig.~\ref{fig:teaser} (Qwen2.5-0.5B-Instruct, philosophy direction at layer 12;
float64): a $640\times640$ map of the window centered on the body (steps $4.4\times10^{-6}$ in $\alpha$ and
$1.25\times10^{-5}$ in $\log_{10}T$, about seven times finer than Fig.~\ref{fig:teaser}; colored as there), with the
continued curves of the period-6 orbit
``$\hookleftarrow\hookleftarrow$The question$\hookleftarrow\hookleftarrow$The existence''. R (cyan diamond): the
crossing of the two flip curves, where two distinct period-6 orbits are found (a pseudo-crossing); C (gold star): the
kite's cusp; the kite lies between them (Fig.~\ref{fig:p6kite}). In both panels the fold curve bounds the main cell
(in (b), the body) above and on the right, flip curves bound it on the left and below, and at the lower left two of its legs cross in the kite. The dark ditches in (b) were measured on a $31\times31$ grid of the continued period-6 orbit over the body, with a
finer patch and a line across the top: along them the orbit's leading multiplier changes sign between neighboring
cells or becomes a complex pair, and the orbit is attracting throughout. The multiplier never vanishes, so these
ditches are pseudo-superstable lines. They meet below C, where the
superstable curves of (a) cross at the tail T, but approach and turn away at the top, where (a) has its head H. In a two-dimensional map the one-dimensional skeleton of (a) generically unfolds in one of two ways: the crossing in
the kite remains, and at the head the curves either close into a loop (Type A) or approach and turn away (Type B, ``a
crossing without a loop'') \cite{facanha2013endoskeletons}. The period-6 shrimp follows the Type B pattern, so it has
no head; we call it Type-B-like because its lines are where the leading multiplier of an 896-dimensional map changes
sign, not the trace-zero curves of a two-dimensional map that define the types (Sec.~\ref{sec:shrimp}).}
\label{fig:anatomy}
\end{figure*}

The chaotic attractors can be drawn, as numerically observed attractors: sets that the tested trajectories approach
and stay near over the recorded horizon, with invariance and attraction resting on finite-time evidence. At the
strongest chaos on the row $\log_{10}T=-0.148$ ($\alpha=0.07248$), 498 of 512 trajectories approach one such set and 14
reach a coexisting period-4 orbit. Projected on principal axes of its states, it shows thin folded filaments in
layers (Fig.~\ref{fig:attractor}); its four leading Lyapunov exponents are $0.274$, $-1.052$, $-1.486$ and $-1.856$ per step, so its
Lyapunov (Kaplan--Yorke) dimension estimate is $1.26$, while the correlation sum has no scaling range. The loop is a map, not a flow: a typical step moves the state by
0.30, more than the mean distance (0.21) of the states from their center (four fully recorded trajectories), so
consecutive states usually lie far apart on the attractor rather than along one filament.

A delay reconstruction
from one output scalar, the log of the largest next-token probability, has 3.4\,\% false nearest neighbors with
three delay coordinates. Past the cascade ($\alpha=0.06626$) the chaos is banded, and in 12 principal coordinates every
recorded state of one band lies within $9.8\times10^{-6}$ of sampled points of the unstable manifold of the period-3
saddle left by the cascade (band width 0.028), which supports an unstable-manifold description of the band at that
resolution.

\begin{figure*}[tb]
\centering
\includegraphics[width=\textwidth]{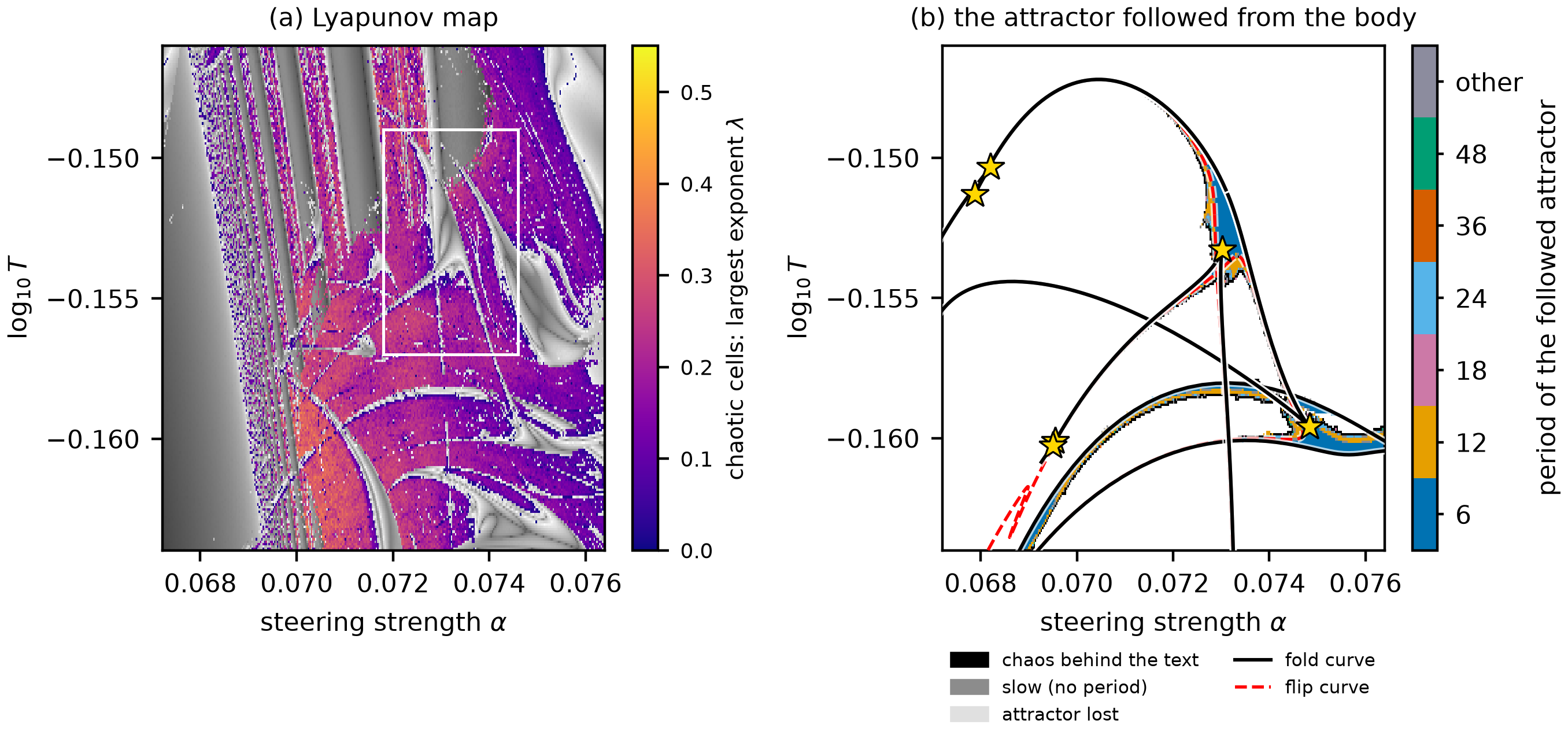}
\caption{The period-6 shrimp in a wider window of the plane of Fig.~\ref{fig:teaser} (Qwen2.5-0.5B-Instruct,
philosophy direction at layer 12; float64). (a) The map of Fig.~\ref{fig:teaser} in this window, colored as there;
white rectangle: the window of Fig.~\ref{fig:anatomy}(b). (b) The attractor followed from the body, on a grid with the
steps of Fig.~\ref{fig:teaser}'s, colored by its period (black: chaos behind the repeating text, an exactly repeating
text with no period up to 64 and an exponent of at least 0.005; gray: slowly converging cells, the same with an
exponent below 0.005; light gray: attractor lost), with the continued curves of Fig.~\ref{fig:anatomy}(b); gold stars:
the located cusps of the fold curves: the kite's cusp C, the one of Gallas's anatomy, and five further ones, in pairs
near the ends of the upper-left and lower-left
legs (the lower pair's stars overlap at this scale) and the tip of the small window beyond the neck. The fold curve
through C is that of the kite (Fig.~\ref{fig:p6kite}). The map starts every cell from the same state and the followed
attractor from the attractor of a neighboring cell, so where attractors coexist the two panels can differ.}
\label{fig:p6shrimp}
\end{figure*}

\section{Shrimp-shaped windows}
\label{sec:shrimp}
The scans show periodic windows with bodies, legs, hooks and crescents: shrimp-shaped windows, which Gallas classified
as the signature structure of the parameter planes of low-dimensional maps \cite{gallas1993structure,gallas1994dissecting}. We study one of them in detail, in the plane of Fig.~\ref{fig:teaser}: along its body it meets our numerical criteria
for a shrimp, and its continued curves are consistent with Gallas's anatomy, with a kite that begins at a located cusp of
its fold curve. In a map of 896 dimensions the name describes this organization of continued curves; it is not a local
normal form. None of the windows we analyzed has the head of a one-dimensional map.

\begin{figure*}[tb]
\centering
\includegraphics[width=0.85\textwidth]{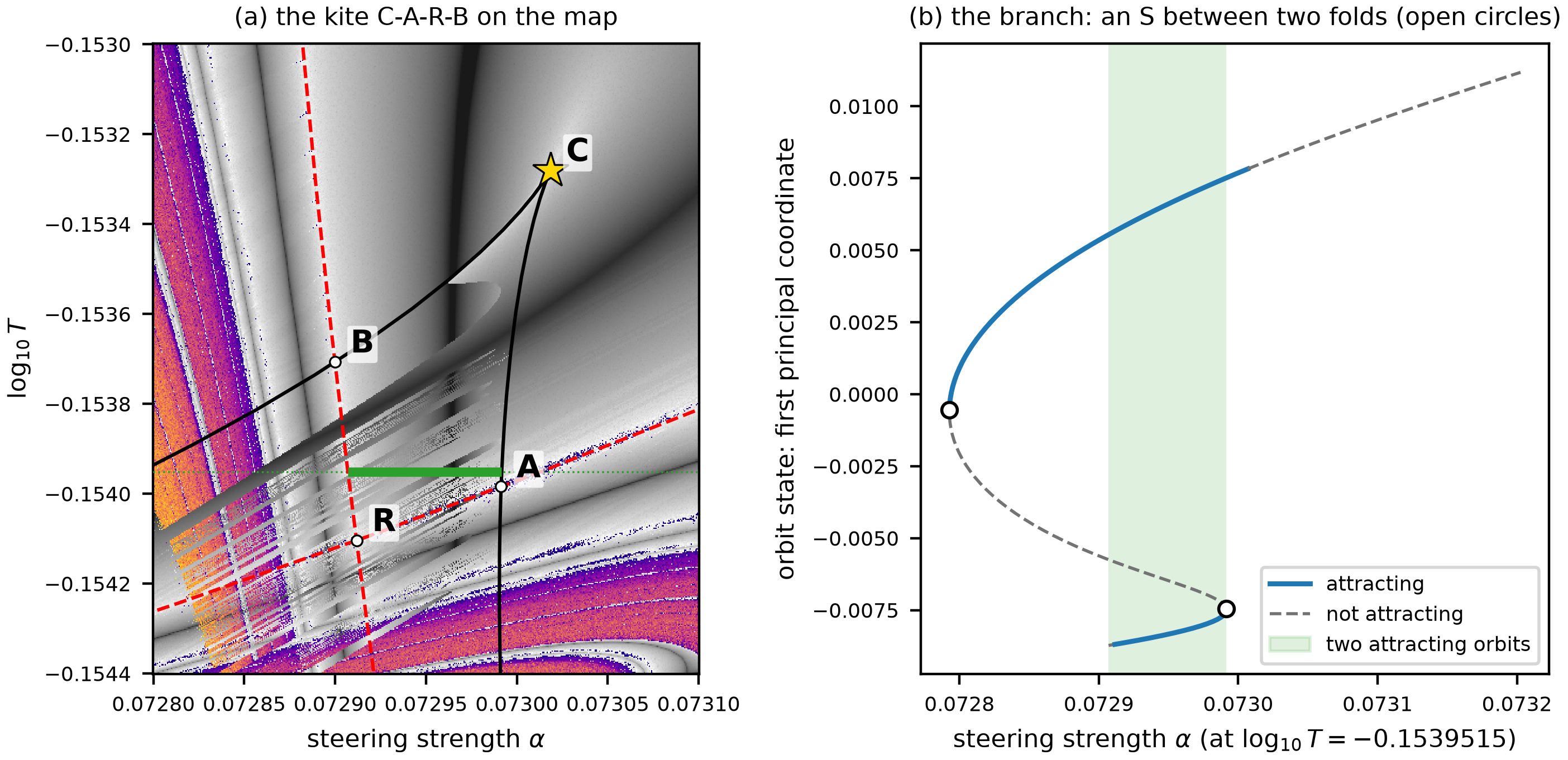}
\caption{The kite of the period-6 shrimp (Qwen2.5-0.5B-Instruct, philosophy direction at layer 12; float64). (a) A $512\times512$ float64 map of the lower-left end of the body, with cells 7.5 times finer in $\alpha$ and 4.6
times finer in $\log_{10}T$ than in Fig.~\ref{fig:anatomy}(b) (gray: periodic, darker where more contracting; colors: chaotic), with the body's two flip
curves (red, dashed), the fold curve continued from the branch of (b) (black), and the corners of Gallas's kite: R,
where the flip curves cross (two distinct orbits, one on each curve); A and B, where the fold curve meets them; C, where
the fold curve turns back (gold star). Green: the line $\log_{10}T=-0.15395$ of the branch, thick
where the two attracting orbits coexist. (b) The branch, computed by numerical continuation of the period-6 orbit in $\alpha$ along the green line of (a):
every point is an orbit converged by Newton's method, solid where its multipliers show it attracting and dashed where
not, and the two open circles are folds, where a multiplier reaches $+1$; they make the branch an S, and the two
attracting sheets overlap in the shaded interval. The vertical axis is a display coordinate: one point of each orbit,
a vector of 896 numbers, centered over the branch and projected on the direction in which these points vary most,
their first principal component. Inside the kite, the straight edge where the gray changes abruptly (near $\log_{10}T=-0.15353$) is a basin boundary
of the fixed start, not a bifurcation, where we checked it: at two neighboring cells across it both attracting
period-6 orbits exist (leading multipliers $0.18$ and $0.60$ to $0.61$), the fixed start reaches one above the edge
and the other below, and at one steering strength, by strong numerical evidence, the start on the edge lies on the stable manifold of a saddle orbit between them.}
\label{fig:p6kite}
\end{figure*}

\begin{figure*}[tb]
\centering
\includegraphics[width=\textwidth]{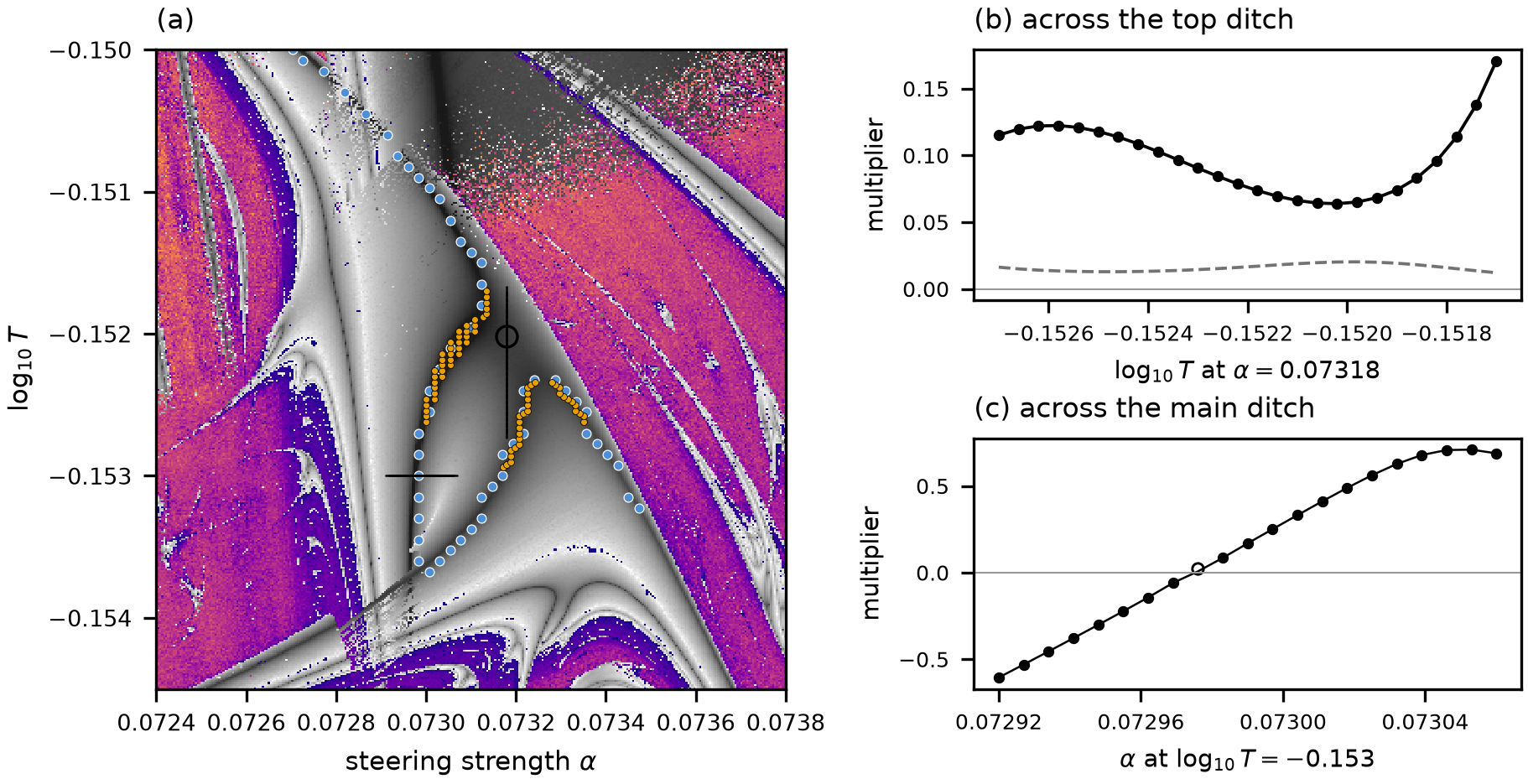}
\caption{Where the period-6 orbit's leading multiplier changes sign over the body of the shrimp (float64). (a) The map
of Fig.~\ref{fig:anatomy}(b) over the body, with three measurements of the continued orbit: a $31\times31$ grid (blue
dots: midpoints of neighboring solved cells whose real leading multipliers have opposite signs, and cells where the
leading multiplier is a complex pair), a patch across the top of the lighter region with steps $1.8\times10^{-5}$ in
$\alpha$ and $4\times10^{-5}$ in $\log_{10}T$ (orange dots, the same rule), and two lines (black ticks): line A across
the dim top ditch at $\alpha=0.07318$, with its lowest leading multiplier circled, and line B across the main ditch at
$\log_{10}T=-0.153$. (b) Along line A the leading multiplier (solid) dips to $0.064$ at $\log_{10}T=-0.15202$ but stays
positive, and the next modulus (dashed) stays at most $0.021$: no sign change, a valley of the same sign between the
main ditch and the arc. (c) Along line B, the control, the leading multiplier changes sign through a complex leading
pair (open marker), from $-0.605$ to $+0.710$. The orbit is attracting at every solved cell.}
\label{fig:p6signs}
\end{figure*}

\begin{figure*}[tb]
\centering
\includegraphics[width=0.86\textwidth]{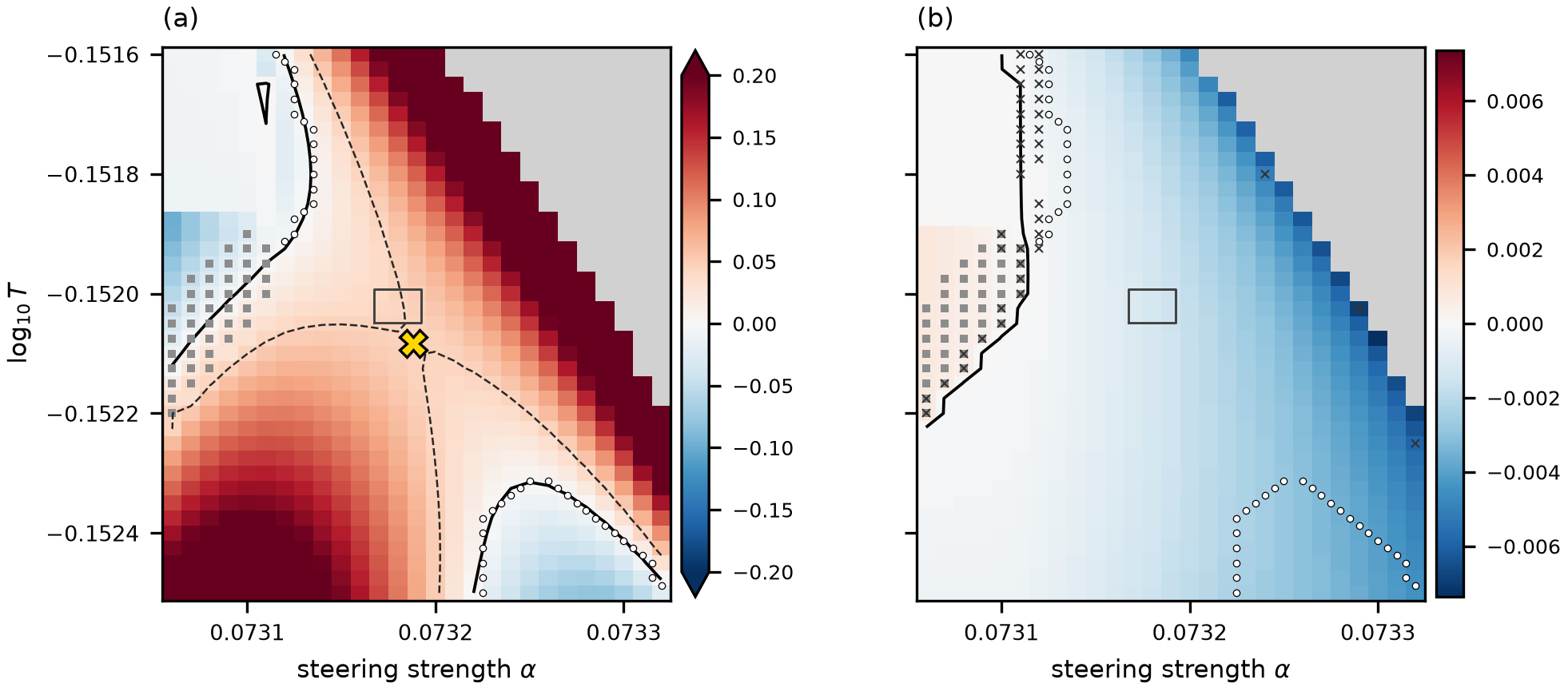}
\caption{Trace geometry associated with the separation of the two pseudo-superstable lines at the top of the period-6 body. A two-dimensional block of the monodromy, initialized from its leading Ritz directions, was tracked from cell to
cell by subspace overlap over 852 continued orbits (float64; steps $10^{-5}$ in $\alpha$ and
$2.5\times10^{-5}$ in $\log_{10}T$; gray: no converged orbit on the followed branch). (a) The trace $\tau$ of the tracked block (colors clipped at $\pm0.2$), its zero curves (black), the measured sign changes of the real leading
multiplier (white dots) and the cells with a complex leading pair (gray squares), and the numerically fitted saddle of $\tau$ (gold cross,
$\tau=0.045$) with the level curve through it (dashed). (b) The determinant $\delta$, with its zero curve and the same
measurements; small crosses mark cells where the tracking overlap falls below 0.9, indicating possible mixing with a third direction. The rectangle marks the patch around the minimum on line A, refined by a factor of ten in each parameter relative to the orange patch in Fig.~\ref{fig:p6signs}(a).}
\label{fig:delta2d}
\end{figure*}

\paragraph{What a shrimp is.} A shrimp is a periodic window with a characteristic shape, which Gallas found in the parameter plane of the H\'enon map \cite{gallas1993structure,henon1976two}: a main body of
period $k$ with thin legs, followed by period-doubled domains of periods $k\cdot2^n$. Its outline is drawn by
bifurcation curves \cite{gallas1994dissecting} (Fig.~\ref{fig:anatomy}(a)): fold curves, where the window's orbit is born or dies together
with a saddle (multiplier $+1$), and flip curves, where it doubles (multiplier $-1$), with the period-doubled shells
beyond the flips. Each leg is a thin strip between a fold and a flip curve, and two legs cross in a kite that begins at
a cusp of the fold curve and in which two attracting orbits coexist. In a one-dimensional map the multiplier passes
through $0$ between a fold edge and a flip edge, on superstable curves, the dark lines of strongly negative exponent in
the scans, and two of them cross at the shrimp's center, its head. In a two-dimensional map whose Jacobian never
vanishes, such as H\'enon's, no multiplier can vanish; curves on which the trace of the two leading multipliers
vanishes take the place of the superstable curves, and at the head they either close into a loop (Type~A) or approach
each other and turn away (Type~B) \cite{facanha2013endoskeletons}.

\paragraph{A shrimp in the plane of Fig.~\ref{fig:teaser}.} The plane of Fig.~\ref{fig:teaser}, a single
steering direction against the temperature, holds a shrimp-shaped window: the period-6 loop
``$\hookleftarrow\hookleftarrow$The question$\hookleftarrow\hookleftarrow$The existence'' (Figs.~\ref{fig:anatomy}(b) and \ref{fig:p6shrimp}).
At a strongly contracting cell of the map, $(\alpha,\log_{10}T)=(0.073097,-0.15204)$, Newton's method in float64
converges to a primitive period-6 orbit (residual $1.6\times10^{-14}$) whose leading multipliers form a complex pair of
modulus $0.016$. Followed from there through neighboring cells, the attractor fills 4224 cells of a grid with
the steps of Fig.~\ref{fig:teaser}'s (Fig.~\ref{fig:p6shrimp}(b)). They cover the shrimp's body and its legs, and below the body they continue through a narrow neck into a smaller,
distorted window at the lower right of the panel. Of these cells, 1946 have period 6, 945
along the edges have its doublings (periods 12, 24 and 48), 35 have other periods, 811 show chaos behind the
repeating text (an exactly repeating text, no period up to 64 and an exponent of at least 0.005), and 487 hold an
orbit that converges too slowly for a period to be detected (the same, but with an exponent below 0.005, next to a
fold or a flip; with the maps' threshold of $10^{-3}$ the split would be 842 and 456). Continued from the body in four
directions, up and down in $\alpha$ and in $T$, the period-6 orbit ends at a fold above and on the right and at flips
on the left and below.
Continued in the plane, with every point checked (residual at most $10^{-10}$, critical multiplier real and within
$10^{-7}$ of $\pm1$ on the body and kite curves; four points of the extended leg curves reach $2.9\times10^{-7}$), the curves outline the anatomy of Gallas's canonical cell \cite{gallas1994dissecting}
(Fig.~\ref{fig:anatomy}). The fold curve bounds the body, the main cell, above and on the right. One flip curve
bounds it on the upper left and a second along its bottom, and the doubled shells lie beyond both. The two flip curves
cross at the lower-left end of the body, near $(0.072912,-0.154104)$, where two of its legs leave the body. They are
flip curves of two different orbits. Newton's method at that point, started from the orbit stored on each curve,
converges to two distinct period-6 orbits (residuals $3\times10^{-14}$ and $2\times10^{-13}$). They share their
six-token text but differ by 6.4\,\% of the mean state norm (the largest point distance after the best of the six
cyclic alignments), and their leading multipliers are $-0.9952$ and $-1.0005$ at this approximate crossing of the two traced curves,
whose nearest saved points lie 1.7 and 0.3 steps of the grid of
Fig.~\ref{fig:teaser} away. The crossing is therefore a pseudo-crossing of curves on two sheets of the orbit family, the
configuration at the corner R of Gallas's kite, where the flip curves of two coexisting fixed points cross
(Fig.~7 of Ref.~\onlinecite{gallas1994dissecting}). The rest of the kite follows (Fig.~\ref{fig:p6kite}). Continued in $\alpha$ at
fixed $\log_{10}T=-0.15395$ from the first orbit's flip, the branch is S-shaped: the orbit stays attracting up to a
fold at $\alpha=0.072992$, the branch returns along the saddle to a second fold at $0.072793$, and it continues on
the second orbit, which flips near $0.07301$. At that temperature the two attracting orbits coexist for $\alpha$
between 0.072907 and 0.072992. The fold curve continued from the first fold, and independently from a fold of the
second orbit, turns back at the same point, the kite's cusp C, located at $(0.0730186,-0.1532803)$, on the side of R toward the body (cubic coefficient $c=-1.33\times10^{4}$; critical multiplier within $7.7\times10^{-9}$ of $+1$, next modulus
$0.00095$). The two parameters unfold it regularly \cite{kuznetsov2019maps}: the matrix $K$ of derivatives of the coefficients $\beta_0,\beta_1$ of
the reduced equation $0=\beta_0+\beta_1u+bu^2+cu^3$ with respect to $(\log_{10}T,10\alpha)$ has determinant $-2.6\times10^{3}$ and singular values 672 and 3.9. The fold curve's two branches meet the two flip curves at
A $=(0.072991,-0.153984)$ and B $=(0.072900,-0.153707)$, completing the kite C--A--R--B of Gallas's cell. A and B are pseudo-crossings too: at the continued fold points
nearest them the orbit's other multipliers are at most 0.0022 in modulus, none near $-1$, and at the flip points nearest
them the others are at most 0.0042, none near $+1$. A fold--flip point would need one orbit with both multipliers; here,
as at R, the curves belong to different period-6 orbits. Only C belongs to Gallas's anatomy of a shrimp. Besides it, the fold curves have five more located cusps
(Fig.~\ref{fig:p6shrimp}(b)), which show how the ends of the legs and the small window beyond the neck close. At each of the six, the quadratic coefficient of the fold's normal form \cite{kuznetsov2019maps}, of magnitude between 0.17 and 39.11 at
the twelve bracketing curve points, falls to at most $9.5\times10^{-5}$ at the located cusp, while the cubic coefficient stays large ($|c|$ from $2.0\times10^{3}$ to
$4.0\times10^{4}$) and the unfolding is regular ($|\det K|$ from $2.6\times10^{3}$ to $2.2\times10^{5}$, smallest singular value 1.0 to 13); the critical multiplier equals $+1$ to within
$8.2\times10^{-9}$, and the next has modulus at most 0.74. Four of these five come in two pairs, near the ends of the upper-left and lower-left legs: the two cusps of each pair
are joined by a short segment of the fold curve and have cubic coefficients of opposite sign, as in a plane section of
a swallowtail \cite{arnold1985singularities}. Below the body the second flip curve runs down a
narrow neck, where the orbit has doubled; beyond it the orbit is stable again in a small separate window, whose tip
is the sixth cusp; two branches, run independently, trace that fold curve and its turn to within one curve step. Within the window of Fig.~\ref{fig:anatomy}(b), the edge of the followed domain lies within two grid steps of the
curves for all 232 neighboring pairs of periods 6 and 12 and all 459 pairs of a period-6 cell and a cell where the followed attractor was lost (median
distances 0.19 steps); over the whole followed domain the shares fall to 388 of 916 and 1{,}720 of 1{,}888 (medians 4.3
and 0.24 steps), because the far parts of the legs leave the traced curves, so their boundary is not resolved.
On a $31\times31$ grid of the continued orbit over the body (529 cells solved; the run reached 582 of the 961 cells
before its time limit), the leading multiplier is nonzero at
every solved cell. Its smallest modulus on the grid, $0.0063$ at the cell $(0.073053,-0.15120)$ on the line of strongest
contraction, belongs to a nearly imaginary leading pair,
$0.0004\pm0.0063i$, with a third multiplier of modulus $0.0051$: three multipliers of nearly equal small modulus, a
near-nilpotent point (the leading eigenvalues of the monodromy matrix nearly vanish, although the matrix itself does not). The dark line of strongest contraction runs from the flip crossing up the
body and bends to the upper left, and the nine cells with a complex leading pair all lie on its upper part. Such dark ditches take the place of Gallas's superstable curves (Fig.~\ref{fig:anatomy}(b); the measurements are drawn in
Fig.~\ref{fig:p6signs}). A leading multiplier, the largest in
modulus, can vanish only if all multipliers do, so instead of passing through zero it changes sign across the ditches,
negative toward the flip curves and positive toward the fold, in one of two ways: it meets the next multiplier, the two leave the real axis as a complex pair and return to it with
the opposite sign, or a multiplier of the opposite sign grows past it in modulus and becomes the leading one, a
hand-over. Its modulus is smallest there. The main ditch and a second such line, an arc around the lighter region to
its right, meet just below the kite's cusp C, as Gallas's superstable curves meet in his kite, at T, but at the top
they approach and turn away without crossing. On a finer patch of the continued orbit (steps $1.8\times10^{-5}$ in $\alpha$ and $4\times10^{-5}$ in $\log_{10}T$, 595
cells solved; Fig.~\ref{fig:p6signs}(a)) the arc closes over a pocket of negative leading multiplier. Between the arc and the main ditch runs the dimmer ditch,
and there the leading multiplier keeps its sign: along 26 cells across the dimmer ditch at $\alpha=0.07318$ it falls to
0.064 and rises again, while the next multiplier stays below 0.021 in modulus (Fig.~\ref{fig:p6signs}(b)). A patch ten times finer around
that minimum (225 cells, steps $1.8\times10^{-6}$ in $\alpha$ and $4\times10^{-6}$ in $\log_{10}T$) finds no sign change
either: the leading multiplier stays real and positive, between 0.056 and 0.075, and at $\alpha=0.07318$ reaches the
same minimum, 0.0638. It falls toward smaller $\alpha$, where the valley runs toward the main ditch. The dimmer
ditch is thus a valley in the size of the leading multiplier, not a line where it changes sign. Across the patch and the lines the continued orbit is attracting and keeps its six-token text, with Newton residuals
below $10^{-12}$, and fixed-start scans at the nearest parameters find period six in all but two cells, which are
unresolved; whether the fixed start reaches this very orbit was not checked at every continued point. We examined this separation on 852 continued orbits over the top of the body (steps $10^{-5}$ in
$\alpha$ and $2.5\times10^{-5}$ in $\log_{10}T$), on which a two-dimensional block of the monodromy, initialized from its leading Ritz directions, was tracked from cell
to cell by subspace overlap (Fig.~\ref{fig:delta2d}). The tracked $2\times2$ block has trace
$\tau$ and determinant $\delta$; within this block, the eigenvalue of larger modulus, or the real part of a complex pair, has the sign
of $\tau$. On this patch, every measured sign change of a real leading
multiplier lies within one cell of a sign change of $\tau$. This resembles the trace-zero skeleton that Fa\c{c}anha, Oldeman
and Glass use for shrimps in two-dimensional maps \cite{facanha2013endoskeletons}. Between the main
ditch and the arc, a local quadratic fit to $\tau$ gives a saddle at $(\alpha,\log_{10}T)=(0.073188,-0.152083)$, where it is still positive,
0.045: along the channel that joins the two lines the fitted trace is largest there, and across it smallest. The fitted saddle and the measured zero curves support a local trace-based explanation of why the main ditch and the arc
approach and turn away without crossing. The leading multiplier has a fitted saddle of the same kind nearby (0.067
at $(0.073195,-0.152138)$), and the finer patch lies just above it, on the flank toward the main ditch. The test we had
fixed before computing this grid asked instead for a saddle of $\delta$; none was found on the sampled patch under the specified fitting and filtering procedure. The determinant follows the steady
drift of the second multiplier, and its zero set misses the arc (Fig.~\ref{fig:delta2d}(b)); the trace was examined
after the grid was computed. In 38 of the 852 cells the tracking overlap falls below 0.9, indicating possible mixing with a third direction; the sign of $\tau$ matches that of the leading multiplier in all but three
cells. This local evidence does not establish an exact two-dimensional reduction of the full map.

Near the kite,
where two orbits exist, curves that belong to different orbits can cross in the parameter plane without any
degeneracy, as the superstable curves of Gallas's two coexisting fixed points do at the tail T of his cell (Fig.~\ref{fig:anatomy}(a)). The period-6 window thus
meets along its body the three numerical criteria we require before calling a window a shrimp (primitive periodic
orbits of the full state, their stability, and the organization of their boundary by continued fold and flip curves,
matched by the edge of the followed domain), with its kite and the kite's cusp located: a shrimp-shaped window of the
full map; the far parts of its legs remain unresolved.

\paragraph{A basin edge inside the kite.} The straight edge inside the kite (Fig.~\ref{fig:p6kite}(a)) is where the
fixed start crosses the boundary between the basins of the two attracting orbits. At $\alpha=0.072975$, where we
bisected it in $\log_{10}T$ to the float64 limit, the loop started on the edge comes close to a third period-6 orbit
and lingers there: Newton's method from that stretch converges to a saddle with a single multiplier outside the unit
circle, $\mu_u=1.2625$ (next modulus $7.8\times10^{-4}$). A start a distance $\delta$ off the saddle's stable manifold
should come within a distance that scales as $\delta^{0.97}$ of the saddle (the exponent $\chi_s/(\chi_s+\chi_u)$ of a
passage near a linear saddle, from its slowest stable multiplier and $\mu_u$) and leave it after a time that grows by
$6/\ln\mu_u=25.74$ steps for every factor $e$ by which $\delta$ shrinks. Started at 17 temperatures on each side of the
edge, from $10^{-6}$ to $10^{-14}$ away in $\log_{10}T$, the loop comes within a distance that scales with $\delta$
(log-log slopes 0.986 and 0.995) and leaves after times that grow by 25.78 and 25.75 steps per factor $e$, and it ends
on the strongly contracting orbit below the edge and on the other orbit above it. The edge also depends on where the loop starts: a $128\times128$ map of the same region, in which every cell starts
from the embedding of one randomly chosen token instead of the fixed start, shows no such edge. Bifurcation curves are properties of the map, whereas a fixed-start scan shows only the attractors reached from that
start; together with the coexisting orbits and the saddle found on the edge, its disappearance from the random-start map
supports a boundary between basins of attraction, which moves or disappears when the start changes. This is strong numerical evidence that, at this steering strength, the edge is where the fixed start lies on the
saddle's stable manifold; the region above it that reaches the weaker orbit is bounded by the fold curve on which that
orbit disappears. The edge meets the fold where the fixed start lands exactly on the orbit that disappears there: at
fold points located by Newton's method, the start's coordinate along that orbit's critical direction changes sign at
$\alpha=0.0729655$ (the tip to leading order in the saddle-node normal form \cite{kuznetsov2023elements}), and the boundary between the two outcomes, followed toward the fold, extrapolates to
$\alpha=0.0729651$ to $0.0729657$.

\paragraph{Heads.} We found no one-dimensional head, two superstable curves crossing, in any window we analyzed. As in the period-6 window, whose pseudo-superstable lines approach and turn away
(above), in the period-7 window under hostility steering (Qwen2.5-0.5B-Instruct, layer 12), the curves on which the continued orbit's
leading real multiplier changes sign, candidate superstable curves (a sign change between cells can also mean that
two real multipliers exchange dominance), form two arcs that approach each other near $(\alpha,\log_{10}T)=(0.0700,0.206)$
and turn away without crossing, a Type-B-like geometry \cite{facanha2013endoskeletons}. Where a one-dimensional map
would have its head, the leading multiplier becomes as small as the next and the two form a complex pair, so the
orbit's linearization there has two comparable leading directions\smref{Steering toward hostility}.

\begin{figure}[tb]
\centering
\includegraphics[width=\columnwidth]{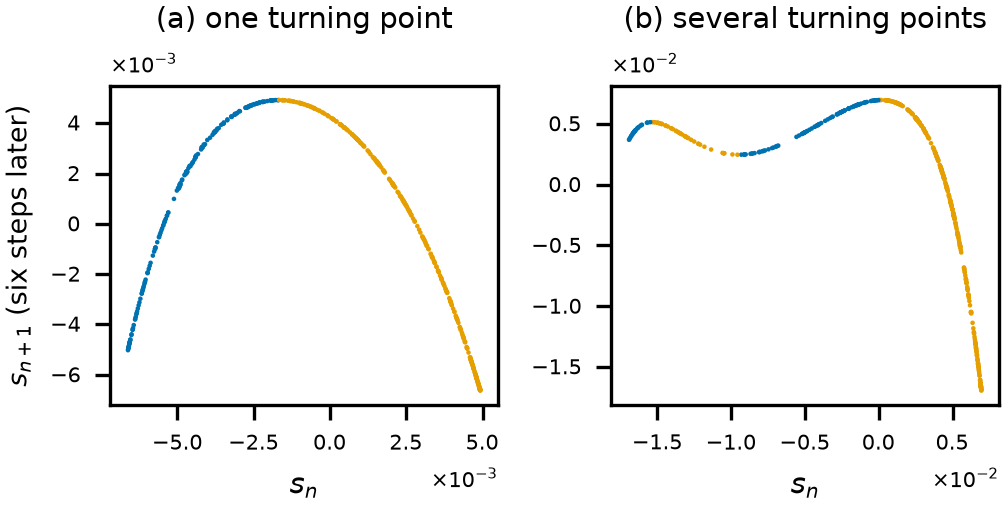}
\caption{Return maps at two cells of the six-band chaos around the period-6 shrimp (Qwen2.5-0.5B-Instruct, philosophy
direction at layer 12, float64; 1000 transient and 3000 recorded steps per cell): the position $s_{n+1}$ at a visit to
one band (along its first principal axis) against the position $s_n$ at the previous visit to it, six steps earlier; blue: visits
where the six-step derivative along the band is positive (L), orange: negative (R). (a) A cell of the window,
$(\alpha,\log_{10}T)=(0.07228,-0.14934)$, largest exponent 0.093: one turning point, as in a one-hump map.
(b) A cell beside the body's left flip curve, $(0.07269,-0.15274)$: more than one turning point.}
\label{fig:returnmaps}
\end{figure}

\paragraph{A one-hump return map.} Where the attractor writes the body's six-token text, the loop visits six thin
bands of the state space in turn, one per token, and returns to each band every six steps. Plotted along a band's
first principal axis, the position at one visit against the position at the previous visit often traces a single
hump, as in a one-dimensional map (Fig.~\ref{fig:returnmaps}(a)); return maps of this kind are how Jones traced the
shape of a periodicity hub of the R\"ossler system to unimodal maps, in part of its parameter plane
\cite{jones2012rossler}. We code each visit L or R by the sign of the six-step derivative along the band. At 48 of
the 58 chaotic six-band cells of a $48\times48$ map of the window, fewer than 5\,\% of the rarer symbol's visits are
misplaced; the other 10 cells (3 beside the body's left flip curve, 7 below the kite) fail this test, and at 7 of
them the return map rises and falls more than once (Fig.~\ref{fig:returnmaps}(b)). These maps support a scalar
picture of the window; they do not prove a reduction to one dimension.

\section{Synchronization of two coupled loops}
\label{sec:sync}

The prefix tells the model that it is talking with another AI, and the simplest dynamical version of two such agents
couples two copies of the loop, each reading a mixture of its own soft token and the other's. The theory of coupled chaotic maps gives, from the single loop's largest Lyapunov exponent alone, the coupling above
which the synchronized chaotic state is stable along a typical orbit. The measured growth of the difference between the
copies follows that theory closely; whether a given pair actually locks together also depends on how long it is
watched, how small a difference counts as none, where it starts and which attractor it reaches.

\begin{figure}[tbp]
\centering
\includegraphics[width=\linewidth]{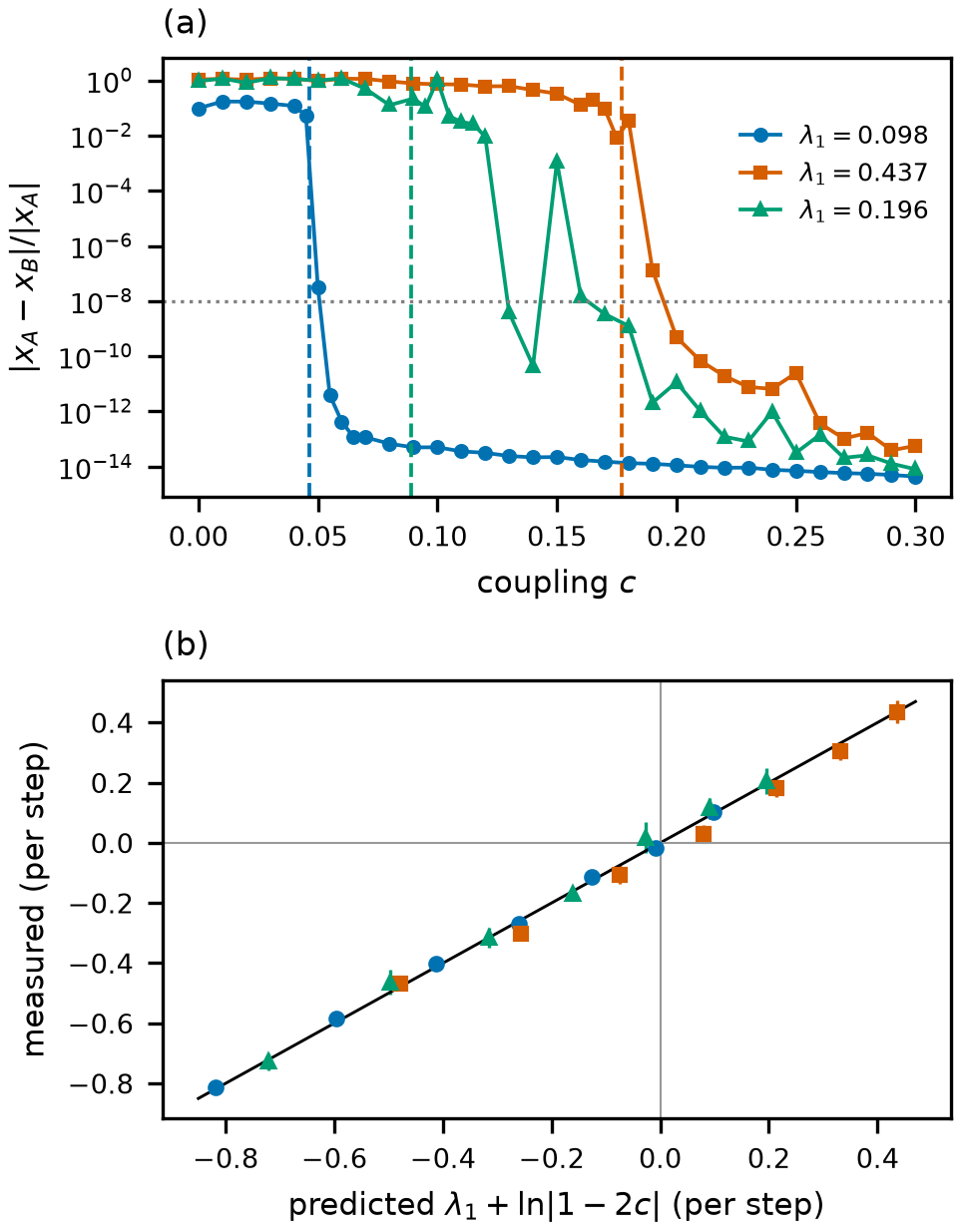}
\caption{Synchronization of two coupled copies of the loop, Eq.~(\ref{eq:coupled}), at three chaotic cells, labeled by
the leading exponent $\lambda_1$ of the single loop (Qwen2.5-0.5B-Instruct, philosophy direction at layer 12,
float64). (a) Mean relative difference $|x_A-x_B|/|x_A|$ over the second half of each run against the coupling $c$ (near starts;
3000-step runs, and 10\,000-step runs at the refined couplings); dashed lines, the predicted thresholds $c^*$ of
Eq.~\eqref{eq:cstar}; dotted line, the reference level $10^{-8}$. A pair counts as synchronized only if the difference
stays below the criterion throughout its last 500 steps, which the plotted mean does not show; the resulting thresholds
are in Table~\ref{tab:sync}. (b) Growth rate of the difference measured with renormalized twins on the coupled map
against the prediction $\lambda_1+\ln|1-2c|$, at seven couplings per cell (bars: two standard errors of the twin
estimates, without the uncertainty of $\lambda_1$; line: equality).}
\label{fig:sync}
\end{figure}

Both copies are the map of Sec.~\ref{sec:map}, with the same model, chat prefix, steering direction and cell
$(\alpha,T)$, in float64. They run as two rows of one batch. At every step, after the model
has produced both soft tokens and before they are fed back, the program replaces each by a mixture of its own and the
other's:
\begin{equation}
\begin{aligned}
x_A'&=(1-c)\,F(x_A)+c\,F(x_B),\\
x_B'&=(1-c)\,F(x_B)+c\,F(x_A).
\end{aligned}
\label{eq:coupled}
\end{equation}
At $c=0$ the copies run independently; at $c=1/2$ both receive the same vector.

Where the copies agree, $x_A=x_B$, the update is exactly $F$ for both, so a synchronized pair stays synchronized and writes identical text (the text is the most probable token at each
step, as elsewhere in this article). Whether this synchronized state attracts depends on a small difference $u=x_A-x_B$ between the
copies, which points across the synchronized state, or transverse to it. The difference obeys
$u'=(1-2c)\,[F(x_A)-F(x_B)]$, which for small $u$ becomes $u'=(1-2c)\,DF\,u$. At each step the loop's derivative $DF$
stretches the difference, on average by the factor $e^{\lambda_1}$, where $\lambda_1$ is the largest Lyapunov exponent
of a single loop, and the mixing multiplies it by $1-2c$. The difference therefore grows on average at the rate
$\lambda_\perp=\lambda_1+\ln|1-2c|$ per step, the transverse Lyapunov exponent. Without coupling, $\lambda_\perp=\lambda_1>0$ and the copies drift apart. As $c$ rises from 0 toward $1/2$, the
factor $1-2c$ falls from 1 toward 0, and $\lambda_\perp$ falls with it. It crosses zero at the critical coupling
\begin{equation}
c^*=\tfrac12\bigl(1-e^{-\lambda_1}\bigr),
\label{eq:cstar}
\end{equation}
where the mixing's contraction per step, $1-2c^*=e^{-\lambda_1}$, exactly cancels the loop's average stretching per
step, $e^{\lambda_1}$. For couplings between $c^*$ and $1/2$ (and, by the symmetry $c\to1-c$, up to $1-c^*$) a small difference shrinks on
average, and the synchronized state is stable along a typical orbit. The calculation gives the transverse-stability condition for our symmetric output coupling; it follows the
Lyapunov-stability framework of Fujisaka and Yamada \cite{fujisaka1983stability}, and
Ref.~\onlinecite{pecora1990synchronization} treats synchronization of chaotic systems more broadly. The more chaotic a single loop, the larger $\lambda_1$, and the
larger the share of the other copy's output that each copy must take in for this condition to hold. It concerns the
synchronized orbits that typical starts produce; other invariant sets inside the attractor may be transversely unstable
at the same coupling, and the condition says nothing about which attractor a pair reaches.

We coupled copies at three chaotic cells, one in the crescent's chaotic core ($\alpha=0.0652$, $\log_{10}T=-0.156$) and
two among the chaotic stripes ($\alpha=0.07484$, $\log_{10}T=-0.1226$, and $\alpha=0.0640$, $\log_{10}T=-0.115$). Over
3000 steps their largest exponents are $\lambda_1=0.098$, 0.437 and 0.196 per step, so Eq.~\eqref{eq:cstar} gives $c^*=0.046$, 0.177 and 0.089.
Copy~A started on the attractor after a 1000-step transient, and copy~B at a relative distance $10^{-3}$ from it in a
random direction. Each pair ran 3000 steps at couplings from 0 to 0.30 in steps of 0.01, and 10\,000 steps at eight
couplings near each threshold. A pair counted as synchronized when $|x_A-x_B|/|x_A|$ stayed below $10^{-8}$ over its
last 500 steps. To measure $\lambda_\perp$ directly, copy~B was started at relative distance $10^{-7}$ from copy~A and
reset to that distance after every step of the coupled map for 2000 steps. Far starts, with copy~B 500 steps further
along the attractor than copy~A (2000 steps, couplings in steps of 0.02), give a second kind of start. Other prefixes,
models and directions, larger ensembles of starts, asymmetric or delayed coupling and changes of precision were not
tested.

\begin{table}[tb]
\caption{Synchronization thresholds of the coupled copies for each protocol: the largest tested coupling that did not
synchronize and the smallest from which all larger tested couplings did, $(c_{\rm lo},c_{\rm hi}]$. Near starts: copy~B
at relative distance $10^{-3}$ from copy~A; far starts: copy~B 500 steps further along the attractor. Criterion: the
relative difference stays below $10^{-8}$ or $10^{-6}$ throughout the last 500 steps. $^{*}$A smaller tested coupling
also synchronized. $^{\dagger}$Synchronized pairs here include pairs locked onto a coexisting period-2 orbit of the single loop.}
\label{tab:sync}
\footnotesize
\begin{ruledtabular}
\begin{tabular}{lccc}
start, steps, criterion & cell 1 & cell 2 & cell 3 \\
\hline
$\lambda_1$ (per step) & 0.098 & 0.437 & 0.196 \\
predicted $c^*$ & 0.046 & 0.177 & 0.089 \\
\hline
near, 3000, $10^{-8}$ & $(0.04,0.05]$ & $(0.22,0.23]^{*}$ & $(0.15,0.16]^{*}$ \\
near, 3000, $10^{-6}$ & $(0.04,0.05]$ & $(0.19,0.20]$ & $(0.13,0.14]$ \\
near, 10\,000, $10^{-8}$ & $(0.045,0.05]$ & none up to 0.19 & $(0.12,0.15]$ \\
near, 10\,000, $10^{-6}$ & $(0.045,0.05]$ & $(0.18,0.19]$ & $(0.12,0.15]^{*}$ \\
far, 2000, $10^{-8}$ & $(0.20,0.22]^{*\dagger}$ & $(0.18,0.20]$ & $(0.14,0.16]$ \\
far, 2000, $10^{-6}$ & $(0.20,0.22]^{*\dagger}$ & $(0.18,0.20]$ & $(0.10,0.12]$ \\
\end{tabular}
\end{ruledtabular}
\end{table}

The measured $\lambda_\perp$ approximately followed $\lambda_1+\ln|1-2c|$, which ranged from $+0.44$ to $-0.82$ per
step: the largest discrepancy was 0.051 per step, up to four times the twins' own standard errors, which do not include
the uncertainty of $\lambda_1$ (Fig.~\ref{fig:sync}(b)). Whether pairs locked together depended on the protocol
(Table~\ref{tab:sync}). From near starts and on the $10^{-8}$ criterion, the first cell synchronized between $c=0.045$
and 0.05, at its predicted threshold; the second synchronized between 0.22 and 0.23 in 3000-step runs and not by 0.19 in
10\,000-step runs, but between 0.18 and 0.19 on the $10^{-6}$ criterion; and the third synchronized between 0.15 and
0.16 in 3000-step runs and between 0.12 and 0.15 in 10\,000-step runs, well above its $c^*=0.089$. Just below the second
cell's predicted threshold, at $c=0.16$ to 0.17, the difference showed intermittency of the on-off type
\cite{platt1993onoff}: it stayed below $10^{-8}$ for long stretches and rose above $10^{-3}$ 12 to 14 times in
10\,000 steps. In the third cell, at $c=0.105$ to 0.115, above $c^*$, the synchronized pair still burst from about
$10^{-14}$ to above $10^{-3}$, 5 to 10 times in 10\,000 steps. Bursts of this kind above the typical-orbit threshold are
known as attractor bubbling \cite{ashwin1994bubbling}: $c^*$ describes a typical orbit, but a chaotic attractor also
contains unstable periodic orbits, and along some of them differences may be stretched more than on average. While the
pair moves near such an orbit, a tiny difference can grow into a burst before the coupling pulls the copies back
together. Our bursts are consistent with this picture; the candidate transversely unstable orbits were not computed.
No noise was added. Two identical inputs in one batch give outputs that differ by up to $9\times10^{-16}$ relative, so
rounding plausibly supplies the tiny differences that grow, but its causal role was not isolated. Some pairs ended
elsewhere: from far starts some first-cell pairs synchronized onto a coexisting period-2 orbit of the single loop instead of its
chaotic attractor, and two first-cell pairs at $c=0.14$ and 0.20, above that cell's $c^*$, settled on
periodic orbits of the joint state without synchronizing. Below the thresholds, many pairs likewise stopped being
chaotic and locked onto periodic orbits of the joint state, of primitive period 6 to 12, on which the two copies write
different or staggered texts.

\paragraph{What the coupling shows.} The number that measures the loop's chaos, its largest Lyapunov exponent, also
sets how fast the difference between two coupled copies grows: the measured growth rate approximately followed the
prediction at every coupling, within 0.051 per step. The predicted threshold for a stable synchronized state rises with
the chaos, from 0.046 in the weakest to 0.177 in the strongest, and in the common near-start protocol the observed
thresholds rank the same way: the more chaotic each copy is alone, the more of the other's output it must take in. The
threshold concerns a typical orbit, however, and synchrony in practice depends on the observation time, the tolerance,
the start and the attractor reached. Near the threshold the copies alternated between long synchronized stretches and
bursts; in one cell bursts from the rounding level continued above the threshold; and from some starts the pair locked
onto a coexisting periodic orbit. Weak coupling, instead of synchrony, often produced order: the copies settled into a
periodic exchange on which they write different texts. These are copies of one fixed map; agents whose contexts grow
form a different system, which these measurements do not describe.

\section{Discussion}
\label{sec:discussion}
\paragraph{What is new.} Folds, flips, cascades, shrimp-shaped domains and Feigenbaum ratios are expected in smooth
dissipative maps and are found in experiments \cite{stoop2010real}, and we add no new bifurcation theory. What is new is the system and the use of its
own derivatives: the inference loops of released language models realize this structure, with concrete texts as
attractors, so that a bifurcation changes what the model writes; the factorization \eqref{eq:gain} gives sufficient
high-temperature contraction and conditional low-temperature bounds, and separates readout-side and network-side
contributions to growth along recorded orbits; and normal forms computed from the network predicted passage times, while window positions and doubling ratios
follow the laws of low-dimensional maps, with the failed criteria and unresolved cases reported alongside.

\paragraph{One picture.} The pieces fit together as follows (computed in float64 along the philosophy direction). As
$T\to0$, wherever each visited state has a unique most probable token, the loop is greedy decoding with a one-token
memory, and its eventual cycles are greedy token cycles. The periodic
orbits that organize the maps are not these cycles. The 2-cycle `` is it'' is a greedy cycle for no $\alpha$ in
$[0,0.12]$ (in steps of 0.0005). It exists because the fed-back vector mixes two competing continuations after each token (`` it'' and
`` there'', then `` is'' and ``'s''). At $\alpha=0.070$ it is stable only for $-0.5148<\log_{10}T<-0.1741$, between a flip
below and the fold whose ghost lines up the stripes (Sec.~\ref{sec:adding}). The crescent's period-3 loop is a
compromise of the same kind, between `` existence'' and `` question''. Its flips start the cascade
(Sec.~\ref{sec:cascade}), and its first doubling, continued to larger $\alpha$, is the band that crosses the stripes.
A single step of the map does not squeeze the state onto one direction: the second-strongest stretch of $DF$, its
second singular value, is 0.11 to 0.90 of the strongest. Composed over a period, however, the steps squeeze all
directions but one: the ratio of the two falls to 0.08 to 0.26, and to $3\times10^{-4}$ over twelve chaotic steps. The map has one Lyapunov exponent for each of its 896 directions, ordered from largest to smallest. The maps of
Sec.~\ref{sec:chaos} give only the largest exponent of each cell, measured from how fast a nearby twin drifts away. To
find the next few, we followed four small perturbations at once with the network's exact derivative, re-orthogonalizing
them at every step so that they do not all turn toward the fastest-growing direction, and averaged their logarithmic
growth rates. We did this at the three chaotic cells of Sec.~\ref{sec:sync}, over 600 steps each. At every one of
these cells the largest exponent is positive (0.11 to 0.42) and the next three are strongly negative ($-0.82$ to
$-2.01$). The finite-time Kaplan--Yorke dimension estimates, one plus the largest exponent divided by the magnitude
of the next, lie between 1.13 and 1.43. These estimates suggest low-dimensional dynamics at the three sampled cells,
with one expanding direction. They support exploring one-dimensional descriptions of these orbits, as Feigenbaum's
ratios also suggest, although we have not established a one-dimensional reduction of the whole map.

\begin{table*}[tb]
\caption{Each bifurcation, crossing and boundary named in this paper: what it requires and what we show.
Multipliers are those of the period-$p$ map $G=F^p$; $a$, $b$ and $c$ are coefficients of the center-manifold normal
forms. The numbers are given in the text.}
\label{tab:signatures}
\begin{ruledtabular}
\renewcommand{\arraystretch}{1.2}
\begin{tabular}{p{0.16\textwidth}p{0.29\textwidth}p{0.37\textwidth}p{0.11\textwidth}}
object & requires & we show & status \tabularnewline
\hline
\raggedright \textbf{fold} & \raggedright one multiplier at $+1$; nonzero coefficients $a$ and $b$ & \raggedright both at the period-2 fold, where a stable and an unstable orbit meet; passage times forecast at four more folds & \raggedright shown \tabularnewline
\colrule
\raggedright \textbf{flip} & \raggedright one multiplier at $-1$; nonzero crossing speed and cubic coefficient & \raggedright multiplier at $-1$ and the doubled orbit at every level of the cascade; cubic coefficient only at two other flips & \raggedright partly shown \tabularnewline
\colrule
\raggedright \textbf{cusp} & \raggedright $+1$ with $b=0$, $c\neq0$ and a regular unfolding & \raggedright the three conditions hold at the kite's cusp C and at five more located cusps & \raggedright shown \tabularnewline
\colrule
\raggedright \textbf{pseudo-superstable line} & \raggedright the leading multiplier changes sign without vanishing & \raggedright the tracked two-dimensional block's trace-zero curves and fitted saddle support a local explanation of the ditches' separation on the sampled patch & \raggedright supported numerically on the patch \tabularnewline
\colrule
\raggedright \textbf{fold--flip point} & \raggedright one orbit with $+1$ and $-1$ at once & \raggedright none at A or B & \raggedright none found \tabularnewline
\colrule
\raggedright \textbf{pseudo-crossing} & \raggedright curves of two different orbits cross & \raggedright two distinct orbits at R; different spectra at A and B & \raggedright shown \tabularnewline
\colrule
\raggedright \textbf{Feigenbaum ratio} & \raggedright ratios approach the constant of a universality class & \raggedright ratios reach 4.6688, against $\delta=4.6692$ & \raggedright consistent \tabularnewline
\colrule
\raggedright \textbf{basin boundary} & \raggedright changes with the initial state; both attractors exist on both sides of it; the start lies on a saddle's
stable manifold & \raggedright absent with random-token starts; both period-6 orbits exist on both sides; from it the loop nears a
period-6 saddle and leaves at the predicted time & \raggedright strong evidence \tabularnewline
\end{tabular}
\end{ruledtabular}
\end{table*}

\paragraph{What is and is not claimed.} Table~\ref{tab:signatures} lists each bifurcation, crossing and boundary named here, what it requires and what we
show. The objects studied here are invariant sets and bifurcations of an exactly
defined smooth map built from a model with fixed weights; we do not claim that ordinary sampled generation is chaotic
in this sense. Chaos labels are finite-horizon estimates: sampled chaos-labeled cells were checked with tangent derivatives,
several precisions and longer horizons (some labels were long transients), but asymptotic chaos has not been
established by a computer-assisted proof; three periodic orbits have such proofs (the period-8 orbit and a coexisting period-2 and period-3 pair), the cusps are float64 computations validated by step-size checks, and the Feigenbaum ratios are
finite-order values. The detailed orbit studies cover Qwen2.5 at 0.5B and 1.5B, and the scans use one prompt, apart from 22 conditions of the 0.5B plane (more in
the size tests\journalonly{; \sm{Limitations in detail}}).

\paragraph{Steering and predictability.} Small changes of the steering strength switch the model between distinct
repeating texts, chaos and coexisting attractors, so in continuous-feedback loops an activation-steering coefficient
\cite{zou2023representation,rimsky2024steering} should be treated like any control parameter of a nonlinear system (whether this carries over to ordinary generation is not tested here): its
bifurcation set should be characterized before an operating point is chosen, history dependence should be expected
near boundaries, and a smooth dose response should not be assumed. Where the loop is chaotic, a perturbation of
$10^{-6}$ grows to order one within roughly 15 to 50 steps at the largest unsteered exponents of the 7B and 8B models
(0.30 to 0.88 per step). Even a complete mechanistic account of what a single forward pass does would then not predict a long trajectory:
that would require knowing the state to far more digits than any interpretability description, which summarizes the
state by a few features, can supply. Step-by-step summaries can also mislead. Along the period-8 orbit of Table~\ref{tab:orbits}, every single step
stretches some direction, by a factor of 1.2 to 21 (the largest singular value of $DF$), yet over the whole cycle every
direction shrinks: the product of the eight step Jacobians has norm 0.059 to 0.644, depending on the point of the cycle
at which it starts. The directions do not line up, so what one step stretches a later step squeezes, and the orbit is
stable although no single step is contracting. What remains predictable are the invariant sets, their statistics and the boundaries between them, which
extends the analysis of trained recurrent networks through their fixed points
\cite{sussillo2013opening,maheswaranathan2019reverse} to the periodic orbits, chaos and bifurcations of a pretrained
transformer. Text can hide the dynamics: the emitted text records only the most probable token, so a monitor that reads it would
call the band chaos of Fig.~\ref{fig:hidden} periodic, although the probabilities that drive the loop do not repeat over the run.
These findings motivate testing whether interventions in continuous-feedback systems, including steering for safety,
alter finite-horizon sensitivity or task-relevant decisions as well as immediate outputs; such tests would complement
research on how latent-reasoning architectures affect chain-of-thought monitorability \cite{korbak2025monitorability}. We tested no safety property: dynamical instability is neither
misalignment nor evidence of it.

\paragraph{Beyond the one-token loop.} Feeding back more context makes chaos more common: when Qwen2.5-7B reads its last 8 or 16 soft tokens rather than
only the last one, it is chaotic at more than four in five of the temperatures tested, against one in three with a
single token\smref{Beyond the one-token loop}. Sampling weakens the
bottleneck: averaging a few sampled token embeddings instead of taking the expected one shortens the mean laminar phase (the nearly periodic stretches of Sec.~\ref{sec:adding}) at the smallest tested
distance from the fold from 120 steps to 7.8 with 8 samples and 10.5 with 64, and with single
samples such phases become rare, one to three per distance\smref{Bridge to sampled decoding}. Recurrent-depth models such as Huginn \cite{geiping2025recurrent}, which apply one shared block repeatedly until
the latent state settles, are a different kind of loop, without the readout factor of Eq.~\eqref{eq:gain}; there we
found convergence and an oscillatory instability, but no chaos (full report\smref{A second system: recurrent depth}). Feedback alone guarantees neither convergence nor chaos: the energy argument of Sec.~\ref{sec:theory} covers only
the tied readout, not the complete loop, which shows fixed points, cycles and chaos.

\paragraph{Conclusion.} A language model that reads its own soft output is a smooth discrete-time dynamical system, a smooth map iterated
step by step, and it behaves as one. Temperature and steering carry it from repeating text through period-doubling cascades into chaos with a positive
Lyapunov exponent. Its periodic windows open and close in folds and flips whose normal forms, computed from the
network's own derivatives, forecast how long trajectories linger near a vanished orbit; its cascades approach
Feigenbaum's constant; and its parameter plane holds shrimp-shaped windows organized as in the low-dimensional maps
where such windows were first found. Two coupled copies drift apart at approximately the rate the loop's Lyapunov exponent predicts.
None of this rests on a simplified model of the network: the orbits, multipliers and normal forms are those of the full
transformer in float64, and three of its periodic orbits are proved to exist with interval arithmetic. For the expected-embedding
feedback studied here with a fixed prompt, the derivative factorization applies to smooth networks with a linear output
head followed by softmax: its factors are inverse temperature, an embedding covariance for tied embeddings or a
cross-covariance for untied embeddings, and the network's Jacobian. Chaos with embedded periodic windows appears in all
three model families we scanned, showing that these feedback dynamics are not confined to one family of models.

\section*{Supplementary Material}
The full report of this study is available online, in the repository given under Data Availability
(\path{paper/llm/main.pdf}). It gives the details of every result summarized here, with its methods, checks, test designs fixed in advance and failed attempts; the levels of evidence and the
computer-assisted proofs; the further results on the hostility direction, the cusps and the steering controls; the
replication in other models and the size study; and the results on memory, sampled decoding, two coupled loops and the
recurrent-depth model Huginn.

\begin{acknowledgments}
The author thanks Jennifer Hoffman and Emin Arakelian for stimulating discussions about large language models and
for encouraging the publication of this work.
This work received no external funding; the author paid for all computation. This work was carried out with AI coding and research agents (Anthropic's Claude Code and OpenAI's Codex) working under
the author's direction. The agents wrote and ran the code, executed the experiments, checked results against the raw
data and drafted text; other AI systems made independent read-only checks of the results and the manuscript. Test
designs fixed in advance, failed attempts, audits and checksummed records of every computation are kept in the project
repository. The
author, whose doctoral research in nonlinear dynamics included work on a periodicity hub of the R\"ossler attractor
\cite{jones2012rossler}, directed the study, reviewed the work and takes full responsibility for its content.
\end{acknowledgments}

\section*{Author Declarations}
\subsection*{Author Contributions}
Timothy D. Jones is the sole author.

\section*{Data Availability}
The full report of the study (\path{paper/llm/main.pdf}), the code, the test designs fixed before data collection with
their audits, and checksummed records of all computations are openly available at \url{https://github.com/tdj28/neural-shrimp}. The
interval-arithmetic certificates are in the same repository, with the verification code, the exact map and input
specifications, and the release records and hashes of the native certificate evidence: the period-8 certificate in
\path{codex/certify-llm} (release record \path{certificates/llm-p8-certificate-final-001.json}), and the coexisting
period-2 and period-3 orbits, their token loops and basin neighborhoods in \path{codex/coexistence-task33}.

\bibliography{references}

\end{document}